\documentclass[sigconf, nonacm]{acmart}

\usepackage{graphicx}
\usepackage{subcaption}
\usepackage{multirow}
\usepackage{booktabs} 
\usepackage{comment}
\usepackage{enumitem}

\definecolor{amazon}{rgb}{0.0,0.42,0.31}

\usepackage[normalem]{ulem}
\newcommand\rsout{\bgroup\markoverwith{\textcolor{red}{\rule[0.5ex]{2pt}{0.4pt}}}\ULon}

\AtBeginDocument{%
  \providecommand\BibTeX{{%
    \normalfont B\kern-0.5em{\scshape i\kern-0.25em b}\kern-0.8em\TeX}}}

\setcopyright{acmcopyright}
\copyrightyear{2022}
\acmYear{2022}
\acmDOI{10.xxxx/xxxxx.xxxxx}

\acmConference[AsiaCCS '22]{AsiaCCS '22: 17th ACM ASIA Conference on Computer and Communications Security (ACM ASIACCS 2022)}{May 30--June 3, 2022}{Nagasaki, Japan}
\acmBooktitle{17th ACM ASIA Conference on Computer and Communications Security (ACM ASIACCS 2022, May 30--June 3, 2022, Nagasaki, Japan}
\acmPrice{15.00}
\acmISBN{978-1-4503-XXXX-X/XX/XX}

\begin{document}

\title{User configurable 3D object regeneration for spatial privacy}

\author{Arpit Nama}
\authornote{The first two authors contributed equally to the paper.}
\affiliation{%
  \institution{University of Sydney}
  \city{Sydney}
	\country{Australia}
}
\email{anam9745@uni.sydney.edu.au}

\author{Amaya Dharmasiri}
\authornotemark[1]
\affiliation{%
  \institution{University of Moratuwa}
  \city{Colombo}
	\country{Sri Lanka}
}
\email{170131R@uom.lk}

\author{Kanchana Thilakarathna}
\affiliation{%
  \institution{University of Sydney}
  \city{Sydney}
	\country{Australia}
}
\email{kanchana.thilakarathna@sydney.edu.au}

\author{Albert Zomaya}
\affiliation{%
  \institution{University of Sydney}
  \city{Sydney}
	\country{Australia}
}
\email{albert.zomaya@sydney.edu.au}

\author{Jaybie Agullo de Guzman}
\orcid{0002-2816-7721}
\affiliation{%
  \institution{University of the Philippines Diliman}
  \city{Quezon City}
    \country{Philippines}
}
\email{jaybie.de.guzman@eee.upd.edu.ph}

\renewcommand{\shortauthors}{A. Nama, A. Dharmasiri, K. Thilakarathna, A. Zomaya, and J. A. de Guzman}

\begin{abstract}
\textit{Environmental understanding} capability of \textit{augmented} (AR) and \textit{mixed reality} (MR) applications and devices are continuously improving through advances in sensing, computer vision, and machine learning. Various AR/MR applications have been developed that demonstrate such capabilities: i.e. scanning a space using a handheld or head mounted device and capturing a digital representation of the space that are undeniably accurate copies of the real space. However, these capabilities impose privacy risks to users: personally identifiable information can leak from captured 3D maps of the sensitive spaces and/or captured sensitive objects within the mapped space.  
Thus, in this work, we demonstrate how we can leverage 3D object regeneration for preserving privacy of 3D point clouds. That is, we employ an intermediary layer of protection to transform the 3D point cloud before providing it to the third-party applications. Specifically, we use an existing adversarial autoencoder to generate copies of 3D objects where the likeness of the copies from the original can be varied. To test the viability and performance of this method as a privacy preserving mechanism, we use a 3D classifier to classify and identify these transformed point clouds: i.e. perform \textit{super}-class and \textit{intra}-class classification. To measure the performance of the proposed privacy framework, we define privacy, $\Pi \in [0,1]$, and utility metrics, $Q \in [0,1]$, which are desired to be maximized. Experimental evaluation shows that the proposed privacy framework can indeed variably effect the privacy of a 3D object by varying the privilege level $l \in [0,1]$: i.e. if a low $l < 0.17$ is maintained, $\Pi_1, \Pi_2 > 0.4$ is ensured where $\Pi_1, \Pi_2$ are the super- and intra-class privacy. Lastly, the proposed privacy framework can ensure relatively high intra-class privacy and utility, i.e. $\Pi_2 > 0.63 \text{ and } Q > 0.70$, if the  privilege level is kept within the range of $0.17 < l<0.25$. 
\end{abstract}



\begin{teaserfigure}
  \includegraphics[width=0.9\textwidth]{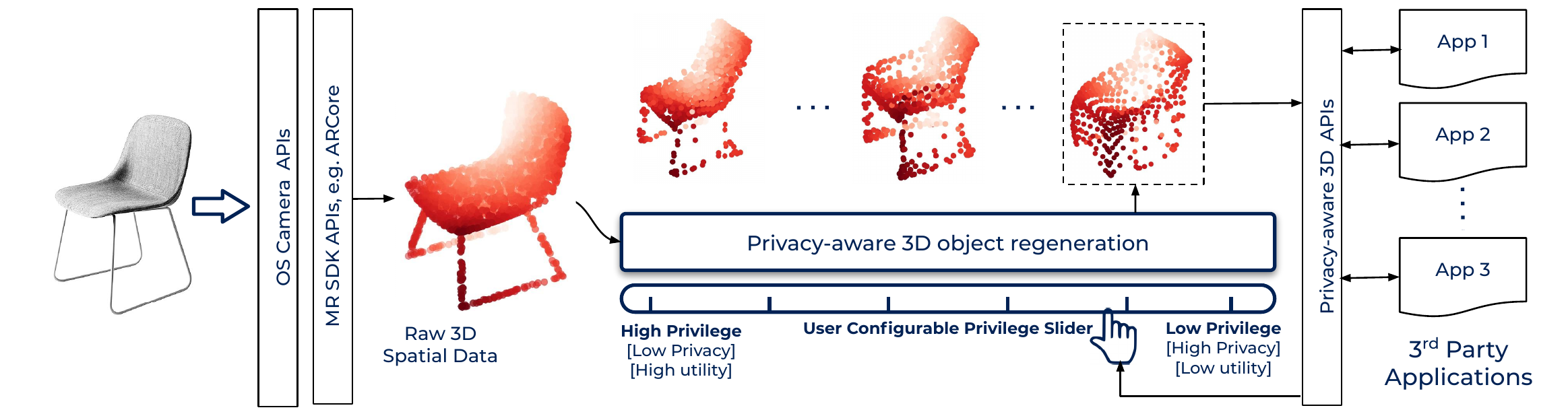}
  \caption{Overview of the proposed user configurable privacy framework for 3D point cloud sharing.} 
  \Description{Teaser Figure description}
  \label{fig:framework-teaser}
\end{teaserfigure}


\maketitle


\section{Introduction}
Advanced camera systems capable of mapping 3D spatial information is increasingly becoming popular among consumer devices.\footnote{For example, Apple Inc. has released the iPad Pro in 2020 with LiDAR, and subsequent iPhone releases in 2021 also has LiDAR.} 
These improved camera systems support the recent developments in \textit{mixed reality} (MR) technology particularly in improving \textit{environmental understanding capability}: i.e. scan a space using a handheld or head mounted device and, then, capture a digital \textit{representation} or \textit{mapping} (i.e. spatial maps or 3D point clouds) of the space that are undeniably accurate copies of the real space. This enables MR applications to deliver realistic MR experiences, i.e. rendering virtual augmentations that seemingly inhabit the real and physical space.\footnote{AR/MR SDKs of popular consumer platforms: Apple ARKit (\url{https://developer.apple.com/documentation/arkit}), Google ARCore (\url{https://developers.google.com/ar/}), and Windows MR API (\url{https://developer.microsoft.com/en-us/windows/mixed-reality}).}

Concurrently, 
3D object detection and classification have also become highly accurate and ubiquitous~\cite{qi2017pointnet, qi2017pointnet++}. While training and running these deep learning models is still compute and resource intensive, it is only a matter of time when such models can be run on mobile devices to be leveraged for environmental understanding and, hence, deliver better MR experiences. Nonetheless, the relatively memory light 3D point clouds can be sent to and processed remotely, i.e. on the edge or cloud, where deep learning models can be run more efficiently.

While these advanced environmental understanding empowers app developers to design innovative MR service, the access to accurate 3D mapping imposes unprecedented risks to user privacy~\cite{deguzman2019firstlook, deGuzman2021unravelling}. With the current access controls in place with these MR devices, regardless of the processing strategy---whether local or remote---the raw 3D point clouds are freely accessed by third party applications. As such, adversarial MR applications can easily access and collect 3D spatial maps which may include sensitive information: whether the captured space itself is sensitive, or portions of the captured space is sensitive. Aside from objects being detected, geometric and photo-metric characteristics of the objects, semantic relationships between objects, etc. may eventually be leveraged by adversaries which can potentially reveal personally identifiable information: whether about the device user, or the user's association, linkage, or relationship to these spaces, objects, or other users. 

Furthermore, the problem of user privacy is exacerbated by the fact that users do not have any control over the visual, and, now, spatial information shared to third parties. Due to binary user permission structure on current devices, the user can either share raw visual and spatial data or completely block permission to the camera. Not just users, device manufactures are also suffering from the lack of finer granular controlled release of spatial data, e.g. Oculus has completely blocked access to 3D point cloud data in order to mitigate unwanted data leakage.\footnote{\url{https://developer.oculus.com/reference/mobile/v29/}} While this is good for user privacy, it significantly limits innovation and hinders the true potential of hardware capability. 






In light of that, we propose a \emph{privacy-aware 3D spatial data sharing framework} providing the user with the ability to control the level of privacy protection required. We introduce a \textit{3D object regeneration} mechanism as an \textit{intermediary layer} of protection before sharing the 3D point cloud to third party applications as shown in Fig.~\ref{fig:framework-teaser}. 
For the 3D point cloud regeneration, we employ a Hypernetwork-based Adversarial Autoencoder (AAE) and vary the level of \emph{likeness} of the regenerated 3D object point cloud from the original. The level of likeness is obtained from the \emph{privilege} level assigned by the user (cf. privilege slider in Fig.~\ref{fig:framework-teaser}). That is, a higher likeness means a higher privilege is afforded to third-party applications, and lower, otherwise. The challenge is to release minimal amount of uniquely identifiable information regarding the 3D object, while also providing enough spatial information to allow the application to function with high quality of service. We first formulate adversarial models, as well as privacy and utility metrics. We then evaluate the viability of the proposed user configurable 3D regeneration approach as a privacy mechanism using a large 3D object database. 



In this paper, we present the following contributions:
\begin{enumerate}[leftmargin=*]
	\item Formalize the 3D privacy problem against a two-fold attacker which tries to reveal the super-class and intra-class labels of a 3D object; hence, we correspondingly define two privacy metrics, i.e. $\Pi_1 \in [0,1]$ for super-class, and $\Pi_2 \in [0,1]$ for intra-class.
	\item Define the utility metrics on which the privacy frameworks' output will be measured against. We present two utility metrics based on two common MR functionalities: bounding box utility $Q_1$, and plane anchoring utility $Q_2$.
	\item Leverage existing 3D adversarial autoencoders (AAE) as an intermediary privacy-preserving mechanism for reducing 3D privacy leakage from 3D point clouds before exposing or sharing to third party applications.
	\item Evaluated the viability of the AAE-based privacy-preserving 3D regeneration using existing 3D deep neural network classifier to classify and identify the regenerated point clouds at various attacker privilege scenarios.
	\item Experimental evaluation shows that the proposed privacy framework can indeed variably effect the privacy of a 3D object by varying the privilege level $l$. For example, if low privilege level $l < 0.17$ is maintained, both $\Pi_1, \Pi_2 > 0.4$ is ensured.
	\item Lastly, the proposed privacy framework can ensure relatively high intra-class privacy and utility, i.e. $\Pi_2 > 0.63 \text{ and } Q > 0.70$, at a privilege level range of $0.17<l<0.25$.
\end{enumerate}

\vspace{-1mm}\paragraph{Overview} The rest of the paper is organized as follows. \S\ref{sec:related-work} presents the related work with emphasis and focus on recent 3D privacy work. \S\ref{sec:Problem formulation} formalizes the 3D privacy problem followed by the formulation of the proposed 3D privacy framework, i.e. privacy-aware 3D point cloud regeneration, in \S\ref{sec:3D Privacy Framework}. The experimental setup is detailed in \S\ref{sec:Experiment Setup}, while the results are discussed in \S\ref{sec:Results}. We discuss the further implications of our work in \S\ref{sec:future_work}. 
Finally, we conclude the paper in \S\ref{sec:conclusion}. The appendix shows supplementary plots and figures.


\section{Related Work}\label{sec:related-work}

In this section, we present a comprehensive exposition of the recent related work aimed at privacy for 3D data, particularly related to MR. Then, we present a breadth of relevant 3D classification approaches followed by the adversarial attacks towards these classifiers.


\vspace{-1.5mm}\subsection{3D privacy} While most of the earlier work in MR privacy were primarily aimed at visual (i.e. image and video) data protection \cite{deguzman2018security}, a handful of recent work have started to point out the potential privacy risks with the ubiquity of 3D ``scanning'' devices \cite{friedman2000value,Roesner2014, bloom2017self}, and highlighted the ease-of-access of third-party applications and services over these 3D point cloud or spatial data \cite{jana2013enabling, figueiredo2016prepose, deguzman2019safeMR}. 


\vspace{-2mm}\paragraph{3D leakage and further inference} On the MR-specific case, the actual risks brought about by indefinite access to spatial data captured by mobile MR devices has been looked into. Specifically, on how adversaries can easily identify spaces from captured 3D point cloud data (using Microsofot Hololens) \cite{deguzman2019firstlook, deguzman2020conservative}. Other works have employed machine learning to reveal original scenes from 3D point cloud data by leveraging on additional visual information, i.e. SIFT (scale-invariant feature transforms) and other photometric information \cite{pittaluga2019revealing}. Subsequent work focused on a privacy-preserving method of pose estimation to counter the scene revelation \cite{speciale2019privacy, Geppert2020ppSFM}: instead of using 3D point cloud data, 3D line clouds are used during pose estimation to obfuscate 3D structures. The same line cloud approach has also been leveraged to translate the features to affine space to ensure that the features are privacy-preserving \cite{ Dusmanu2020Affine}. 
Another work also followed the same strategy in delivering privacy-preserving SLAM (synchronous localization and mapping) \cite{Shibuya2020ppSLAM}. However, it has also been demonstrated that line clouds are still not immune from further inference as dense line clouds can still reveal 3D structures which allows for original scenes to be approximately revealed \cite{Chelani2021lineclouds}. Moreover, the 3D line cloud approach only addresses specific functionalities such as pose-estimation and does not present its viability for surface or object detection which may be necessary for delivering immersive experiences in MR (e.g. to detect object on which virtual objects can be rendered or ``anchored'' onto). Thus, with current MR processing techniques, it would still be necessary for 3D point cloud data to be exposed but, perhaps, with some privacy-preserving transformations to hide sensitive content and prevent further spatial recognition.

\vspace{-2mm}\paragraph{3D encryption} An earlier approach proposed the encryption of 3D point clouds using chaotic mapping to conserve utility for view point histograms \cite{Jin2016_3Dcrypt}. However, this method is reliant on sharing of keys, and once the keys---and, hence, access---are provided, no further privacy protection is ensured. Furthermore,  this 3D encryption method was not tested against standard encryption approaches which have been employed in secure multi-party computation of 3D user models from encrypted user physiometric information for virtual clothe try-on \cite{sekhavat2017privacy}. Nonetheless, this 3D encryption approach together with other standard encryption methods (e.g. AES, and RSA) are essential part in providing protection along the entire MR service pipeline.

\vspace{-1.5mm}\subsection{3D classification and reconstruction}

\vspace{-1mm}\subsubsection{3D classification}\label{rrw:3d-methods}
Now, as we explore risks associated with exposed 3D spatial data, we looked into the various methods that are utilized for 3D shape analysis, classification, and recognition. Early approaches employ description and inference by, first, computing 3D features and, then, utilizing these features for inference, say, by matching: examples include 3D Shape histograms \cite{ankerst19993d}, parametrized shape distributions \cite{osada2002shape, ohbuchi2005shape}, vector field similarity \cite{dinh2008measuring}, shape histograms \cite{johnson1998surface,johnson1999using}, curvature self-similarity \cite{huang2012point}, or heat-kernels \cite{sun2009concise}. 

On the other hand, the latest 3D recognition approaches have employed machine learning and has been more successful in 3D classification as well as segmentation. PointNet was the first to use deep learning, i.e. deep neural network (DNN), for 3D classification with point clouds as input \cite{qi2017pointnet}. PointNet++ presented some improvements in the underlying DNN architecture \cite{qi2017pointnet++}. 
Latest developments on neural network-based 3D classification include the following: identifying \textit{salient} regions of objects that inform about their class \cite{Zheng2019saliency}, improving robustness to rotations \cite{Kim2020_3Drotation}, leveraging alternating 3D projections for point cloud alignment or registration \cite{Ranade2021mapping}, and improving 3D segmentation processing time through point dropping \cite{Zhang2020Slimmer}.


\vspace{-1mm}\subsubsection{3D adversarial approaches: attacks and regeneration} 

Unsurprisingly, 3D neural networks (NNs) are also susceptible to 3D adversarial point clouds which can confuse 3D NNs. Recent efforts focused on demonstrating 2D adversarial approaches to 3D \cite{Liu2019_2Dto3Dadversarial, Xiang2019_3dadversarial}, and providing defenses against such adversarial attacks \cite{Zhou2019DUPnet, Zhang2019defensepointnet}. These efforts are motivated by mission-critical 3D information which requires that 3D NNs be more resilient to adversarial attacks when classifying 3D objects, e.g. detecting pedestrians for self-driving cars.

However, we argue that for most non-critical 3D applications, users may desire to have less of their environments be recognizable to machines (and, hence, potential adversaries). Adversarial autoencoders (AAEs) can be used to transform and generate 3D point clouds to potentially preserve privacy. Recent works focused on improving AAE using hypernetworks \cite{Spurek2020hypercloud}, or using curvature functions with canonical shapes as reference \cite{Ye2021curvaturedensity}. 
\emph{However, none of these methods are originally designed for privacy-preservation, but can be leveraged and modified to be utilized as such as we propose in this work.}

\section{Problem formulation - 3D object privacy in Mixed reality}
\label{sec:Problem formulation}

\setlength{\belowdisplayskip}{1pt} 
\setlength{\abovedisplayskip}{0pt}
\setlength{\abovedisplayshortskip}{0pt}

In the following sections, we introduce the representation of 3D spatial data used in our analysis, formulate the adversary models, establish the intended utilities of the MR application, and define the privacy metrics that we consolidate in our proposed solution.


\vspace{-1.5mm}\subsection{3D spatial data}
Current commodity MR devices use one or combination of SLAM (simultaneous localization and mapping), SFM (Structure from motion) or visual odometry to capture and represent its surrounding 3D environment \cite{saputra2018visual}.
This 3D structure representation usually comes in the form of a point cloud which contains a set of unordered and permutationally invariant points, each representing a point, i.e. an edge or part of a surface, in space (in terms of x-y-z coordinates) 
with no relationship defined among the points. 
Although the mapped 3D surroundings or objects might subsequently be converted to different representations such as 3D voxel grids, meshes, or oriented point clouds, the 3D point clouds which is often the direct output of the captures still encode substantial amount of data, and can even be directly used for processing.
In this paper, we conduct our analysis on 3D objects while representing them as unoriented point clouds (with no normal vector attached). 



\begin{table}[t]
\caption{Table of notations}
\vspace{-2mm}
\label{tab:notations}
\resizebox{\columnwidth}{!}{%
\renewcommand{\arraystretch}{1.2}
\begin{tabular}{lp{8cm}}
\toprule
\centering\textbf{Notation} & \textbf{Description} \\ \midrule
 $S$ & Point cloud\\
 $k_i$ & $i^{th}$ super-class label\\
 $m_{j|i}$ & $j^{th}$ intra-class label given a super class $k_i$\\
 $l$ & variable privilege level, i.e. $l \in [0,1]$ \\
 $S^*$ & Query point cloud with unknown labels \\
 $\Bar{S}$ & Regenerated Point cloud\\
 $h_{1}(S^*)$ & hypothesis 1 for query $S^*$ \\
 $h_{2}(S^*)|k$& hypothesis 2 for $S*$, given super-class $k$\\
 $\gamma_k^*$ & super-class label subsets formed by hypothesis 1 \\
 $\eta_{m_k}^*$ & intra-class label subsets formed by hypothesis 2\\
 $\rho_1 ,\rho_2$& hypothesis subset sizes as a fraction of reference set size\\
 $\sigma_1, \sigma_2$ & attack score functions for hypothesis labelling\\
 $\mathcal{L}_{h_1\left( S^{*}\right)}( \gamma_k^*)$ & Likelihood function of hypothesis 1 \\
 $\mathcal{L}_{h_2\left( S^{*}\right)}( \eta_{m_k}^*)$ & Likelihood function of hypothesis 2 \\
 $\mathcal{L}_{h_2\left( S^{*}\right) |k_x}( \eta_{m_k}^*)$& Likelihood function for hypothesis 2 given the super-class $k_x$ \\
 $\Pi _{1}\left( S^*\right),\Pi _{2}\left( S^*\right)$ & super- and intra-class privacy metrics for query $S^*$  \\
 $G_{v}(S)$ & $v^{th}$ MR functionality for point cloud $S$ \\
 $Y_{v}(S)$ & output of $v^{th}$ MR functionality for input $S$ \\
 $Er_{v:S,\overline{S}}$ & Transformation error 
 for $v^{th}$ utility  \\
 $Q_{v,S:\overline{S}}$& $v^{th}$ utility metric for regenerated $\Bar{S}$ w.r.t $S$ \\ \bottomrule
\end{tabular}}
\vspace{-5mm}
\end{table}

\vspace{-1.5mm}\subsection{Adversary models} \label{sec:Adversary models} 


To formally define the problem and the privacy and utility metrics used in this work, we develop upon the formalization introduced in \cite{deguzman2020conservative} and \cite{deguzman2019firstlook}. We introduce $S$, a representation of an object captured from an MR user's surrounding as the key entity of the 3D object privacy problem. We identify $S$ with two key attributes; it belongs to a super-class object category based on the practicality (e.g. chair, and table), and it represents a unique member within its super-class in terms of 3D shape, structure, make, brand, etc. (e.g. product code 604.169.25 from IKEA 2021 catalog).

The MR application offers a functionality $G$, in which the object representation $S$ is processed in some way to generate the output $Y$. The utility of the 3D object $Q$ 
gives a measurement of the quality of the produced $Y$ which is required for the functionality $G$ to be derived.
While an adversarial attacker formulates a hypothesis $h$ on a certain attribute of $S$, the privacy preserving regeneration mechanism generates a privacy-preserved version $\Bar{S}$ corresponding to input $S$, expecting to reduce the accuracy of $h$ formulated by the attacker. For the convenience of the reader, we include all the notations used in our work in Tab. \ref{tab:notations}.

\vspace{-1mm}\subsubsection*{Formalizing the attacker} In this paper, our main focus is on object inference attacks. 
More specifically, the adversarial attacker that we present in this work attempts to identify a 3D object represented by a point cloud in two main hierarchical levels: (1) \textit{Super-class inference}, i.e. identifying the high-level category of the object according to its practicality; and (2) \textit{Intra-class inference}, i.e. identifying the unique object within the super-class.
For each scenario, the attacker attempts to narrow down the assumed super-class/intra-class to a subset of previously encountered references.



Let $S$ 
be a known point cloud within the reference set where $k_i$ is the object's super-class label (i.e. $k_i \in K$ where $K$ is the set of all the super-class object labels), and $m_{j|i}$ is its intra-class label within super-class $k_i$ (i.e. $m_{j|i} \in M_{i}$ and $M_{i}$ is the set of all the intra-class object labels of super-class $k_i$). Note that $k_i, m_{j|i}\in \mathbb{Z}^{+}$.

Now, the attacker makes two-fold hypothesis about a query point cloud $S^*$ with unknown super-class label $k_i^*$ and intra-class label $m_{j|i}^*$, i.e. we denote with a $^*$ the unknown labels to the attacker:
$$Attacker:S^{*} \rightarrow h_{1,\rho_{1}}  , h_{2,\rho_{2}}$$\vspace{-2mm}

\paragraph{Super-class hypothesis- $h_1$} The attacker makes a super-class hypothesis $h_{1,\rho_{1}}$ that the unknown super-class label $k_i^*$ is within a ``basket'' subset $\gamma_{k}^* \subset K$ of super-class labels:
\vspace{-1mm}
\begin{equation}\label{Eq:hypothesis 1}
    h_{1,\rho_{1}}: k_i^* \in \gamma_{k}^* = \{k_1, k_2, ...\},\text{ where } \rho_1 = \frac{|\gamma_{k}^*|}{|K|}
\end{equation}

The parameter $\rho_1$ is the ratio of the size of the super-class hypothesis basket $\gamma_{k}^*$ to that of the known super-class set $K$.
When $\rho_1|K|=1$, i.e. $\gamma_{k}^*$= $\{k_x\}$, $k_x$ is the top-1 super-class hypothesis.

\vspace{-1mm}
\paragraph{Intra-class hypothesis- $h_2$} Next, the attacker makes an intra-class hypothesis $h_{2,\rho_{2}}$ given the $h_1$ super-class label $k_x \in K$, that the unknown intra-class label $m_{j|x}^*$ is within a set $\eta_{m_x}^* \subset M_x$ of intra-class labels:
\vspace{-1mm}
\begin{equation}\label{Eq:hypothesis 2}
    h_{2,\rho_{2}} | k_x: m_{j|x}^* \in \eta_{m_{x}}^* = \{m_{1|x}, m_{2|x}, ...\}, \text{ where } \rho_2 = \frac{|\eta_{m_x}^*|}{|M_{x}|} 
\end{equation}

Similarly, $\rho_2$ is the ratio of the size of the intra-class hypothesis set $\eta_{m_x}^*$ of the known intra-class set $M_x$ of super-class $k_x$.

As $\rho$ increases, 
the attacker will identify the query $S^*$ as a member of a larger set of possible super-classes and unique-objects. Therefore as $\rho$ increases, the attacker's ability to precisely reidentify the object decreases, hence the privacy threat that the attacker can pose to the user decreases. Hence, an attacker will ideally try to narrow down the hypothesis to a subset with as small $\rho$ as possible.

\vspace{-1.5mm}\subsection{MR application utility}\label{sec:utility metrics}

Various functionalities are offered by MR applications that leverage 3D data from the user environments. Such functionalities vary in terms of level of details required in the 3D scans to effectively deliver the intended functionality. A given functionality $G_{v}$, requires a particular output $Y_{v}$ from the device obtained by applying some operation on the raw point cloud scans: i.e. $G_{v}( S) \rightarrow Y_{v,S}$.

The privacy mechanism takes original point cloud $S$ as input, and generates a privacy-preserved version $\Bar{S}$ of it. Thus, we can define a utility metric for the regenerated point cloud with respect to the original point cloud by using an \textit{output transformation error} for the particular application functionality, where $Er_v$ is a distance/error function that quantifies the dissimilarity between $Y_{v,S}$ and $Y_{v,\overline{S}}$: 
i.e. $G_{v}: Er_v\left(Y_{v,S},Y_{v,\overline{S}} \right) \rightarrow Q_{v,S:\overline{S}}$. 

In this work, we analyze two main MR application utilities that require varying levels of details in the output $Y_{v,S}$, and we define utility metrics separately for each. 

\vspace{-1mm}\paragraph{Bounding box utility} The most basic and naive MR functionality requires the general location of an object, often to be treated as an obstacle in the 3D environment. We define this functionality as $G_{1}$ whereas the output $Y_{1,S}$ represents the general location and size of the object represented by $S$ in the form of a \textit{3D bounding box}: i.e. $G_{1} \rightarrow Y_{1,S} =( x_{S} ,y_{S} ,z_{S} ,x_{0:S} ,y_{0:S} ,z_{0:S})$, where $x,y,z$ indicate the dimensions of the bounding box in 3 axes, and $x_{0} ,y_{0} ,z_{0}$ indicate the position of the center of the bounding box.

Next, we formulate the utility metric for the MR functionality $G_{1}$. Here, we can skip the intermediate error function and directly formulate the utility metric to quantify the similarity of object bounding boxes in terms of 3D Intersection-over-union (IoU). 

\begin{equation}\label{Utility metric 1}
    Q_{1:S,\overline{S}} = IoU(Y_{v,S},Y_{v,\overline{S}})
\end{equation}

\paragraph{Plane anchoring utility} The second and most widely used MR functionality $G_{2}$ involves anchoring 3D virtual objects onto 2D surfaces of real objects. For this, we define the function output: i.e. $G_{2} \rightarrow \{ Y_{2,S} :p,\ Y_{2,\overline{S}} :\overline{p} \}$, where $p$ and $\Bar{p}$ indicate the point clouds representing the most prominent 2D plane of the object from the original and regenerated point clouds separately.

3D object anchoring in MR could be either \textit{static} or \textit{dynamic}. In static anchoring, anchored object is stationary, and MR functionality only requires the position and normal vector for one point on the plane. Hence, we formulate the \textit{transformation error} in terms of the \textit{angle and distance between the planes} $p$ and $\Bar{p}$. For \textit{dynamic} anchoring, anchored object moves around the plane, therefore functionality requires the area and the point distribution of the 2D plane. Hence, we additionally use \textit{difference in area}, and \textit{difference in point distributions} between $p$ and $\Bar{p}$ to define the error function.  

\vspace{-2mm}
\begin{equation}\label{Utility metric error 2}
 \begin{array}{l}
 \begin{array}{l}
Er_{2:Static} =0.5 |1-\overrightarrow{n_{p}} .\overrightarrow{n_{\overline{p}}} |+0.5 |\underline{r}_{p} -\underline{r}_{\overline{p}} |\\
Er_{2:Dynamic} =0.25 |1-\overrightarrow{n_{p}} .\overrightarrow{n_{\overline{p}}} |+0.25 |\underline{r}_{p} -\underline{r}_{\overline{p}} |+\\
\ \ \ \ \ \ \ \ 0.25|A( p) -A(\overline{p}) |+0.25CD( p,\overline{p})
\end{array}
\end{array}
\end{equation}

Here, $\overrightarrow{n_{p}}$ indicates the unit normal vector to the 2D plane represented by $p$; thus,  $|1-\overrightarrow{n_{p}} .\overrightarrow{n_{\overline{p}}} |$ calculates the angle between the planes. 
$\underline{r}_{p}$ indicates the perpendicular vector from origin to the plane $p$; thus, $|\underline{r}_{p} -\underline{r}_{\overline{p}} |$ measures the distance between planes.  $A(p)$ measures the area of the 2D convex-hull formed by $p$, and $CD(p,\overline{p})$ returns the Chamfer distance between two point clouds $p$ and $\Bar{p}$, reflecting the point distribution discrepancy between them. 


Before deriving the utility function $Q_{2:S,\overline{S}}$ from $Er_{2}$, we separately apply \textit{min-max normalization} for each component 
$$|1-\overrightarrow{n_{p}} .\overrightarrow{n_{\overline{p}}} |,\\\ |\underline{r}_{p} -\underline{r}_{\overline{p}} |,\\\ |A( p) -A(\overline{p}) |,\\\ CD( p,\overline{p})$$
The modified error functions obtained after the said normalization are as follows:
$Er_{2:Static}\rightarrow Er'_{2:Static}$, and 
$Er_{2:Dynamic}\rightarrow Er'_{2:Dynamic}$.
Finally, we formulate the second (twofold) utility metric based on the functionality $G_2$:

\begin{equation}\label{Utility metric 2}
    \begin{array}{l}
    Q_{2:Static} =1-Er'_{2:Static}\\
    Q_{2:Dynamic} =1-Er'_{2:Dynamic}
    \end{array}
\end{equation}






\vspace{-1.5mm}\subsection{Privacy metrics}\label{sec:privacy metrics}

We characterize the privacy introduced by the proposed mechanism for a particular 3D point cloud $S$ with respect to the type of adversarial inference attempted upon by the attacker:
(1) Super-class privacy, and (2) Intra-class privacy.
We formulate the privacy metrics for both of the above scenarios using the expected error in hypotheses $h_{1}$ and $h_{2}$. Furthermore, since the hypotheses are given as sets of labels, i.e. $\gamma_k^*$ and $\eta_{m_k}^*$, the privacy metrics are also dependent on the size of these label sets which are represented as fractions of the reference sets by the parameters $\rho_{1}$ and $\rho_{2}$.

\vspace{-1mm}\subsubsection{Super-class privacy}

In trying to infer a query point cloud $S^{*}$, the attacker generates a likelihood function for $\gamma_k^*$: 

$$\mathcal{L}_{h_1\left( S^{*}\right)}( \gamma_k^* ) =\sum _{\forall {k} \in \gamma_k^* } \sigma_{1}\left( {k} ,S^{*}\right)$$

$\sigma_{1}\left( {k} ,S^{*}\right)$ is an attacker super-class score function for a hypothesis super-class label $k$ and for the unknown point cloud $S^*$, i.e. it represents the inclination of the attacker to match the query $S^*$ with the particular super-class, such that those values for each super-class adds up to 1:

\begin{equation}\label{Eq:scoring function 1}
    \sigma_{1}\left( {k} ,S^{*}\right)\leq 1, \sum_{\forall k \in K} \sigma_{1}({k}, S^*) = 1
\end{equation}






Then, the super-class privacy $\Pi_1$ for a query point cloud $S^{*}$ with actual super class label $k_i$ can be expressed in terms of the expected error in hypothesis set $\gamma_k^*$,

\begin{equation}\label{Eq:privacy metric 1}
\begin{aligned}
    \Pi_{1}\left( S^{*}\right) = & \delta(k_i,\gamma_k^* )\mathcal{L}_{h_1\left( S^{*}\right)}( \gamma_k^* ) \\
    & +( 1-\delta(k_i,\gamma_k^* ))( 1-\mathcal{L}_{h_1\left( S^{*}\right)}( \gamma_k^*))
\end{aligned}
\end{equation}

where $\delta(s, S)$ is a set membership function, i.e. 
$$
\delta(s,S) = \{ 1 \text{ if } s \in S; 0 \text{ if } s \notin S\}$$

\vspace{-1mm}\subsubsection{Intra-class privacy}

Next, given a particular hypothesis super-class label $k_x$, attacker generates a likelihood function for the intra-class hypothesis label set $\eta_{m_x}^*$:

$$\mathcal{L}_{h_2\left( S^{*}\right)}( \eta_{m_x}^*) =\mathcal{L}_{h_1\left( S^{*}\right)}(k_x) \cdot \mathcal{L}_{h_2\left( S^{*}\right) |k_x}( \eta_{m_x}^*)$$

\vspace{-2mm}
$$\text{where } \mathcal{L}_{h_2\left( S^{*}\right) |k_x} ( \eta_{m_x}^*) =\sum _{\forall m \in \eta_{m_x}^*} \sigma_2\left( m ,S^{*} |k_x\right)$$

$\sigma_{2}({m}, S^*|k_x)$ is an attacker intra-class score function for a hypothesis object label $m$ and for the unknown point cloud $S^*$ within a given super-class $k_x$, i.e. it represents the inclination of the attacker to match the query $S^*$ with the particular object m, given that $S*$ belongs to the super-class $k_x$, such that those values for each object within $k_x$ adds up to 1:

\begin{equation}\label{Eq:Scoring function 2}
    \sigma_{2}({m}; S^*|k_x) \leq 1 \text{, and }\sum_{\forall m \in M_x} \sigma_{2}({m}; S^*|k_x) = 1.
\end{equation}




The intra-class privacy $\Pi_2$ for a query point cloud $S^*$ with actual super class label $k_i$ and intra-class label $m_{j|i}$ is also calculated as the expected error of the attacker hypothesis where the error of the super-class hypothesis affects the intra-class hypothesis: 

\begin{equation}\label{Eq:privacy metric 2}
\begin{aligned}
    \Pi_2 \left( S^* \right) & = \sum_{\forall k \neq k_i} \mathcal{L}_{h_1\left( S^* \right)} (k) \\
    & + \mathcal{L}_{h_1(S^*)}(k_i) \{ \delta(m_{j|i}, \eta_{m_i}^*) \mathcal{L}_{h_2\left( S^* \right)|k_i}(\eta_{m_i}^*) \\
    & +  (1-\delta(m_{j|i}^*, \eta_{m_i}^*))(1 -\mathcal{L}_{h_2\left( S^* \right)|k_i}(\eta_{m_i}^*)) \}
\end{aligned}
\end{equation}

We leverage the privacy metrics and the utility metrics formulated in this section to measure and evaluate the performance of the proposed framework in various contexts as given in \S\ref{sec:Results}.



    


\section{3D Privacy Framework}
\label{sec:3D Privacy Framework}

Our framework utilizes an Auto-Encoder based point cloud  
regeneration mechanism to introduce privacy preserving versions for point cloud objects. 

\subsection{AAE-based point cloud regeneration}

The deep neural network used in our framework is called HyperCloud \cite{Spurek2020hypercloud} which consists of an Adversarial Auto-Encoder (AAE) and a target network. The AAE works as a hyper-network ($\mathcal{H}$) for the target network ($\mathcal{T}$) i.e. the AAE returns the weights of the target network as output. The AAE takes a point cloud as the input ($\mathcal{X}$) which passes through an encoder ($\mathcal{E}$) and a decoder ($\mathcal{D}$) to return the weights ($\theta$) of the target network. 

In addition to $\mathcal{E}$ and $\mathcal{D}$, the AAE also includes another neural network called discriminator. The goal of the discriminator is to learn to distinguish between the latent space ($\mathcal{E(X)} \sim \mathcal{Z}$) and the prior distribution ($\mathcal{P}$) by using a regularization loss term in the cost function. The target network with the weights $\theta$ i.e. $\mathcal{T_{\theta}}$ accepts a baseline distribution (3D ball or sphere) as input and generates a reconstruction ($\mathcal{Y}$) of the original 3D point cloud. The cost function includes the regularization term and the Chamfer pseudo-distance between original point cloud and the reconstruction as follows:
\begin{equation} \label{hypercloud_cost}
    cost(\mathcal{X}; \mathcal{E}, \mathcal{Y}) = CD(\mathcal{X};\mathcal{Y}) + Reg(\mathcal{E(X)}; \mathcal{P})
\end{equation}

Where $CD$ refers to the Chamfer pseudo-distance and $Reg$ refers to the regularization loss term. The Chamfer pseudo-distance can alternatively be replaced with the Earth Mover's distance. If $\mathcal{P}$ is chosen to be some known distribution like Gaussian, KL-divergence or adversarial training can be used for the regularization term.

\begin{figure}[t]
	\centering
	\vspace{-2mm}
	\includegraphics[width=0.9\columnwidth]{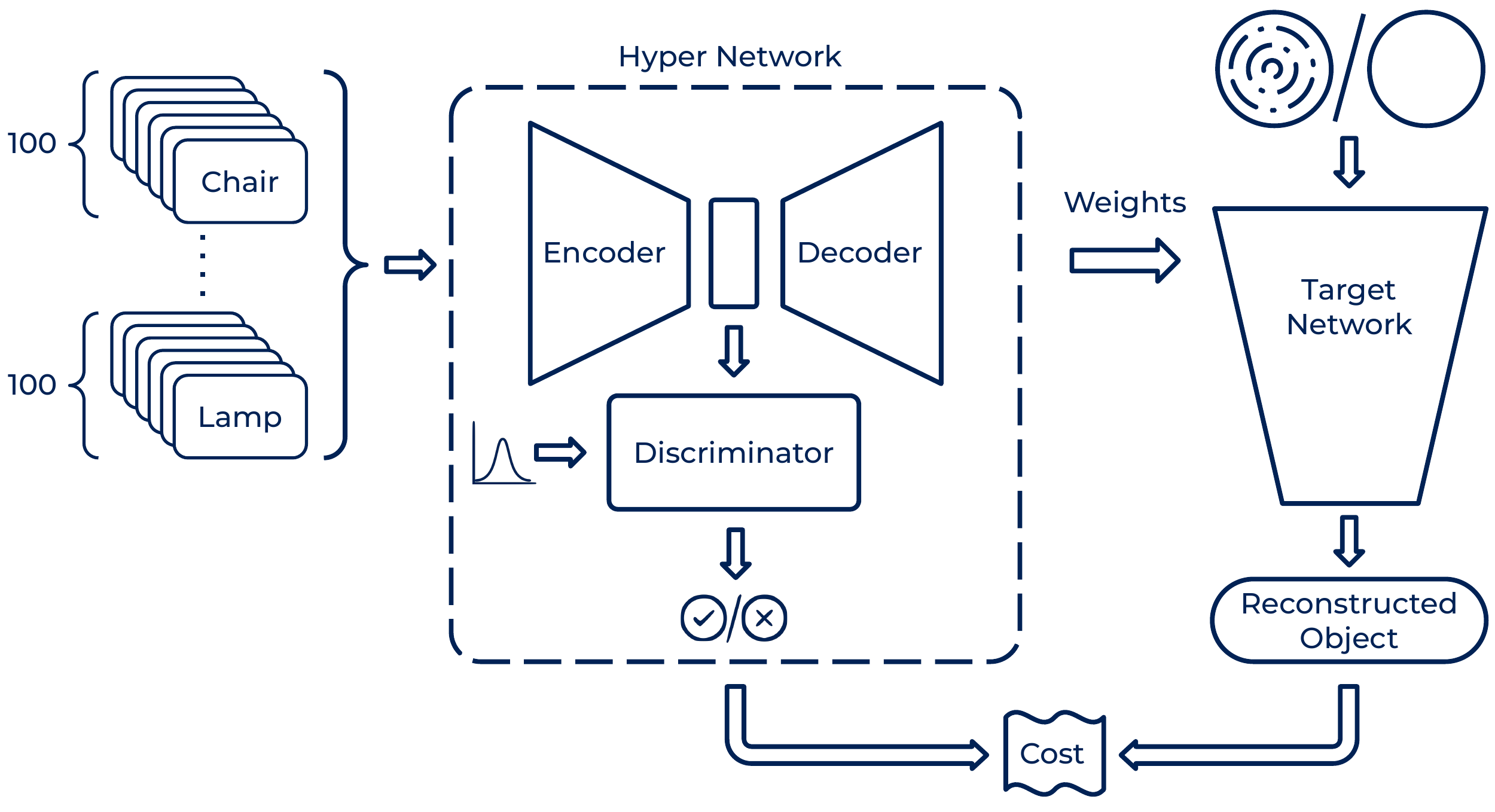}
	\vspace{-3mm}
	\caption{Training the proposed framework}
	\vspace{-3mm}
	\label{fig:framework-training}\vspace{-1mm}
\end{figure}

\begin{figure}[t]
	\centering
	\includegraphics[width=0.9\columnwidth]{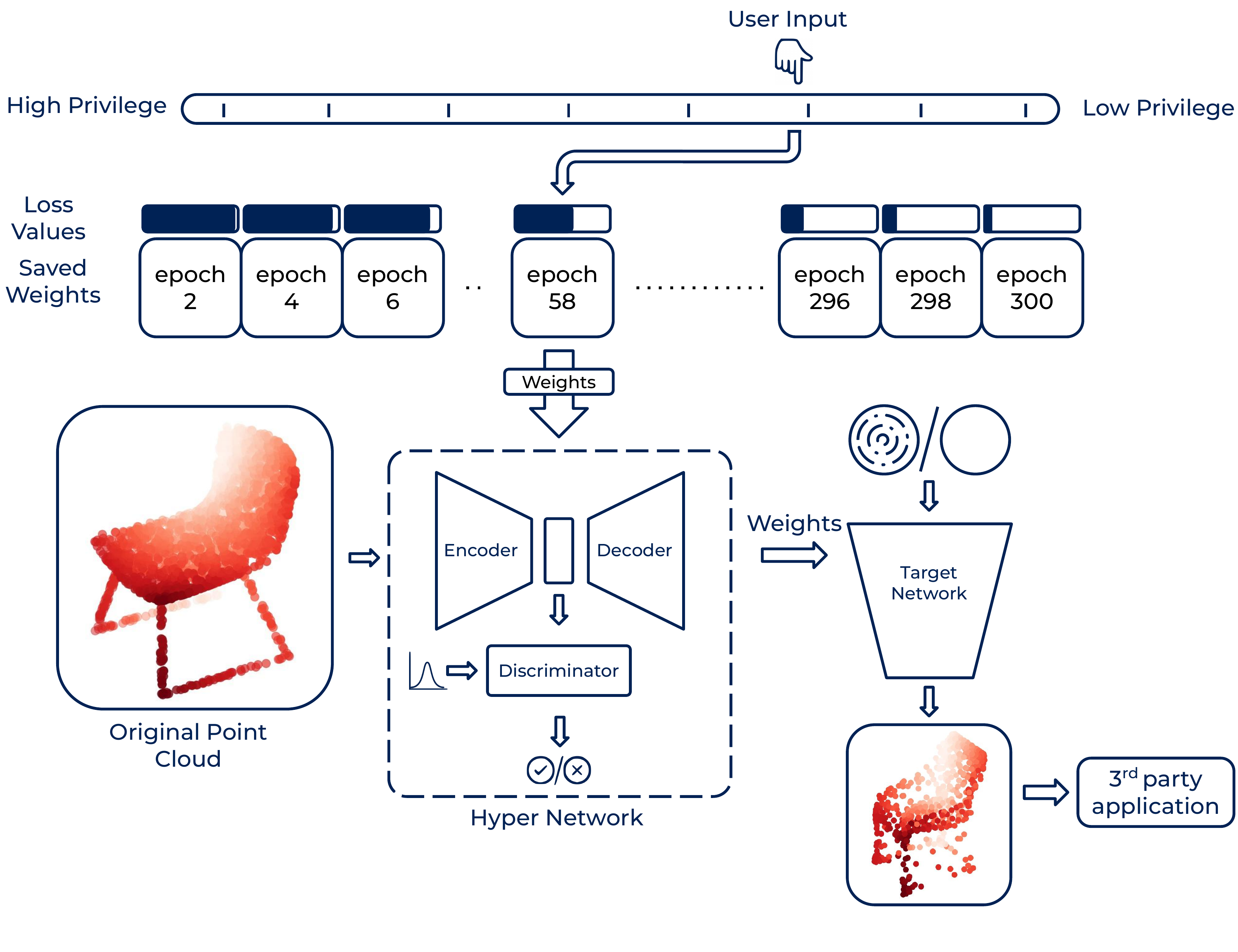}
	\vspace{-3mm}
	\caption{Privacy-preserving regeneration using the proposed framework}
	\vspace{-3mm}
	\label{fig:framework-usage}
\end{figure}

\vspace{-1.5mm}\subsection{User configurable point cloud regeneration} \label{sec:point-cloud-transformations-for-different-privacy-levels}

Our framework allows users to control the privilege, denoted by $l$, given to the MR application to acquire 3D data using a slider mechanism. 
When high privilege is given, the third party MR application can collect 3D information from the user's surrounding with high level of detail, hence can cause larger privacy risk to the user. Similarly, when low privilege is allowed for the application, the amount of detail in the 3D data is limited by the proposed mechanism, thereby minimizing the possible privacy risk. The slider ranges from 0 to 1 where 0 refers to no privilege allowed, while 1 refers to allowing maximum privilege. 

As shown in \autoref{hypercloud_cost}, the cost function directly depends on the Chamfer pseudo-distance between the original point cloud and the reconstruction using the architecture; so, as the loss function decreases, Chamfer pseudo-distance also decreases implying better and more accurate reconstructions. We have used this property of the loss function to generate different $l$ privilege levels allowed by the user to the MR application. 
The regenerations with higher loss (usually referring to lower epoch in the architecture training) are used for lower privilege settings, and those with lower loss (referring to lower epoch in the architecture training) are for higher privilege settings. 

During the training of our privacy framework as shown in Fig. \ref{fig:framework-training}, we save the weights of the hypernetwork to create a weight bank corresponding to each training epoch; which is translated to the allowed privilege level $l$ as explained above.
As we can see in Fig. \ref{fig:framework-usage}, the user input on the privilege slider determines the weights that will be loaded into the hypernetwork from the weight bank. After successfully loading of the weights, the original point cloud is passed through the privacy framework which produces a regeneration with corresponding $l$. This regeneration is shared with the 3rd party application.

\section{Experimental Setup}
\label{sec:Experiment Setup}

\subsection{Dataset}\label{sec:dataset}


For this work, we created a custom dataset by selecting a subset of the dataset introduced by \cite{achlioptas2017latent_pc} where 3D CAD models of ShapeNetCore \cite{shapenet2015} were sampled with (area) uniform sampling. Our dataset has 10 super-classes of household items i.e. \textit{bathtub, chair, monitor, sofa, table, bench, cabinet, bed, bookshelf and lamp}, with 100 unique object instances within each super-class. (We have avoided large non-household objects like airplane, ship, etc. as it would be impractical for MR devices to capture these for a normal user.) Every point cloud object was represented by 2048 points, and was normalized. 


\vspace{-1.5mm}\subsection{Adversarial attacker realization/simulation}\label{sec:Adversarial attacker realization/simulation}

In \S\ref{sec:Adversary models} , we discussed the theoretical framework and the hypothesis formulation used by an adversarial attacker to infer details about a point cloud at both \textit{super-class} and \textit{intra-class} level. In this section, we will discuss the practical simulation of such attackers in various scenarios and using them to evaluate our privacy framework effectiveness. In this work, we used the \textit{PointNet} neural network architecture as the 3D point cloud classifier \cite{qi2017pointnet} to realize the object reidentification attacks posed by an adversarial attacker. 

\vspace{-1mm}\subsubsection{Attacker hypothesis formulation}\label{sec:Attacker hypothesis formulation}

The attackers that we realize in the form of 3D point cloud classifiers formulate two consecutive hypotheses for a query point cloud as explained in \autoref{Eq:hypothesis 1} and \autoref{Eq:hypothesis 2}. In this section, we will elaborate the hypothesis formulation process followed by \textit{PointNet} based classifiers. 

At the last activation function of \textit{PointNet}, the \textit{Softmax} layer outputs the probability distribution over predicted reference classes. We use this \textit{Softmax} output to define the scoring function in \autoref{Eq:scoring function 1}, $\sigma_{1}\left( {k} ;S^{*}\right)$ for each super-class $k$ in the reference set.
Then, the attacker formulates the hypothesis in \autoref{Eq:hypothesis 1} by solving the following optimization. 
\begin{equation}\label{Eq:hypothesis 1 optimization}
    h_{1,\rho_1}= \gamma_k^* =\ \underset{\overline{\gamma_k } }{\mathrm{Argmax}} \ \sum _{k \in \overline{\gamma_k }} \sigma_{1}\left( k,S^{*} \right) 
\end{equation}

Similarly in the next step, we use the \textit{Softmax} output of the intra-class classifier trained on a particular super-class $k_x$ to formulate the scoring function in \autoref{Eq:Scoring function 2}, $\sigma_{2}({m}; S^*|k_x) $ for each object $m$ in the intra-class reference set of super-class $k_x$.
We formulate the hypothesis in \autoref{Eq:hypothesis 2} by solving the following optimization:
\begin{equation}\label{Eq:hypothesis 2 optimization}
    h_{2,\rho_2}|k_x =\eta_{m_x}^* =\ \underset{\overline{\eta_{m_x}}}{\mathrm{Argmax}} \ \sum _{m \in \overline{\eta_{m_x}}} \sigma_{2}\left( m ,S^*|k_x\right) \\ 
\end{equation}

Note that the cardinalities of the subsets $\gamma_k^*$ and $\eta_{m_x}^*$ obtained in \autoref{Eq:hypothesis 1 optimization} and \autoref{Eq:hypothesis 2 optimization} are parameterized by $\rho_{1}$ and $\rho_{2}$ as shown in \autoref{Eq:hypothesis 1} and \autoref{Eq:hypothesis 2}. We evaluate the performance of our attackers with respect to these parameters in \S\ref{sec:Results}. Moreover, for this experimental setup, the super-class reference set size $|K|=10$, and intra-class reference set size $|M_k|=100$ for each $k\in K$.

\vspace{-1mm}\subsubsection{Attacker classification}
Both \textit{super-class} and \textit{inter-class} adversarial inference is done with respect to a set of reference 3D objects that are previously encountered and learnt by the attacker in one or many of the following ways: from publicly available data, from the data intercepted previously from the same user or different users, or data obtained via other potentially colluding applications. In a situation where the query object is not present in the references (i.e. not previously encountered by the attacker), the attacker still attempts to identify the query as the structurally closest (most similar) reference object.


We define the following four types of attackers based on the approach of acquiring and the quality (in terms of the `exposed' privilege level $l$) of 3D data in their reference sets:
\begin{itemize}
    \item $\bf{J_1}$ - \textbf{public-aware attacker} -Using publicly available data as a set of reference 3D objects
    \item $\bf{J_2}$ - \textbf{low-privilege-aware attacker} -Intercepting data from users with low privilege settings 
    \item $\bf{J_3}$ - \textbf{medium-privilege-aware attacker} -intercepting data from users with medium privilege settings 
    \item $\bf{J_4}$ - \textbf{high-privilege-aware attacker} -Intercepting data from users with high privilege settings 
\end{itemize}

In \S\ref{subsec:all-attackers-n-1}, we present our analysis on how the proposed framework effectively preserve privacy against these different attackers. 

\begin{figure}[t]
	\centering
	\vspace{-2mm}
	\includegraphics[width=0.9\columnwidth]{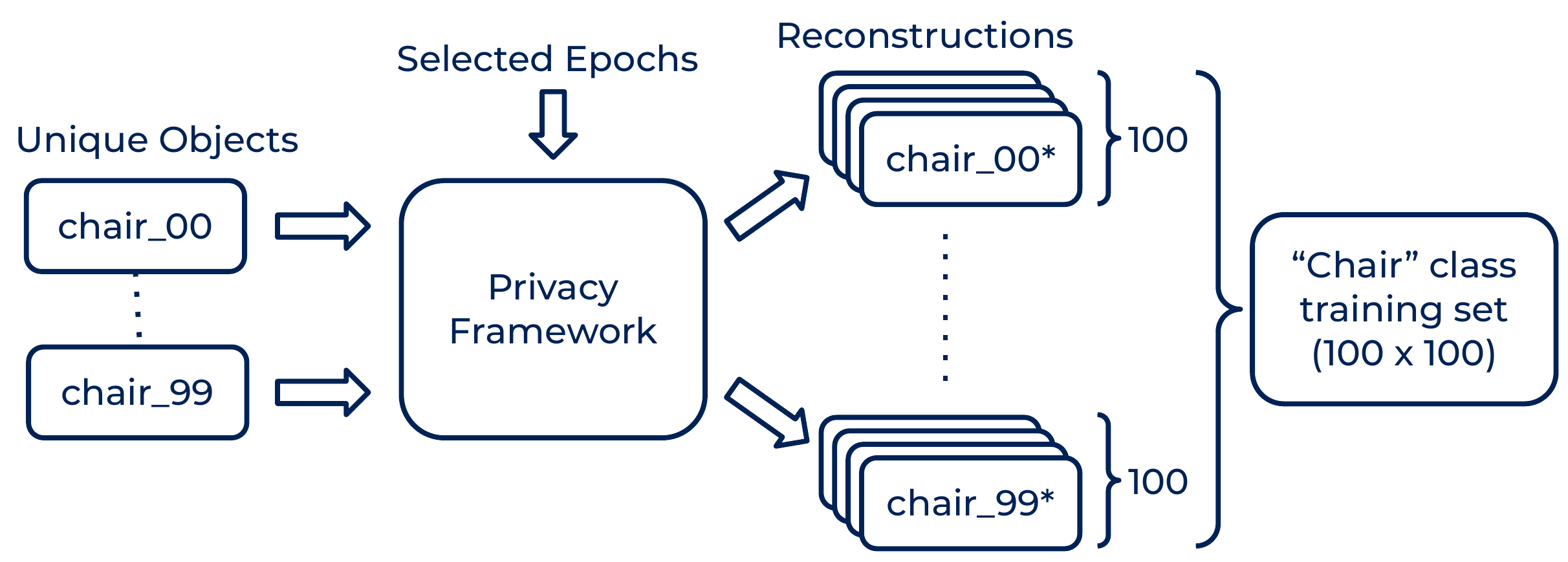}
	\vspace{-3mm}
	\caption{Reference (training) dataset generation for the super-class "Chair"}
	\label{fig:attacker-training-dataset}
	\vspace{-5mm}
\end{figure}

\vspace{-1mm}\subsubsection{Simulating attacker learning process}
We created the reference dataset for the attacker $J_1$ by using publicly available dataset as mentioned in \S\ref{sec:dataset}. 
We created 100 normalized augmentations for each unique object in all the super-classes by adding random rotations and Gaussian noise to complete the reference (training) dataset.
For attackers $J_2, J_3 ~\text{and} ~J_4$, we used our privacy framework to create 100 regenerations for each unique object in every super-class at various $l$ privilege levels (allowed to each attacker) to create the reference (training) datasets.


As explained in \S\ref{sec:point-cloud-transformations-for-different-privacy-levels}, the training epoch number, which the hypernetwork weights are loaded from, has a direct proportional relationship with the provided $l$ privilege level. Therefore, we divided range of $l$ to three abstract privilege settings based on the epoch ranges. The privacy framework based on \textit{Hypercloud} was trained for 300 epochs in this experimental setup.
\begin{itemize}
    \item Low privilege (0- 0.167) - epoch range (0,50]
    \item Medium privilege (0.167- 0.232) - epoch range [50,70]
    \item High privilege (0.232- 1) - epoch range [70,300]
\end{itemize}
This division was done based on experimental results that we obtained, as will be explained in \S\ref{sec:Results}. 
For example, for the attacker $J_2$, we randomly chose 100 epoch numbers within the range (0,50] with replacement and regenerated each unique object $m\in M_k$, for every super-class $\forall k\in K$, using the \textit{hypernetwork} loaded with weights from that particular epoch number. A similar procedure was followed for the attackers $J_3$ and $J_4$, except the epoch ranges were changed to [50,70] and [70,300] respectively to reflect each privilege setting. The reference dataset generation for a single super-class (e.g. "Chair") is shown in the Fig. \ref{fig:attacker-training-dataset}.

Now, for each attacker, we have a reference dataset of size $10 \times 100 \times 100$, i.e. 100 reconstructions of each of the 100 unique objects for all the 10 super-classes. The attacker uses this reference data for training and validation of their classification models, and uses them to  infer the super-class and intra-class attributes of a query point cloud $S^*$. 

\begin{figure}[t]
	\centering
	\vspace{-2mm}
	\includegraphics[width=0.9\columnwidth]{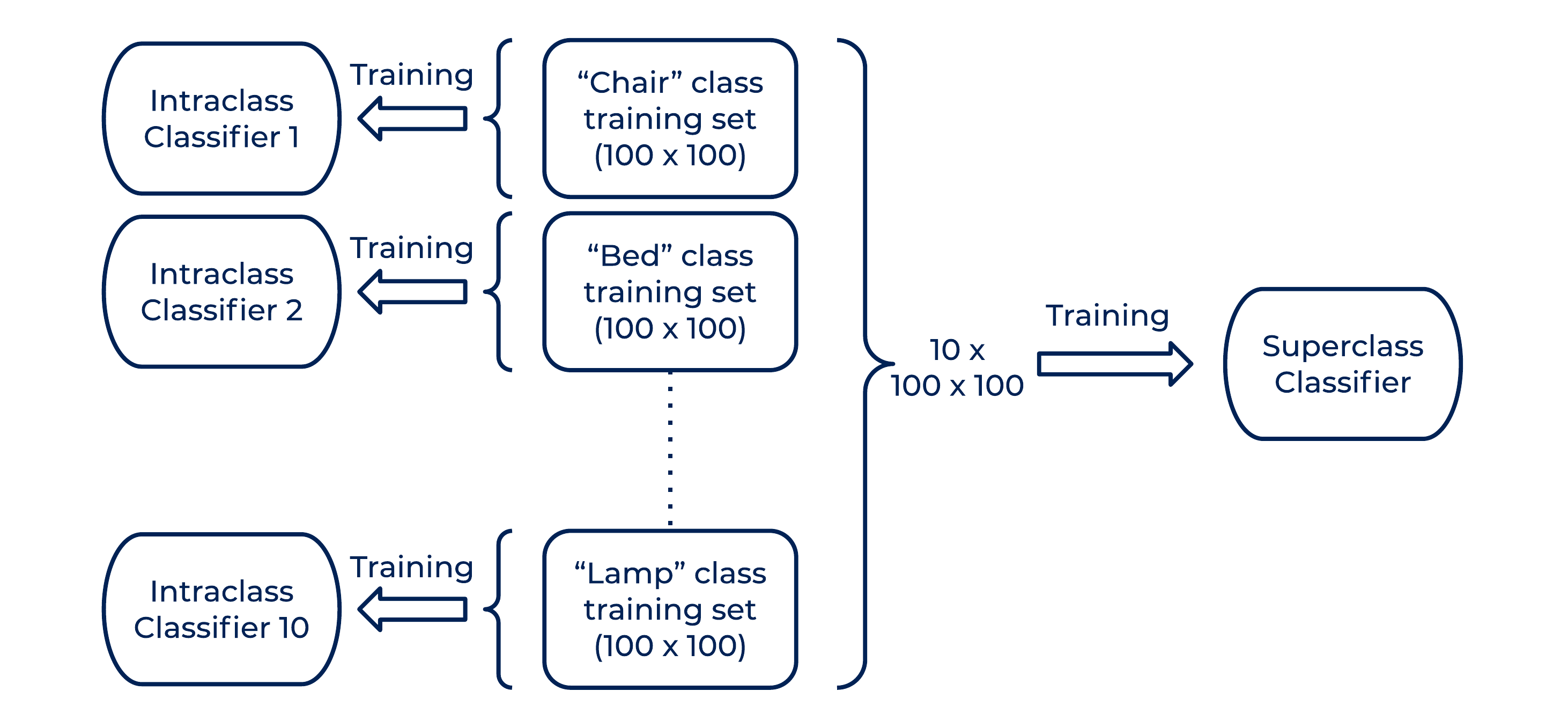}
	\vspace{-3mm}
    \caption{Classifier training process for an Attacker $J$: $1\times$ super-class classifier, and $10\times$ intra-class classifiers (i.e. one for each super-class). Each attacker generates their own set of classifiers.
    }
	\label{fig:attacker-classifier-training}
	\vspace{-3mm}
\end{figure}

For super-class classification, each attacker trains one classification model which uses the labeled dataset of size $10 \times 100 \times 100$ where we have $100 \times 100$ samples for each super-class. For intra-class classification, each attacker trains 10 different classification models, i.e. one for each super-class. Each of these 10 models uses the dataset of size $100 \times 100$, where we have 100 labeled samples (regenerations) for each unique object. An attacker's classifier training process is shown in Fig. \ref{fig:attacker-classifier-training}.

\begin{figure}[t]
    \centering
    \vspace{-2mm}
	\includegraphics[width=0.9\columnwidth]{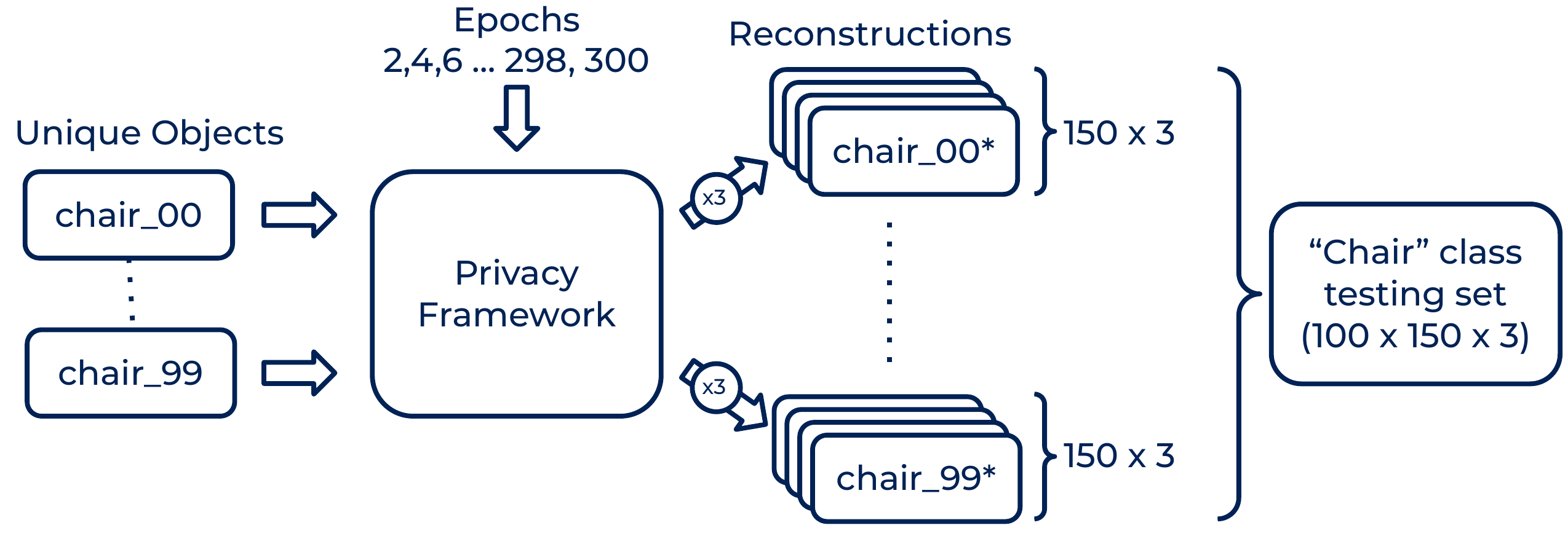}
	\vspace{-3mm}
	\caption{Testset generation for the super-class "Chair"}
	\label{fig:attacker-testing-dataset}
	\vspace{-5mm}
\end{figure}

\vspace{-1mm}\subsubsection{Evaluating the attacker performance}
To assess each attacker's success, we test all of their classification models against a scenario where it attempts an object reidentification attack (following the attacker hypothesis formulation in \S\ref{sec:Attacker hypothesis formulation}) while using the proposed privacy framework.
As discussed in \S\ref{sec:point-cloud-transformations-for-different-privacy-levels}, according to the user input on the privilege slider, we choose its corresponding epoch number from within the range (0,300] and load the corresponding weights into our framework. This network is used to create a regeneration of the original point cloud with the privilege allowed by the user. 

For each of the 100 unique objects from all the 10 super-classes, we created 3 reconstructions at every other epoch, i.e. all the even numbered epochs in the range (0,300]. This labeled dataset of size $10 \times 100 \times 150 \times 3$ was used to test all the 11 classifiers trained by each attacker. Our testing dataset will allow us to assess the attacker's success at inferring both the super-class and the intra-class attribute at various levels of privilege set by a user. 
The testing dataset generation for a single super-class is shown in Fig. \ref{fig:attacker-testing-dataset}.

\vspace{-1mm}\subsubsection{Baselines to evaluate the privacy preservation of proposed framework}\label{sec:privacy-baselines}

We simulated an object reidentification attack attempted by Attacker $J_1$ on a user who does not use our framework, but releases the original point clouds of the objects. In such case, the point clouds released by the user resembles the publicly available point clouds for those objects. We run the classifier inference on the original point clouds from our dataset in \S\ref{sec:dataset} to simulate this scenario. The calculated privacy metrics $\Pi_{1,B}$ and $\Pi_{2,B}$ are used as baselines to evaluate the performance increase of the proposed framework in terms of privacy preservation against Attacker $J_1$. 

\begin{figure}[t]
    \vspace{-2mm}
    \begin{subfigure}{0.5\columnwidth}
    \centering
        \includegraphics[width=0.99\textwidth]{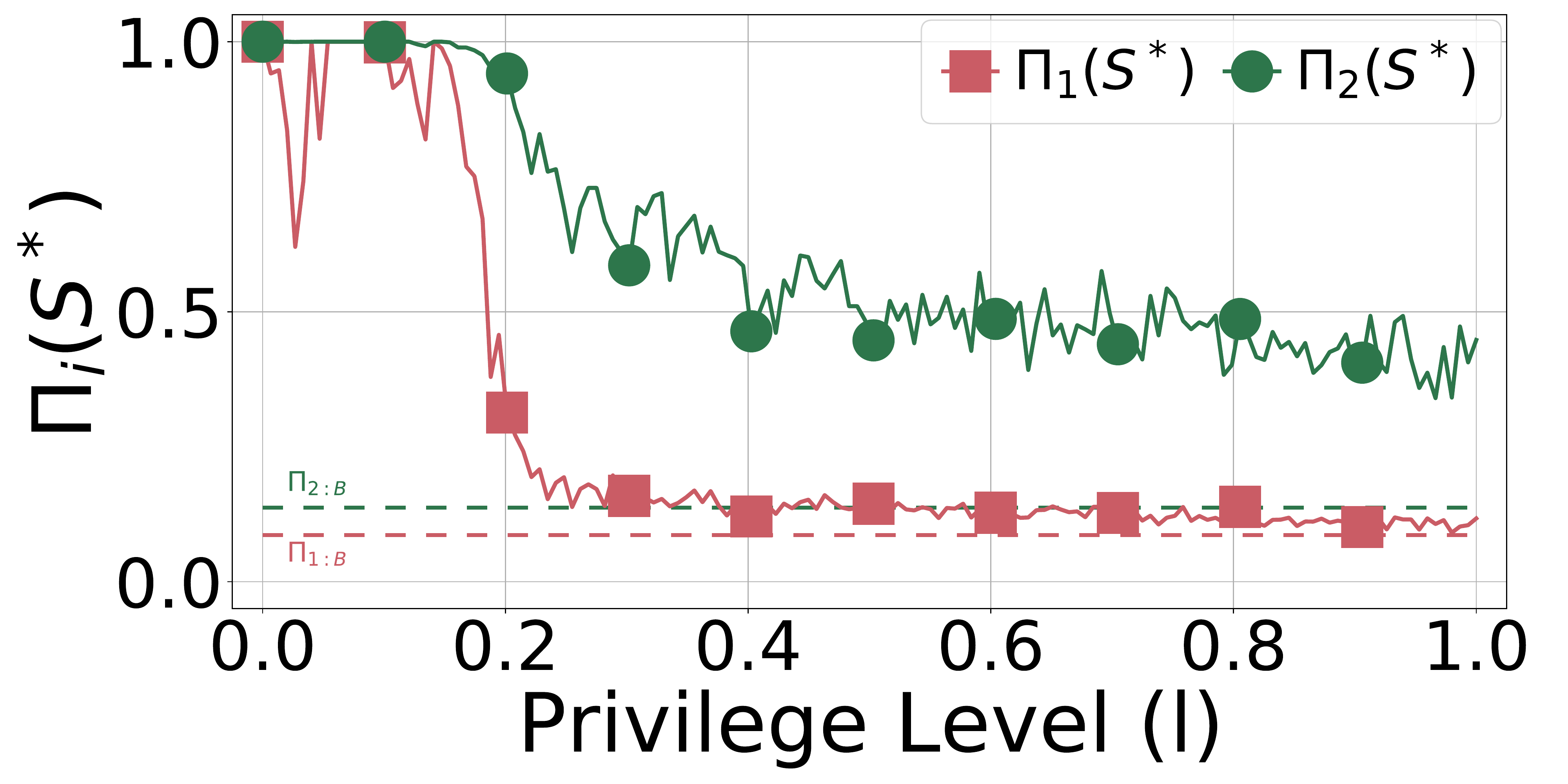}
        \caption{Chair}
        \label{fig:privacy-chair}
    \end{subfigure}%
    \begin{subfigure}{0.5\columnwidth}
    \centering
        \includegraphics[width=0.99\textwidth]{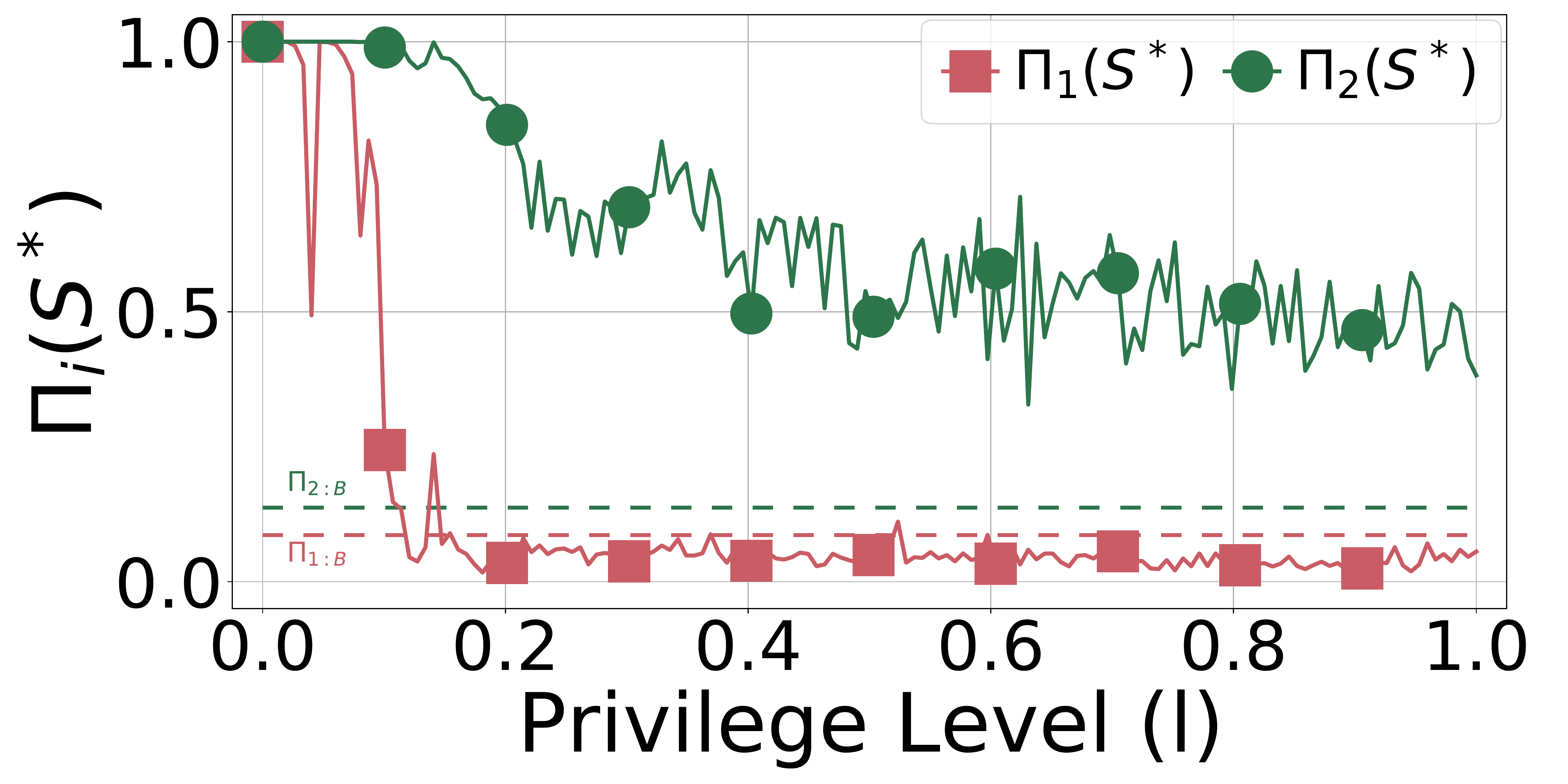}
        \caption{Table}
        \label{fig:privacy-table}
    \end{subfigure}%
    
    
    \begin{subfigure}{0.5\columnwidth}
    \centering
        \includegraphics[width=0.99\textwidth]{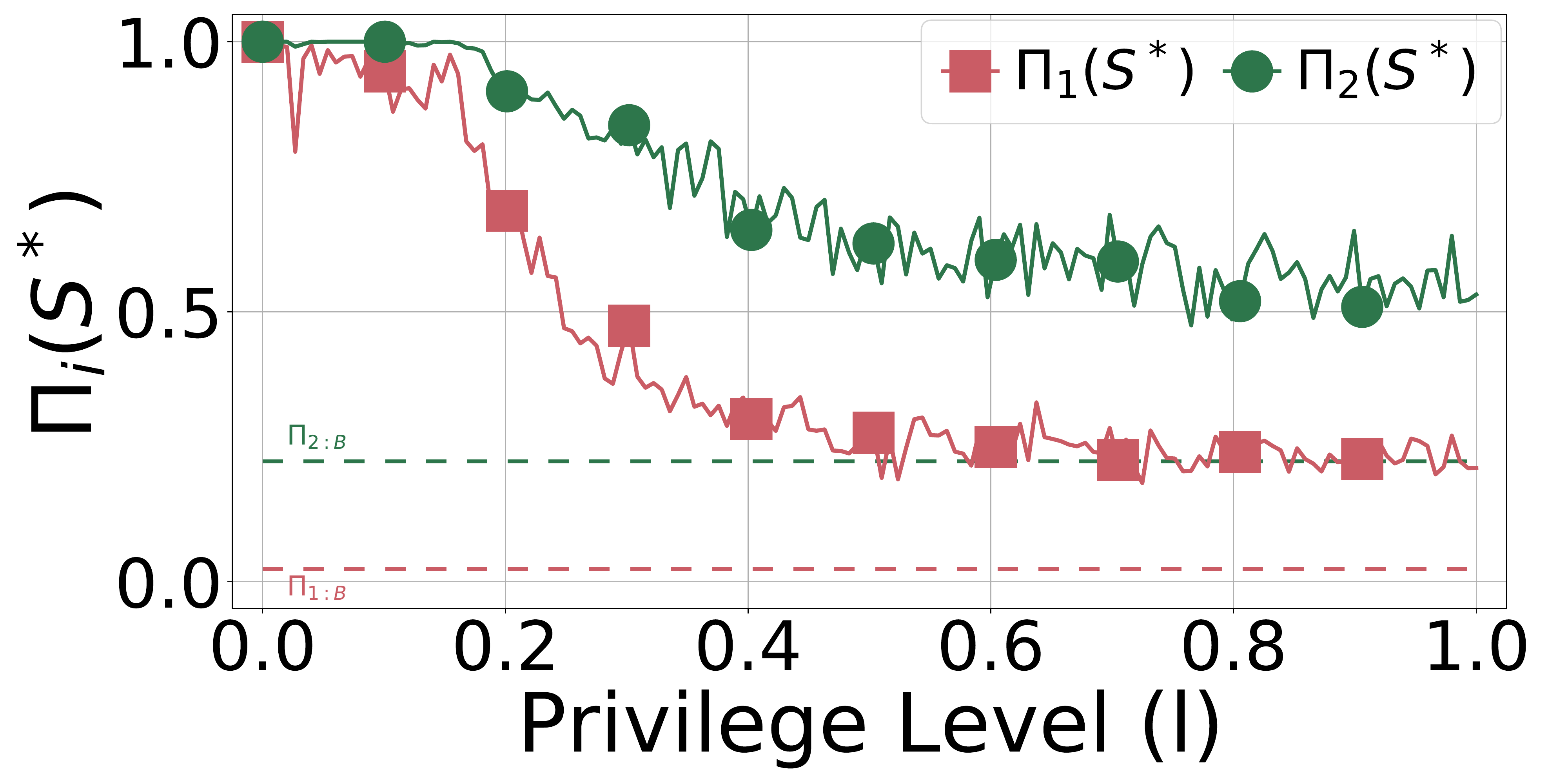}
        \caption{Bed}
        \label{fig:privacy-bed}
    \end{subfigure}%
    \begin{subfigure}{0.5\columnwidth}
    \centering
        \includegraphics[width=0.99\textwidth]{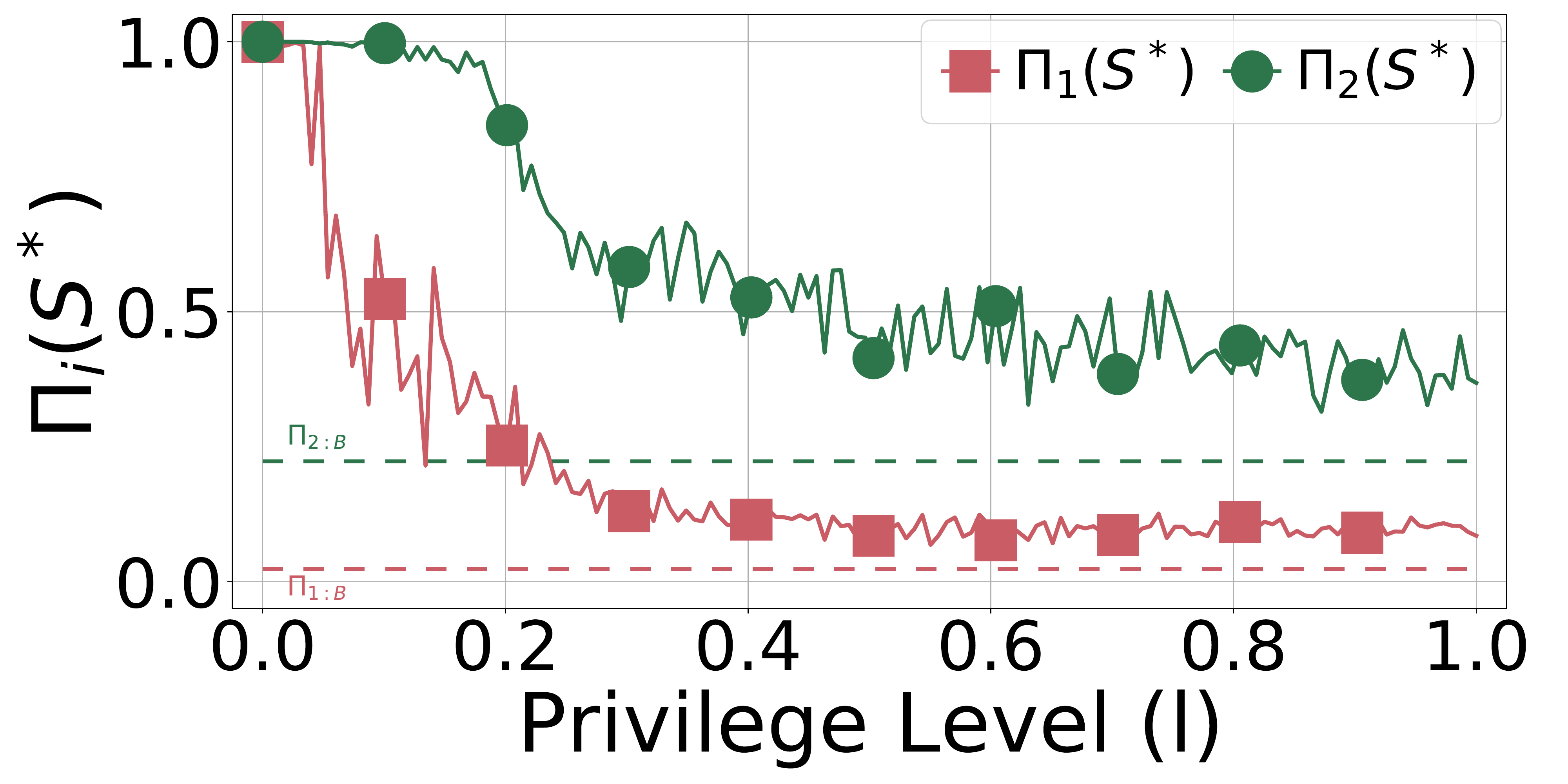}
        \caption{Bench}
        \label{fig:privacy-bench}
    \end{subfigure}%
    
    \vspace{-3mm}
    \caption{Super-class and intra-class ($\Pi_1$ and $\Pi_2$) privacy metrics for the top-1 success of Attacker $J_1$ for selected 3D objects: (a) chair, (b) table, (c) bed, and (d) bench.} 
    \label{fig:privacy-metric-n1-attacker1}
    \vspace{-2mm}
\end{figure}

\vspace{-1.5mm}\subsection{Utility evaluation for regenerated point clouds }\label{sec:Evaluation of point cloud Utility}


Higher privacy usually comes at a cost of quality of service. The goal is for the privacy framework to still allow MR applications to successfully deliver their intended functionalities while preserving the privacy of the point cloud objects.
To this end, we utilize two utility metrics defined in \S\ref{sec:utility metrics}, along with an additional Chamfer distance measurement to evaluate the regenerated point clouds.

In order to realize the utility metric $Q_{1:S,\Bar{S}}$ from \autoref{Utility metric 1}, we calculated the axis-aligned \textit{bounding boxes} for each original point cloud $S$ and compared it with the bounding boxes of regenerated point clouds $\Bar{S}$ from each privilege level $l$. We use the standard 3D \textit{Intersection-over-union -- IoU} as the comparison metric in $Q_1$. 

Next, to realize the utility metric $Q_{2:S,\Bar{S}}$ from \autoref{Utility metric 2}, we used a modified implementation of RANSAC \cite{RANSACFischlerandMartin} algorithm to extract the most prominent 2D plane of each point cloud object. Since our dataset consists primarily of household objects, we conducted our analysis for 2D horizontal planes, e.g. sitting plane of chairs where virtual objects are most likely to be anchored on. But the utility metric $Q_{2:S,\Bar{S}}$ can be applied to any type of plane of choice. We obtain a set of points $p$ that are fitted to the most prominent 2D plane of the object by the RANSAC algorithm, and calculate the metrics given in the \autoref{Utility metric 2} separately for \textit{static} and \textit{dynamic} anchoring cases. 


Additionally, some MR functionalities might require near-truth representation of the complete structure of the object. Therefore its also important to investigate how the object structure of $S$ is preserved in privacy-preserving regenerations $\Bar{S}$ created from $S$.
To this end, we calculate the well known \textit{Chamfer distance} between the original point cloud $S$, and its regenerated point clouds $\Bar{S}$ at each $l$ privilege level. According to \autoref{hypercloud_cost}, \textit{Chamfer distance} is also a part of the cost function in our framework. It essentially measures the squared distance between each point in one point cloud to its nearest neighbor in the other point cloud.








\section{Results and Discussion}\label{sec:Results}

\subsection{Privacy evaluation for the top-1 hypothesis of Attacker $J_1$}
\label{subsec:attacker-1-n-1}

\begin{figure}[t]
    \vspace{-2mm}
    \includegraphics[width=\columnwidth]{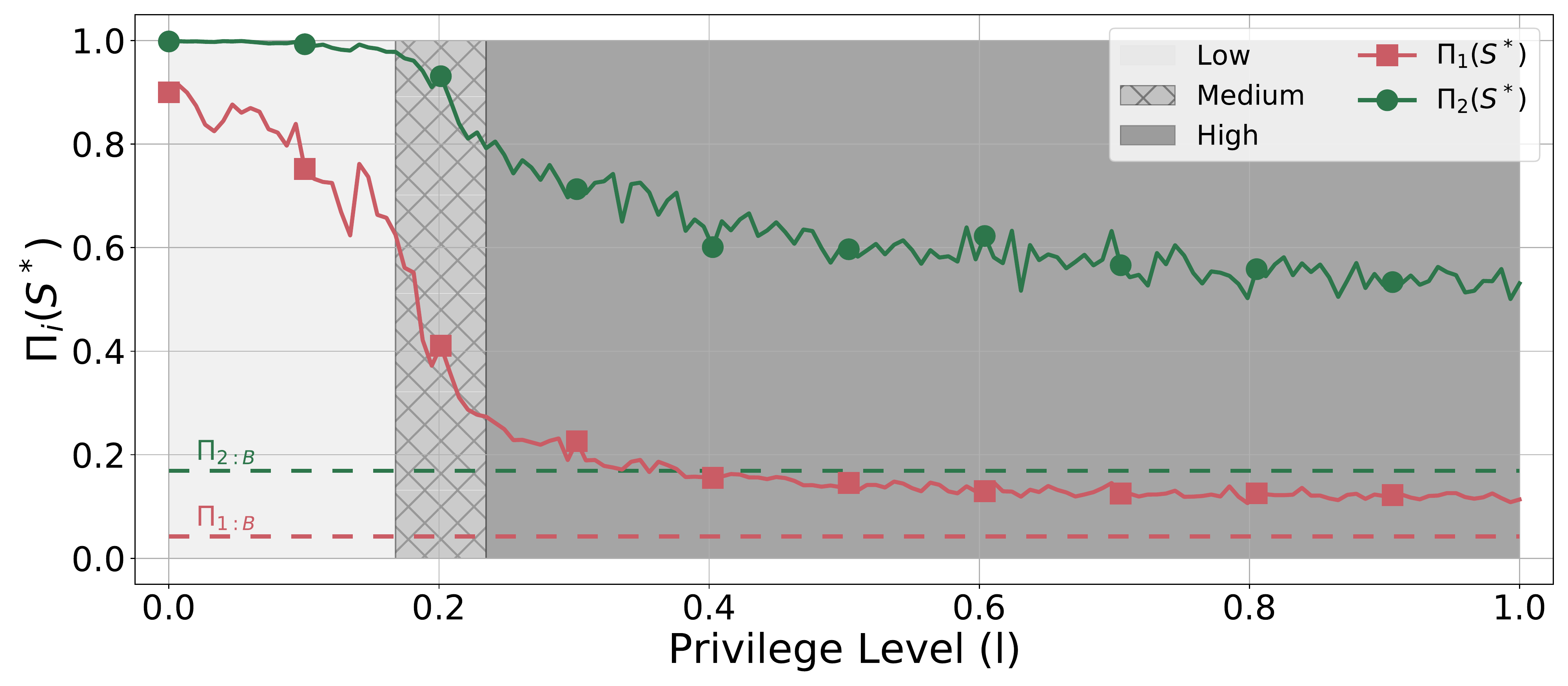}
    \vspace{-5mm}
    \caption{Aggregated $\Pi_1$ and $\Pi_2$ metrics for the top-1 success of Attacker $J_1$ for all test objects}
    \label{Fig:Data2}
    \vspace{-5mm}
\end{figure}

We evaluated the performance of our proposed privacy-preserving regeneration separately for each super-class. 
In Fig. \ref{fig:privacy-metric-n1-attacker1}, x-axis indicates the privilege level $l$ at which a query point cloud has been regenerated. 
Each plot indicate the variation of the two privacy metrics $\Pi_1$ and $\Pi_2$ from \autoref{Eq:privacy metric 1} and \autoref{Eq:privacy metric 2} for different regenerations. For this analysis, the parametric values $\rho_{1}$ and $\rho_{2}$ are set such that $|\gamma_k^* |=\rho_{1} |K |=1, |\eta_{M_k}^* |=\rho_{2} |M_{k} |=1 $; that is, the attacker is narrowing down its hypothesis to a top-1 super-class and intra-class label. The attacker used in this analysis is Attacker $J_1$ according to the categorization in \S\ref{sec:Adversarial attacker realization/simulation}, who builds the reference point cloud object sets using publicly available data. 

As $l$ increases, the regeneration becomes increasingly similar to the original point cloud: eventually decreasing the privacy for both super-class and intra-class attack scenarios. One notable observation is that the privacy for a particular regeneration is higher for the intra-class case than for the super-class case. This is due to the penalization in \autoref{Eq:privacy metric 2} for failed super-class hypothesis in the first step of hypothesis 2 formulation. 


In Fig. \ref{fig:privacy-metric-n1-attacker1} and Fig. \ref{Fig:Data2}, the baselines calculated in \S\ref{sec:privacy-baselines} are shown in dashed lines. The privacy metric $\Pi_2$ for the proposed framework is significantly large in comparison to baseline values in all shown super-classes, while privacy metric for $\Pi_1$ is large in comparison to baselines for relatively low $l$ values, and gets closer to the baseline value for higher $l$ values. For example, according to Fig. \ref{Fig:Data2}, proposed framework adds 0.58 increase in $\Pi_1$ privacy, and 0.81 increase in $\Pi_2$ privacy at privilege level $l= 0.167$, and respectively 0.23 and 0.62 increase in $\Pi_1$ and $\Pi_2$ at $l= 0.232$. 

A significant change in both $\Pi_1$ and $\Pi_2$ can be observed within the approximate range of $l=0.167$ and $l= 0.4$ followed by a slow decay thereafter, indicating the potential of the proposed regeneration mechanism to be used as a \textit{tunable} privacy encoding for point cloud objects.


\begin{figure}[t]
\centering

    \vspace{-2mm}
    \begin{subfigure}{0.5\columnwidth}
    \centering
        \includegraphics[width=0.99\textwidth]{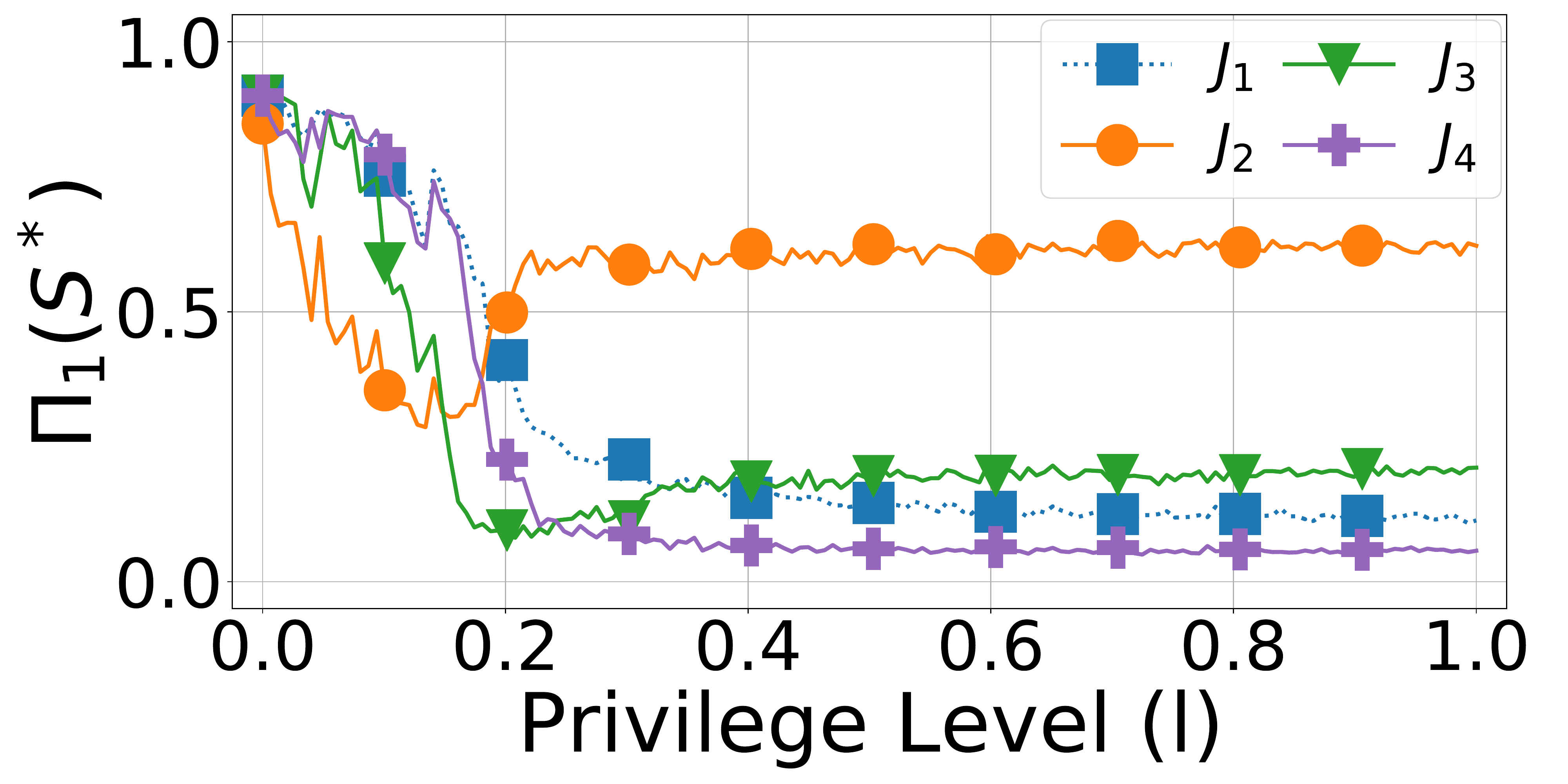}
        \caption{Superclass Privacy $\mathbf{\Pi_1(S^*)}$}
        \label{fig:pi-1-n1-varn-attacker}
    \end{subfigure}%
    \begin{subfigure}{0.5\columnwidth}
    \centering
        \includegraphics[width=0.99\textwidth]{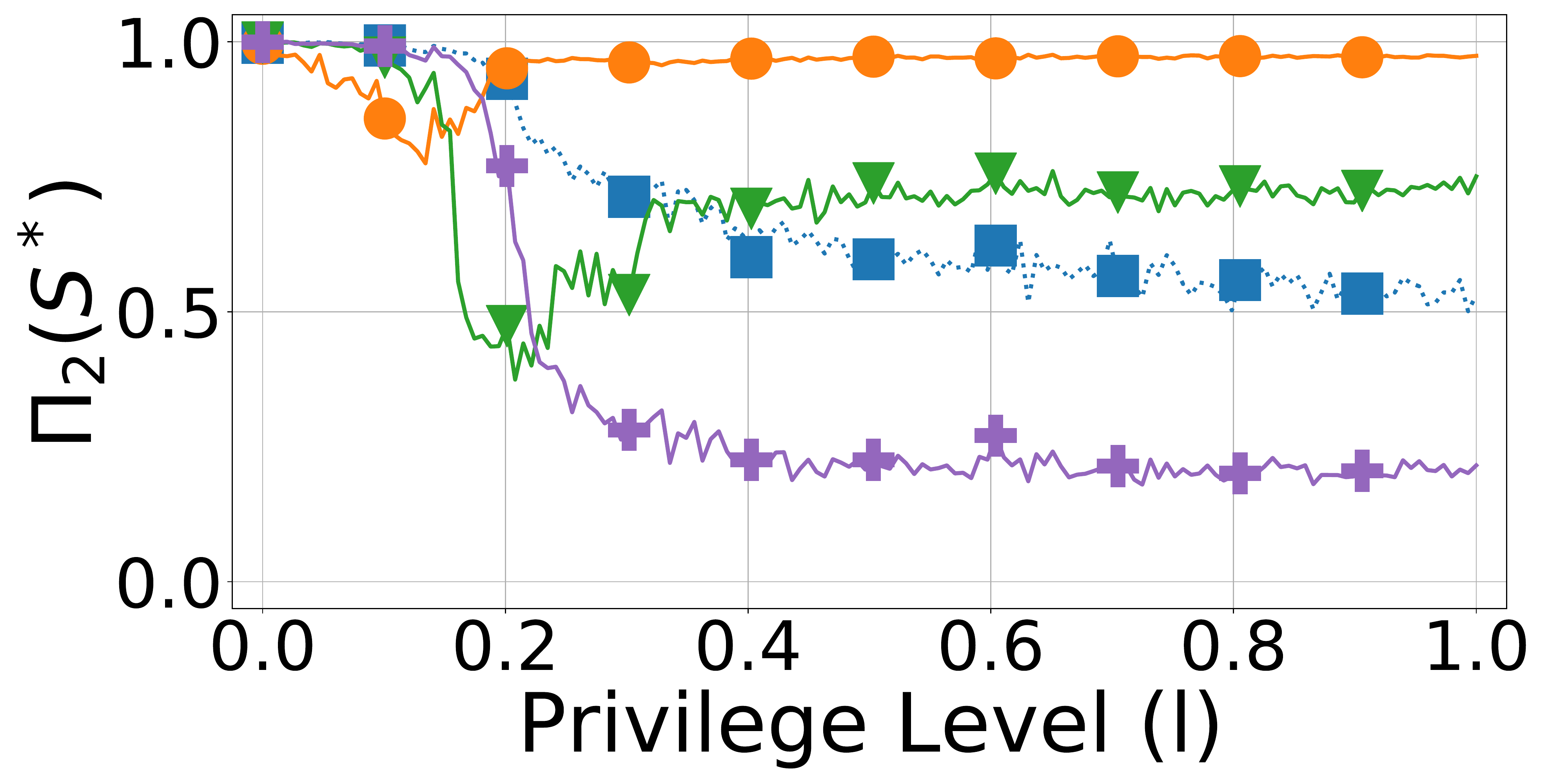}
        \caption{Intraclass Privacy $\mathbf{\Pi_2(S^*)}$}
        \label{fig:pi-2-n1-varn-attacker}
    \end{subfigure}%
    \vspace{-3mm}
    \caption{Aggregated $\Pi_1$ and $\Pi_2$ metrics for the top-1 success of all Attackers: $J_1, J_2, J_3 \text{and} J_4$.}
    \label{fig:privacy-metric-n1-varn-attacker}
    \vspace{-5mm}
\end{figure}

\vspace{-1.5mm}\subsection{Privacy performance against various attackers with different privilege levels}\label{subsec:all-attackers-n-1}

In \S\ref{sec:Adversarial attacker realization/simulation}, we introduced 4 different attackers based on how each attacker obtain the reference data for its hypothesis formulation. Out of them, Attackers $J_2$,$J_3$, and $J_4$ can be analyzed separately since they use regenerated point clouds 
to form its reference set. This resembles a practical situation where a user is already using the proposed framework to transform the 3D objects prior to being released to the MR application, an adversarial attacker can still intercept these transformed data and learn to formulate hypotheses. In Fig.
\ref{fig:privacy-metric-n1-varn-attacker}, we present the analysis of the variation of privacy metrics for different Attackers $J_2$, $J_3$ and $J_4$. The results are aggregated for all super-classes and all unique-objects within each super-class.

One notable observation is the clear difference of privacy metric values for different attackers. The regenerated point clouds show very high privacy with respect to Attacker $J_2$, which is trained only on low privilege data, while showing significantly low privacy metric values for Attacker $J_4$, which is trained on high privilege data. Similar to the results in \S\ref{subsec:attacker-1-n-1}, intra-class privacy metric values are comparatively higher than that of super-class for all three attackers. 
On the other hand, the privacy of regenerated point clouds with respect to Attacker $J_4$ is even lower than that for Attacker $J_1$ that we discussed in \S\ref{subsec:attacker-1-n-1}. The potential reason is because the features learnt by Attacker $J_4$ from the regenerated point clouds with highest privilege are more similar to the query point clouds than the features of original point clouds (from public datasets). The dip in the plots for Attackers $J_2$ and $J_3$ indicate that the attackers can pose a higher privacy threat for point clouds which are regenerated in the similar $l$ range as their reference sets, but the threat does not generalize well for other $l$ ranges: e.g. trained at lower $l$ but does not perform well against query objects released at higher $l$.



Given the difference in how Attacker $J_1$ and the rest of the attackers ($J_2$,$J_3$ and $J_4$) learn their reference sets, two types of privacy threats can be identified. 
Since Attackers $J_2$, $J_3$, and $J_4$ are only exposed to regenerated point clouds from the proposed framework, they lack knowledge about the unique objective features of the original point cloud object. The inference of a query point cloud $S^*$ by such attackers can be identified as an object reidentification attack; where the query can be matched with one of its reference super-classes/objects, but the attacker cannot derive further information or pose additional threats since the regenerated objects do not necessarily resemble their original counterparts. On the other hand, attacker $J_1$ learns its reference set from publicly available data, where additional object, semantic, and contextual features are also possibly available for a given object. Once Attacker $J_1$ infers a query point cloud $S^*$ with respect to such a reference set, in addition to an object reidentification threat, revealing the true identity of the object can lead to multiple other privacy threats. Although not reflected in the privacy metrics $\Pi_1$ and $\Pi_2$ in Fig. \ref{fig:privacy-metric-n1-varn-attacker}, it is noteworthy that $J_1$ has comparatively larger potential to pose additional privacy threats. 


\vspace{-1.5mm}\subsection{Privacy metric evaluation for varying sizes of hypothesis sets}\label{subsec:priv-vary-subset-size}

\begin{figure}[t]
        
    \vspace{-2mm}
    \begin{subfigure}{0.95\columnwidth}
    \centering
        \includegraphics[width=\textwidth]{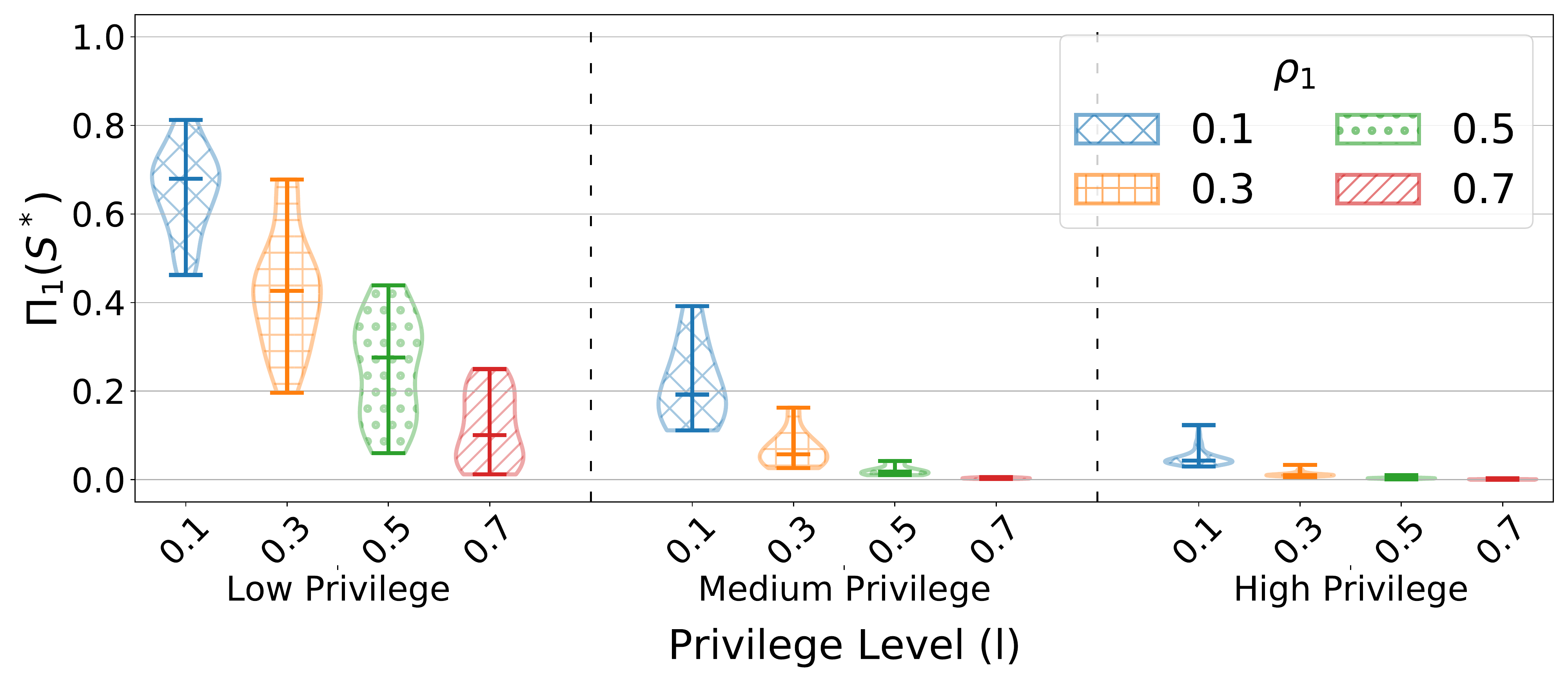}
        \caption{Superclass Privacy $\mathbf{\Pi_1(S^*)}$}
        \label{fig:pi-1-n1-varn-vplot}
    \end{subfigure}%
    
    \begin{subfigure}{0.95\columnwidth}
        \centering
        \includegraphics[width=\textwidth]{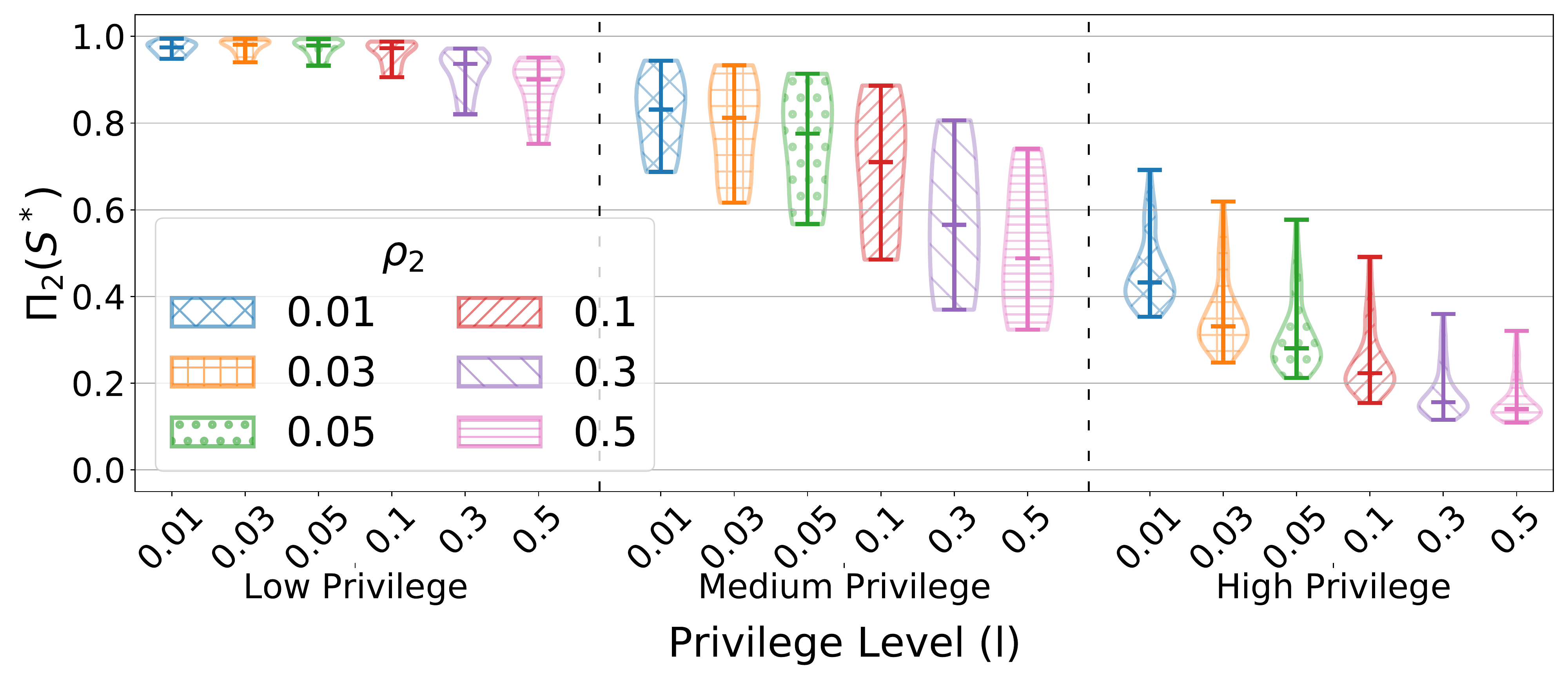}
        \caption{Intraclass Privacy $\mathbf{\Pi_2(S^*)}$}
        \label{fig:pi-2-n1-varn-vplot}
    \end{subfigure}%
    \vspace{-3mm}
    \caption{$\Pi_1$ and $\Pi_2$ metrics aggregated for varying hypothesis subset sizes $\rho$, and privilege levels $l$}
    \label{fig:privacy-vs-subset-size}
    \vspace{-5mm}
\end{figure}
 
In \autoref{Eq:hypothesis 1} and \autoref{Eq:hypothesis 2}, we introduced the hypothesis formulation in terms of the parameters $\rho_1$ and $\rho_2$. In simple terms, given a query point cloud $S^*$, the attacker attempts to narrow down the possible identity of it to a portion of its reference set, and $\rho$ parameterizes the size of the hypothesized subset as a portion of the reference set.  Intuitively, when $\rho$ increases, the hypothesis subset size increases, hence its becomes increasingly probable that the query point cloud $S^*$'s identity is included in the hypothesis subset. This behaviour is reflected in privacy metrics $\Pi_1$ and $\Pi_2$ as shown in Fig. \ref{fig:privacy-vs-subset-size}. But at the same time, for large $\rho$ values, it becomes increasingly harder for the attacker to pose a unique privacy threat to the user. 

According to the violin plots in Fig. \ref{fig:privacy-vs-subset-size}, for low privilege point clouds, the increase in $\rho_1$ drastically drops the $\Pi_1$ privacy, whereas for high privilege point clouds, $\Pi_2$ is already low, therefore the increase in $\rho_1$ only induces a small drop in $\Pi_2$ privacy. On the other hand $\Pi_2$ for low privilege point clouds is already $\sim1$, therefore $\rho$ should be increased by a significant fraction to induce a decrease $\Pi_2$ privacy. Another notable behaviour is that the privacy metrics $\Pi_1$ or $\Pi_2$ vary non-linearly with $\rho$ in almost all different privilege regions.

\vspace{-1.5mm}\subsection{Utility results}\label{subsec:utility-results}

\begin{figure}[t]
\centering

    \vspace{-2mm}
    \begin{subfigure}{0.5\columnwidth}
    \centering
        \includegraphics[width=0.99\textwidth]{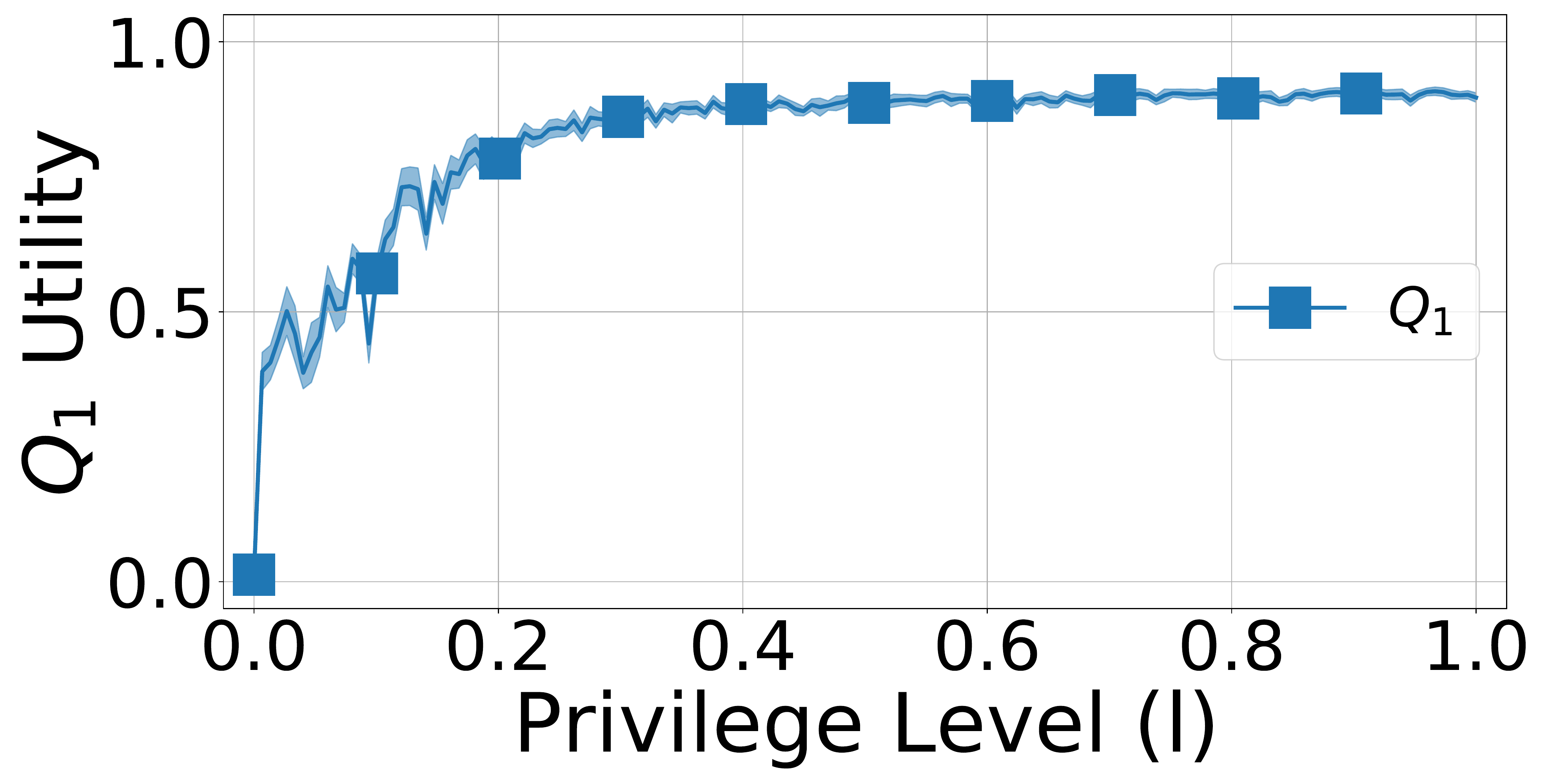}
        \caption{Bounding Box $\mathbf{Q_1}$}
        \label{fig:bbox_Q_1}
    \end{subfigure}%
    \begin{subfigure}{0.5\columnwidth}
    \centering
        \includegraphics[width=0.99\textwidth]{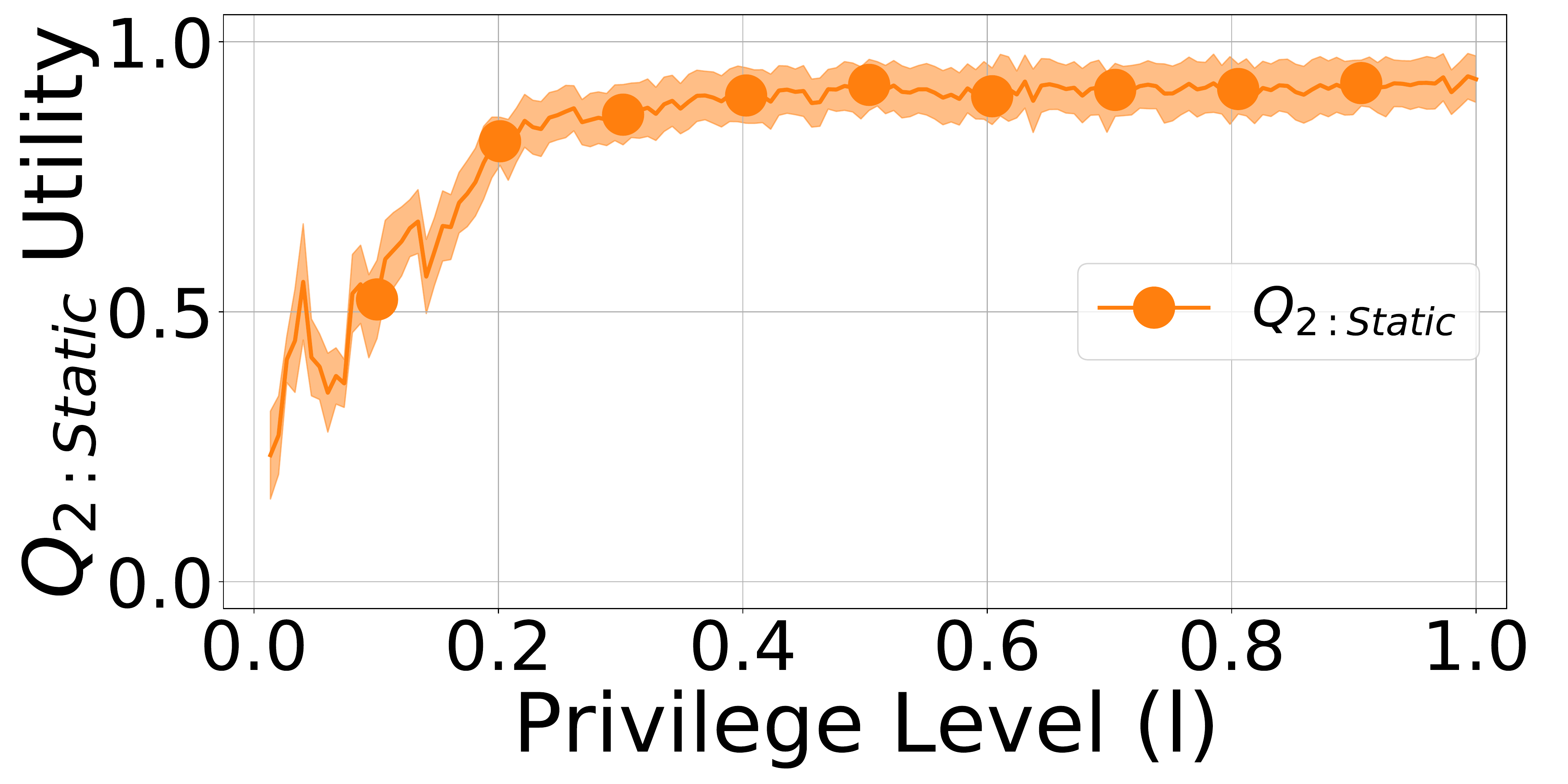}
        \caption{Static Anchoring $\mathbf{Q_2}$}
        \label{fig:static_Q_2}
    \end{subfigure}%

    \begin{subfigure}{0.5\columnwidth}
    \centering
        \includegraphics[width=0.99\textwidth]{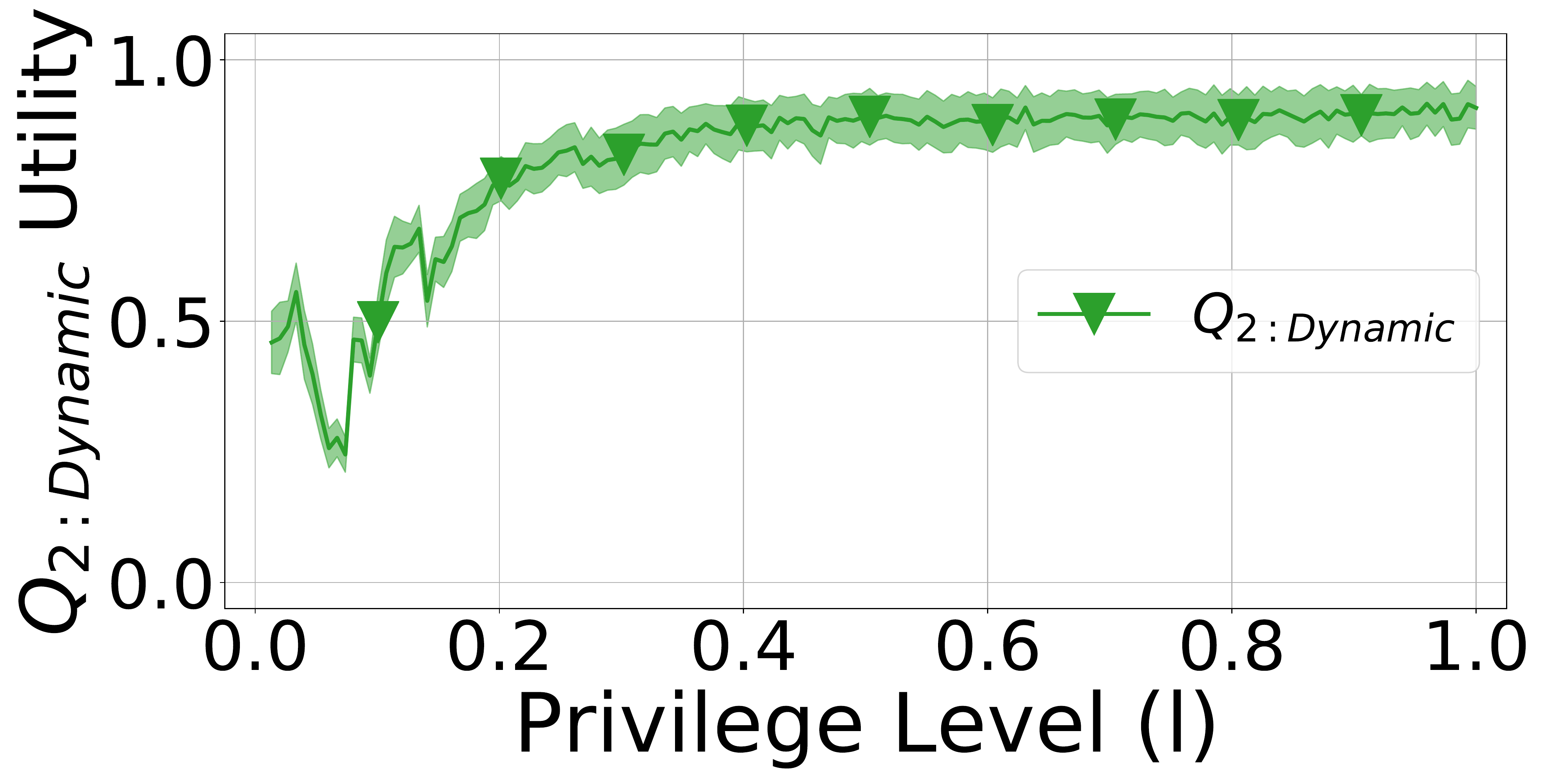}
        \caption{Dynamic Anchoring $\mathbf{Q_2}$}
        \label{fig:dynamic_Q_2}
    \end{subfigure}%
    \begin{subfigure}{0.5\columnwidth}
    \centering
        \includegraphics[width=0.99\textwidth]{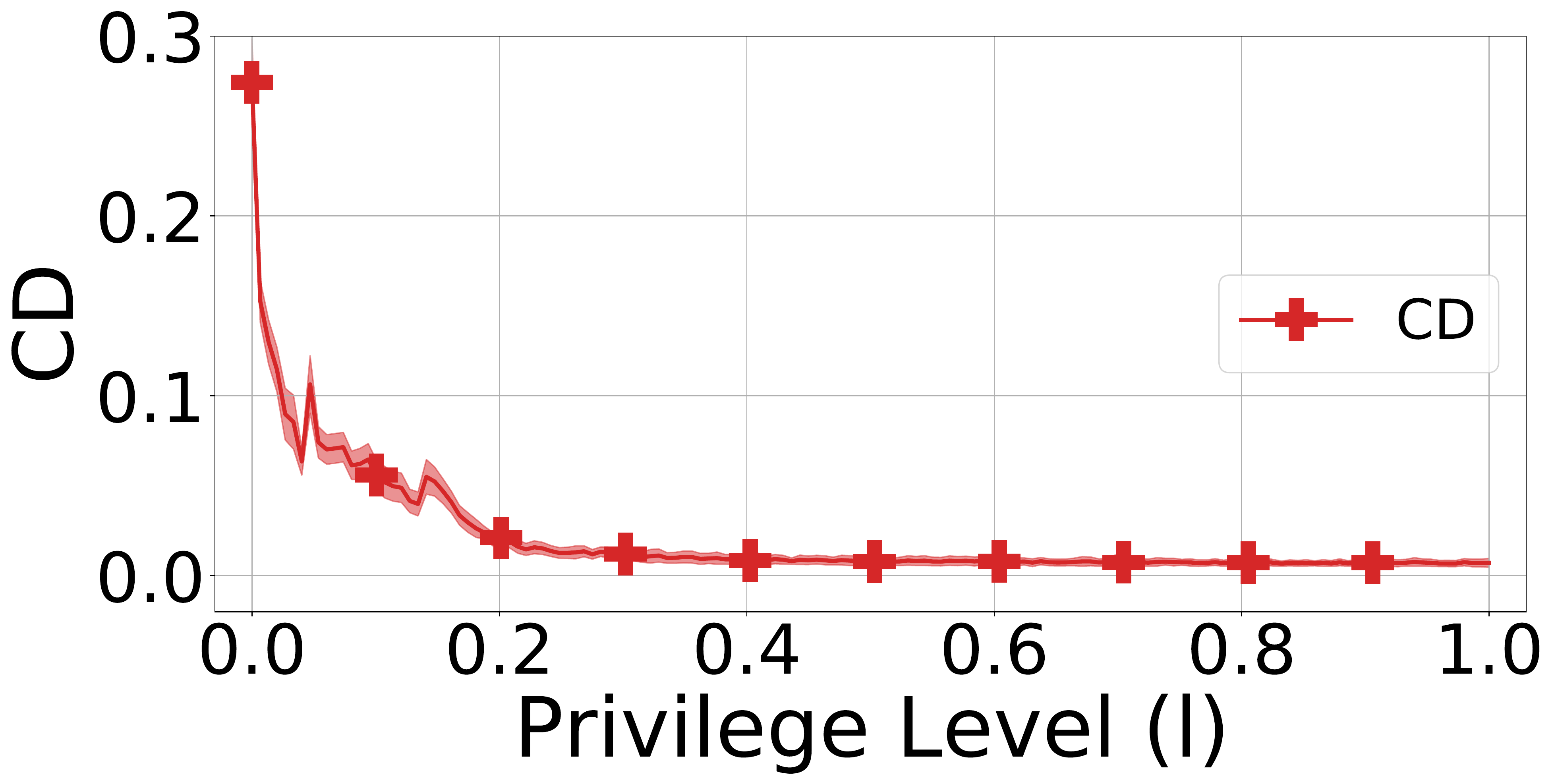}
        \caption{Chamfer Distance}
        \label{fig:c_d_utility}
    \end{subfigure}%
    \vspace{-3mm}
    \caption{Utility metrics aggregated for all test point clouds at varying privilege levels}
    \vspace{-5mm}
\label{fig:utilit-metric-comparison}
\end{figure}

In this work, we leveraged two utility metrics based on two generic MR application functionalities that require 3D information in different levels of details as explained in \S\ref{sec:utility metrics}. Fig. \ref{fig:utilit-metric-comparison} shows the aggregated utility metrics values (averaged across 10 super-classes and 100 unique objects per each super-class) from the test dataset. 
Additionally, we also use \textit{Chamfer distance} as a measurement of structural dissimilarity between the original point cloud and its regenerated counterparts at different privilege levels. This measurement gives insight into the success of regenerated point clouds in delivering MR functionalities which require near-truth representations of objects.  

Utility metric $Q_1$ given in \autoref{Utility metric 1} calculates the IoU between the bounding boxes that encompasses the original point cloud and the regenerated point cloud. A higher IoU implying that the \textit{dimensions} and \textit{location} of the regenerated point cloud is more similar to that of the original. As shown in Fig. \ref{fig:bbox_Q_1}, $Q_1$ reaches 0.78 by 0.167 percentage privilege (low privilege), and 0.83 by 0.232 percentage privilege (medium privilege), and maintains a $Q_1$ of $\sim 0.9$ between percentage privileges 0.332 to 1 (high privilege). It is evident that the utility of the proposed regeneration mechanism with respect to $Q_1$ utility converges quickly, indicating a reasonably good utility guaranteed even for point clouds released with lower privileges. 

Figures \ref{fig:static_Q_2}-\ref{fig:dynamic_Q_2} shows the aggregated results for the test dataset with respect to
$Q_2$ given in \autoref{Utility metric 2}. We implemented a RANSAC based plane fitting to find the point clouds corresponding to the most prominent 2D horizontal plane in each transformed object, and compared its attributes with the corresponding plane from the original object. 
Since our dataset is limited to household objects, and anchoring virtual objects predominantly happen for horizontal planes, we conducted the analysis for horizontal planes.
By $l=0.167$, $Q_{2:Static} \text{and } Q_{2:Dynamic}$ reach 0.70 and 0.69 respectively, and by  $l=0.232$, $Q_{2:Static} \text{ and } Q_{2:Dynamic}$ reach 0.84 and 0.79.

The unexpected variations in the $Q_2$ utilities in the very low $l$ range, $l< 0.1$, can be accounted for the abrupt variation of point cloud re-generations with high loss values.

Next, Fig. \ref{fig:c_d_utility} shows the \textit{Chamfer distance} between original and regenrated point clouds (at varying privilege levels) averaged across all super-classes unique objects in each super-class. 
As shown in the graph, by 0.167 percentage privilege, the average chamfer distance has already decreased to 0.03, and by 0.232 percentage privilege, to 0.01. 
This indicates that the regenerated point clouds from the proposed framework are able to effectively retain the structural properties and its associated utilities even in the point clouds released with lesser privilege. Finally, the error margins in every utility metric in Fig. \ref{fig:utilit-metric-comparison} makes it evident that the proposed framework allows the discussed utilities consistently across a large number of super-classes and objects despite the large variation in their shapes and structures.









\vspace{-1.5mm}\subsection{Privacy-utility trade-off}\label{sec:Privacy-utility-trade-off}

\begin{figure}[t]
    \centering
    \vspace{-2mm}
    \includegraphics[width=\columnwidth]{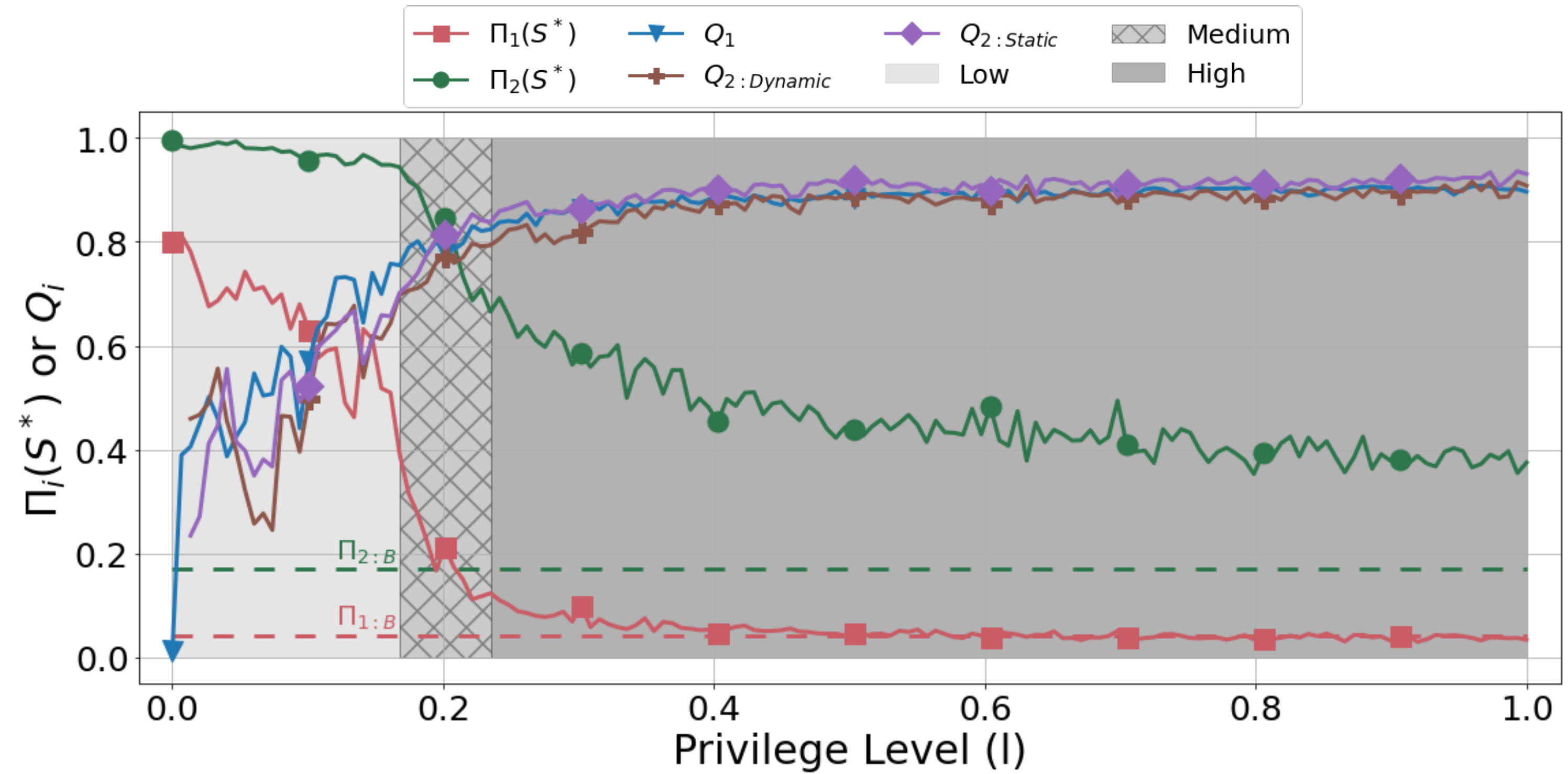}
    \vspace{-5mm}
    \caption{Privacy and Utility metrics vs privilege level. Indicates the tradeoff between offered privacy and utility levels}
    \label{fig:privacy-vs-utility-tradeoff}
\end{figure}

\begin{figure}[t]
\centering


    \begin{subfigure}{0.5\columnwidth}
    \centering
        \includegraphics[width=0.99\textwidth]{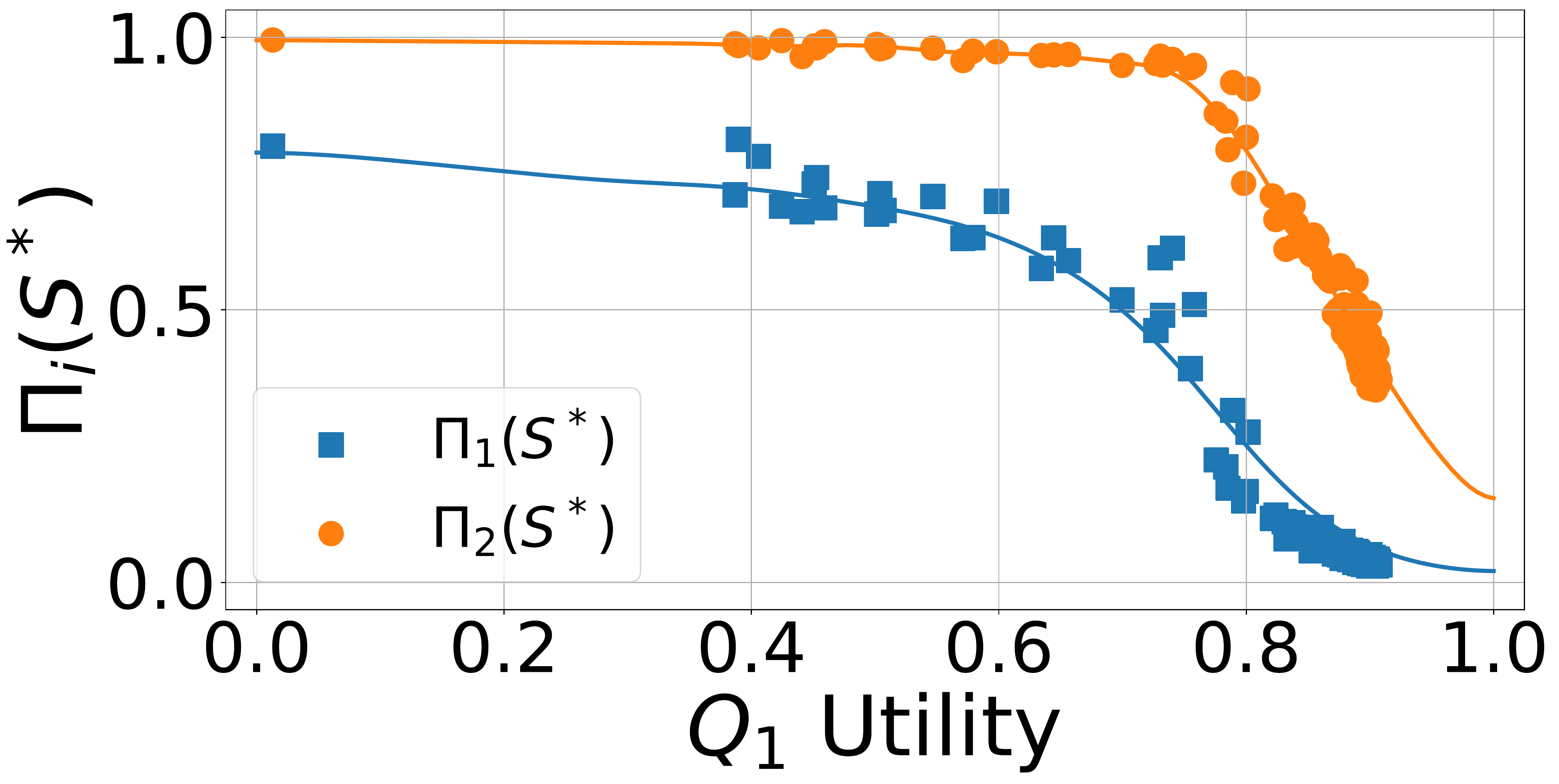}
        \caption{$\Pi$ vs $\mathbf{Q_1}$}
        \label{fig:pi_vs_bbox_Q_1}
    \end{subfigure}%
    \begin{subfigure}{0.5\columnwidth}
    \centering
        \includegraphics[width=0.99\textwidth]{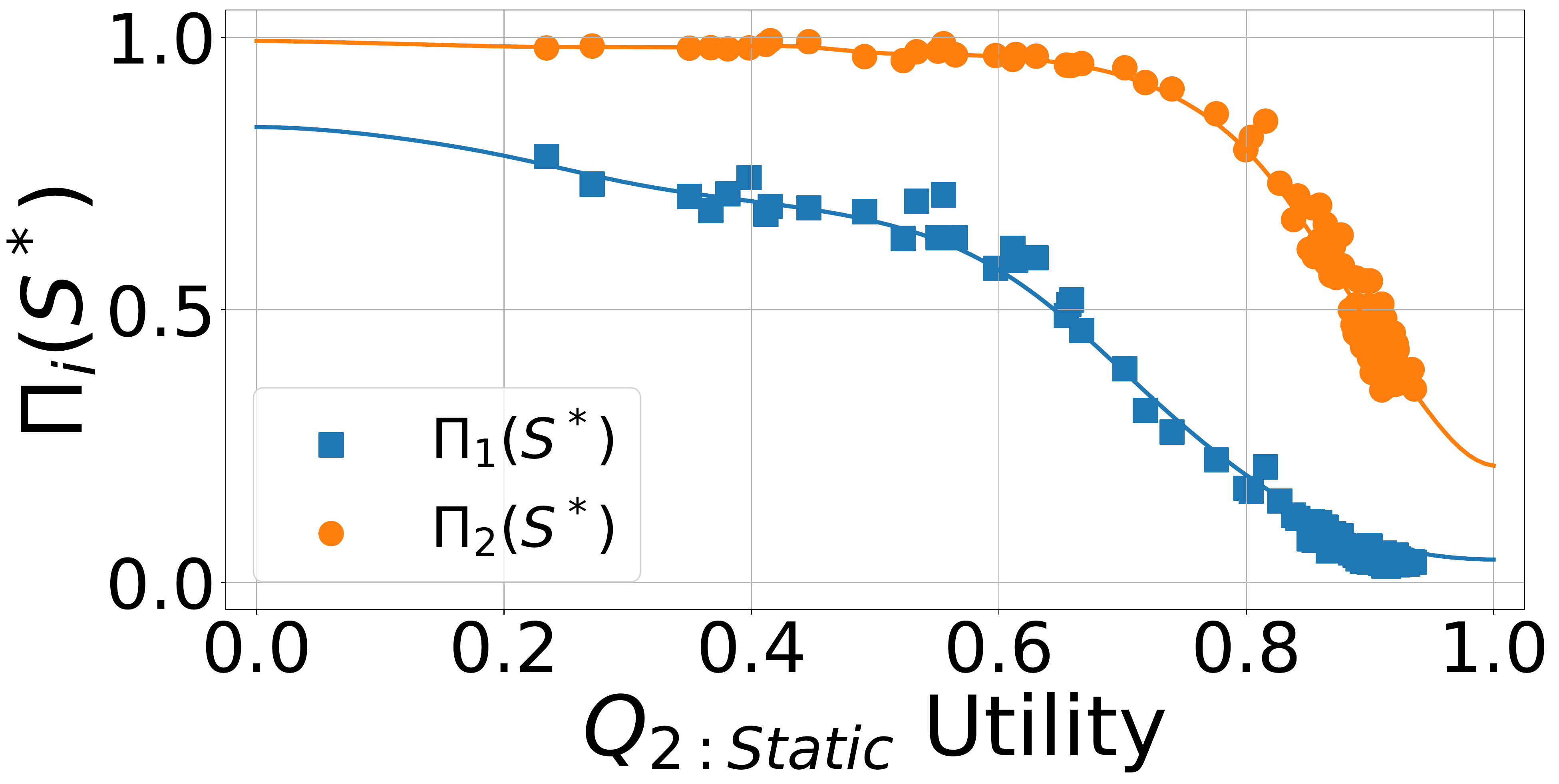}
        \caption{$\Pi$ vs static $\mathbf{Q_1}$}
        \label{fig:pi_vs_static_Q_2}
    \end{subfigure}%


    \begin{subfigure}{0.5\columnwidth}
    \centering
        \includegraphics[width=0.99\textwidth]{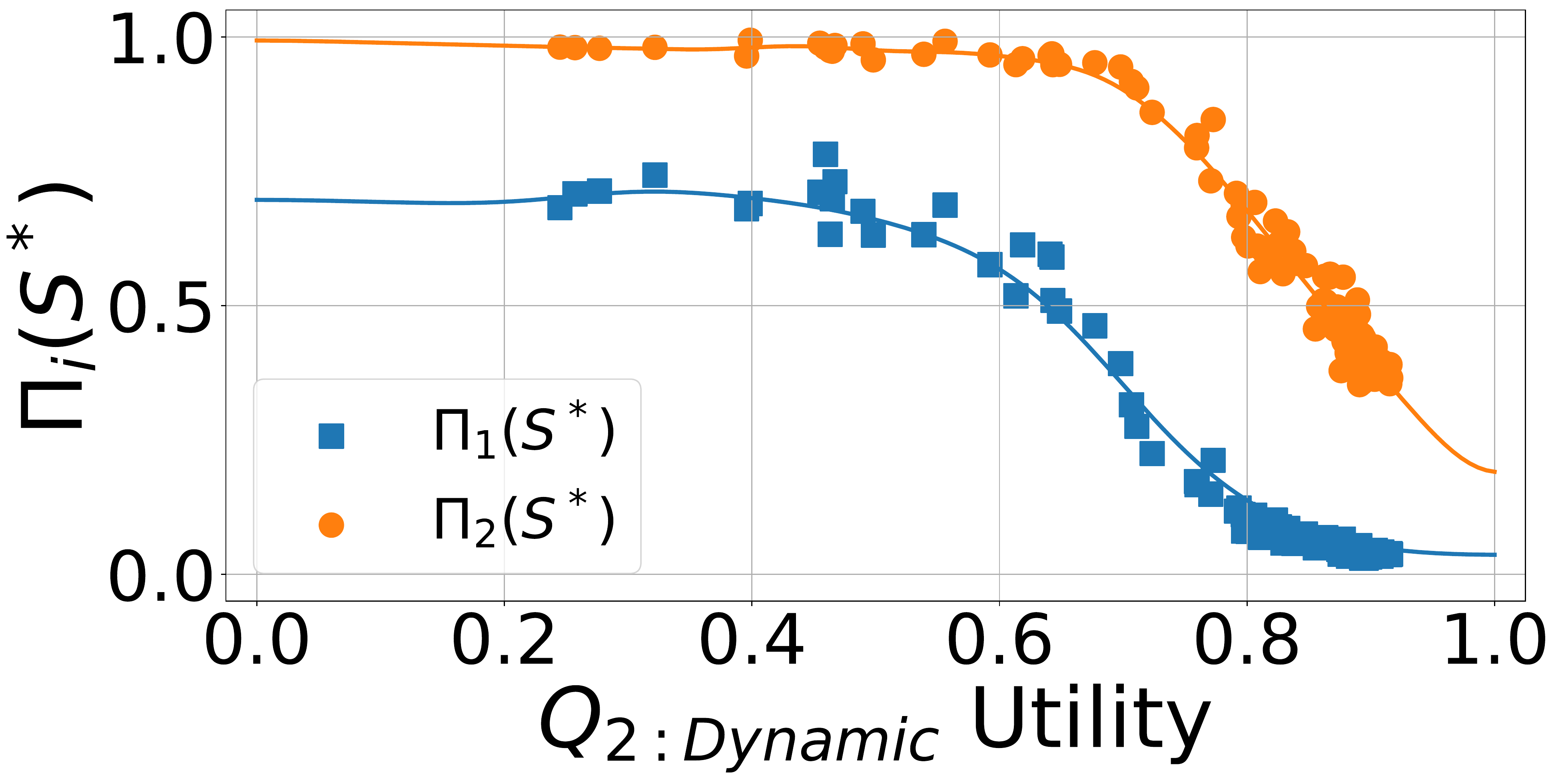}
        \caption{$\Pi$ vs dynamic $\mathbf{Q_2}$}
        \label{fig:pi_vs_dynamic_Q_2}
    \end{subfigure}%
    \begin{subfigure}{0.5\columnwidth}
    \centering
        \includegraphics[width=0.99\textwidth]{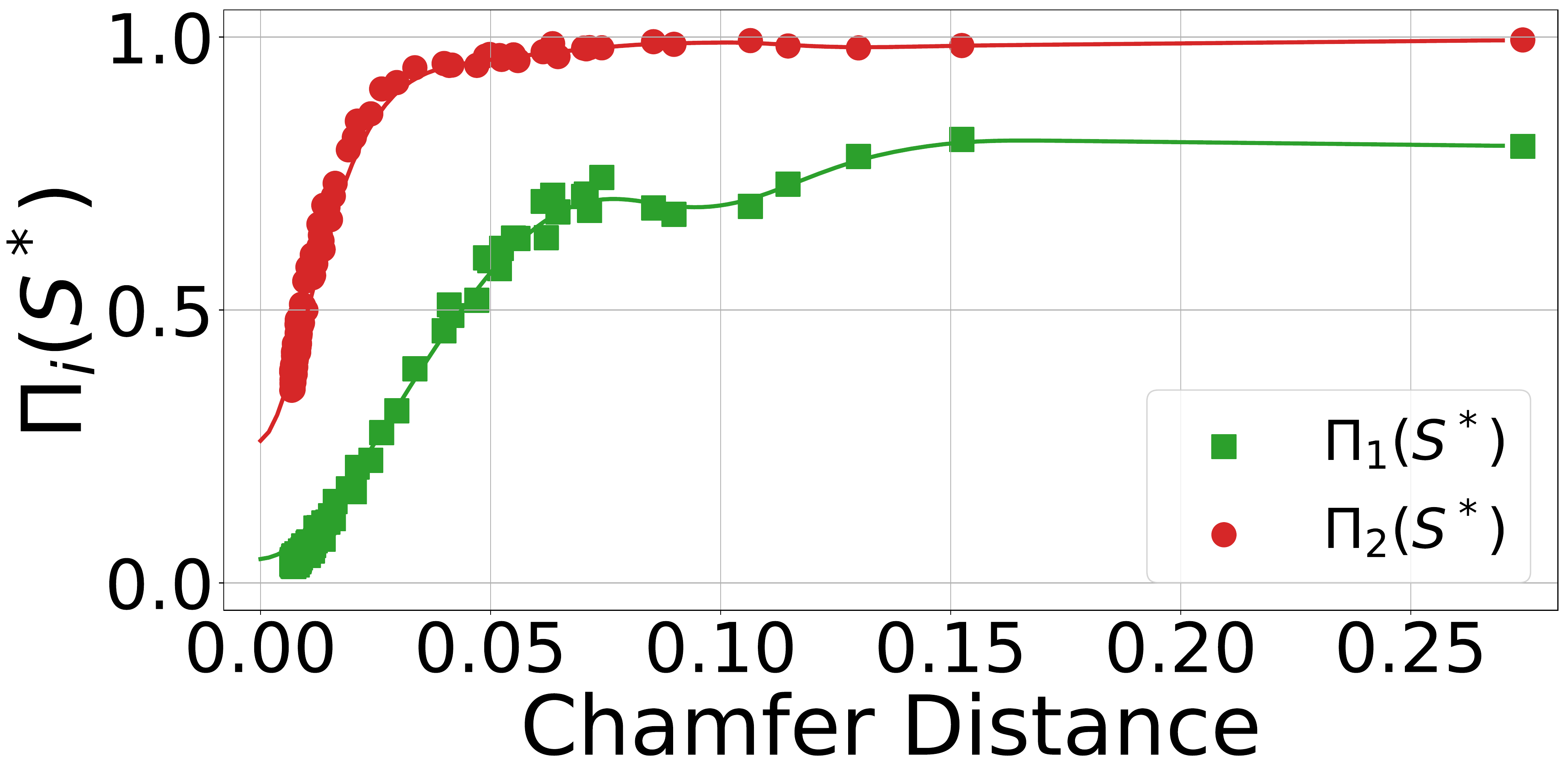}
        \caption{$\Pi$ vs Chamfer distance}
        \label{fig:pi_vs_c_d}
    \end{subfigure}%
    \vspace{-3mm}
    \caption{Privacy metrics $\Pi_1$ and $\Pi_2$ vs the 3 Utility metrics and Chamfer distance. Area under the plots indicates the effectiveness of the framework.}
    \label{fig:privacy-vs-utility-plots}
    \vspace{-5mm}
\end{figure}

Fig. \ref{fig:privacy-vs-utility-tradeoff} shows how the privacy metrics $\Pi _{1}$ and $\Pi _{2}$, and the utility metrics $Q_1$, $Q_{2:Static}$, and $Q_{2:Dynamic}$ vary with different privilege levels $(l)$ allowed by the privilege slider. This analysis is done over the test dataset explained in \S\ref{sec:dataset} using the Attacker $J_1$ which is trained on publicly available data, for hypothesis subset size=1.
Consistent with the results from previous sections (\S\ref{subsec:attacker-1-n-1},  \S\ref{subsec:all-attackers-n-1}, and \S\ref{subsec:utility-results}), the privacy metrics drop (with the super-class privacy dropping faster than the intra-class privacy) while the utility metrics increase drastically as the exposed privilege level $l$ is increased. Baseline privacy metrics $\Pi_{1:B}$ and $\Pi_{2:B}$ are shown in dashed lines for comparison.

At $l = 0.10$, a relatively high \textit{super-class} privacy, i.e. $\Pi_1 > 0.63$, can only be maintained at low utility values, i.e. $0.50<Q < 0.57$. On the other hand, the same relatively high \textit{intra-class} privacy, i.e. $\Pi_2 > 0.63$, can be maintained while providing potentially higher utility, i.e. $0.83 < Q < 0.87$, at $l =0.25$. 
For regenerations with $l \leq 0.25$, the lower bound of $\Pi_1$ is 0.09, and the lower bound of $\Pi_2$ is 0.63. It implies that for privilege levels $l \leq 0.25$, although there's a high chance of the object's \textit{super-class} label to be exposed, it's uniqueness within that super-class cannot be ascertained by an attacker.

In the medium privilege region (e.g, $l = 0.20$), privacy metric values $\Pi_1 = 0.21$ ($ 0.17$ improvement from baseline) and $\Pi_2 = 0.85$ ($0.68$ improvement from baseline) can be guaranteed for a utility $0.77 < Q < 0.81$.
Moreover, we can draw a ``goldilocks'' privilege level range of $0.17<l< 0.25$ which allows us to really ensure both high intra-class privacy, i.e. $\Pi_2>0.63$, and utility, i.e. $Q_1, Q_2 >0.70$.

\vspace{-2mm}\paragraph{AUC as privacy mechanism efficiency} Fig. \ref{fig:privacy-vs-utility-plots} shows the variation of the privacy metrics vs. different utility metrics including the Chamfer distance. The objective of the proposed privacy framework is to regenerate point clouds so that the utility is maximized (transformation error minimized) while the privacy against object inference/reidentification attacks is also maximized. For the case of \textit{Chamfer distance}, structual similarity of regenerations should be maximized by minimizing the \textit{Chamfer distance}, while maximizing ensured privacy. Hence, we formulate a generic measurement for the efficiency of any privacy mechanism in terms of the \textit{Area under the curve - AUC} of the Privacy-Utility graph. Note that the AUC measurement for \textit{chamfer distance} in Fig. \ref{fig:pi_vs_c_d} is normalized by the total area of the plot.

As shown in \autoref{tab:AuC}, using the AUC to measure the efficiency of our proposed mechanism, we can see that the AUC range is at the medium level with $0.49 < AUC < 0.64$ for super-class privacy, while the AUC range is at the high level with $0.84 < AUC < 0.89$ for intra-class privacy. 
This corroborates our earlier discussion that the proposed privacy framework can maintain high \textit{intra-class} privacy while ensuring high utility despite that an equally high \textit{super-class} privacy isn't guaranteed. 
Again, the privilege level range of $0.17 <l< 0.25$ ensures this performance as shown in Fig. \ref{fig:privacy-vs-utility-tradeoff}.


\begin{table}[t]
\caption{AUC Scores for Privacy-Utility curve}
\vspace{-2mm}
\label{tab:AuC}
\begin{tabular}{ccp{1cm}p{1.25cm}p{1.25cm}p{1cm}}
\toprule
&&\multicolumn{4}{c}{\centering\textbf{Utility or distance}} \\ \cline{3-6}
                                           &              & $Q_{1}$ & $Q_{2:Static}$ & $Q_{2:Dynamic}$ & CD   \\\midrule
\multirow{2}{*}{\textbf{AuC}}&$\Pi_1(S^*)$ & 0.55    & 0.54           & 0.49            & 0.64\\ 
&$\Pi_2(S^*)$ & 0.87    & 0.87           & 0.84            & 0.89 \\ \bottomrule
\end{tabular}
\vspace{-5mm}
\end{table}


\vspace{-1.5mm}\section{Limitations \& Future Work}\label{sec:future_work}

Despite not being able to ensure the super-class privacy, the proposed privacy framework can ensure relatively high intra-class privacy and utility, i.e. $\Pi_2 > 0.63, Q > 0.70$, at a privilege level range of $0.17 < l <0.25$. This is arguably acceptable as most non-critical MR functionalities may not necessarily require a unique object but only requires that the type or super-class of object be known. Moreover, since the unique identity (i.e. intra-class label) of the object is not exposed, the associated personally identifiable information with the unique object identity remains unexposed. Following the attacker descriptions from \S\ref{sec:Adversarial attacker realization/simulation}, as long as the user maintains a relatively low privilege level of $l < 0.23$, even with a high-privileged attacker like $J_4$, we can expect that the intra-class privacy of the regenerations are kept relatively high $\Pi_2 > 0.67$ with a high upper bound for both utilities $Q_1, Q_2 < 0.83$.

\vspace{-2mm}\paragraph{Temporal relationships of attacker success} So far, we have primarily quantified the object reidentification privacy in terms of an attacker formulating a hypothesis to match a query point cloud object $S^*$ with one or set of reference point cloud objects. Quantification was done based on the expected error in this hypothesis. On the other hand, another type of privacy threat could be posed in a situation where the attacker's hypothesis on the query doesn't correctly match with the corresponding original point cloud, but \textit{consistently} matches with the same mismatching class over a large number of $S^*$ queries from the same object. Here, although the object is not correctly reidentified, the attacker can generate a unique reference with respect to its previous queries and locate that object in a user's surrounding. A successful privacy-based regeneration framework should be able to regenerate the privacy preserved version of the same object with enough variation between each iteration such that the possible misclassifications at the attacker's end would not be consistent. We hope to do this analysis and evaluate the performance of our proposed framework in this context in future work.

\vspace{-2mm}\paragraph{Point cloud segmentation-then-transformation} Given a captured spatial mapping represented as a point cloud, relevant features in the point cloud needs to be detected and separated in order to process it further. For the proposed privacy framework, it is also necessary for the point cloud to be segmented and, then, separately transform the different objects to improve their privacy. In this work, we have directly explored and investigated the viability of the framework over already separated object point clouds. We intend to incorporate the actual segmentation task within a future iteration of this framework and apply the privacy transformations to these segmented point clouds.

\vspace{-1.5mm}\section{Conclusion}\label{sec:conclusion}




Mixed reality and 3D processing technologies are steadily developing in parallel: advances in MR aim to improve the environmental understanding of the various viable devices that can be used for MR, while advances in 3D, particularly in machine learning, enables better processing of 3D data (i.e. 3D segmentation, classification, and identification). The interplay of the two technologies allows the rapid development and delivery of better and immersive MR experiences. However, at the same time, there is a rapid increase in the level of privacy risks that MR users are exposed to. These 3D processing algorithms along their with ever improving accuracy and discriminative power can be utilized by adversaries to reveal personally identifiable information from the 3D data that is captured and collected by MR devices and platforms.

In this work, we leverage these state-of-the-art 3D algorithms (1) to design a privacy framework for regenerating 3D point cloud data, and (2) to evaluate the viability of this framework using the latest 3D neural network-based classifiers. 
With the proposed privacy framework, varying the privilege level allows us to variably transform and, hence, effectively control the privacy exposure of the shared 3D data to potentially malicious third party applications.
%
That is, as we increase the allowed privilege level $l$, the privacy $\Pi$ decreases, while the utility $Q$ increases. Nonetheless, the proposed privacy framework is still able to ensure a relatively high intra-class privacy and utility if the privilege level is kept within $0.17 <l<0.25$. Therefore, despite not being able to always ensure the super-class privacy of a 3D object, the proposed privacy-aware 3D regeneration can prevent unique intra-class reidentification of 3D objects while maintaining high utility for common MR functionalities.



\bibliographystyle{ACM-Reference-Format}
\bibliography{bib_all.bib}

\clearpage
\newpage
\appendix

\newpage

\section{Appendix}

\subsection{Privacy metric results for the top-1 success of Attacker $J_1$}

\begin{figure}[H]
    \centering
    \begin{subfigure}{0.495\columnwidth}
    \centering
        \includegraphics[width=0.95\linewidth]{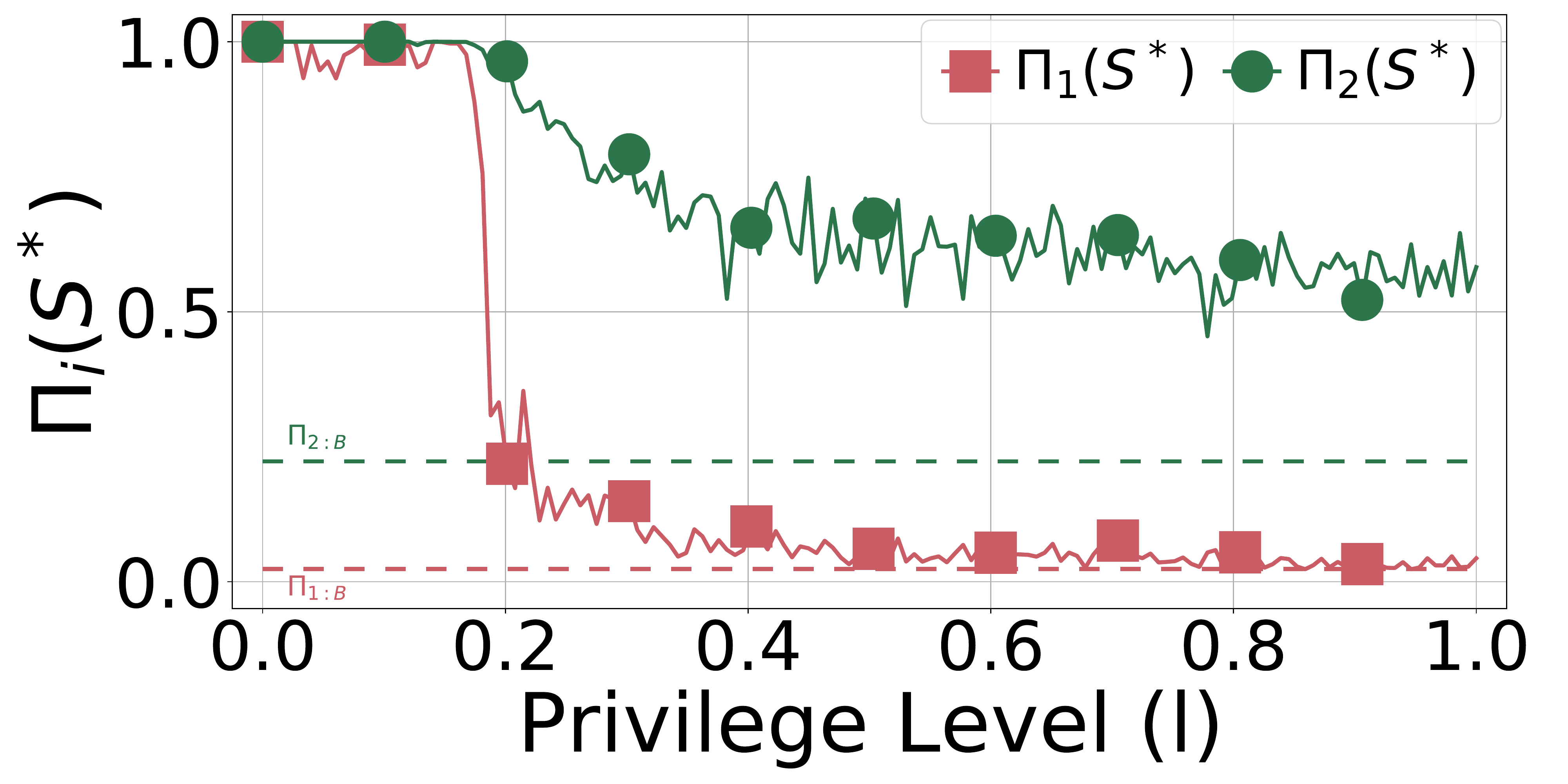}
        \caption{Bathtub}
    \end{subfigure}%
    \begin{subfigure}{0.495\columnwidth}
    \centering
        \includegraphics[width=0.95\linewidth]{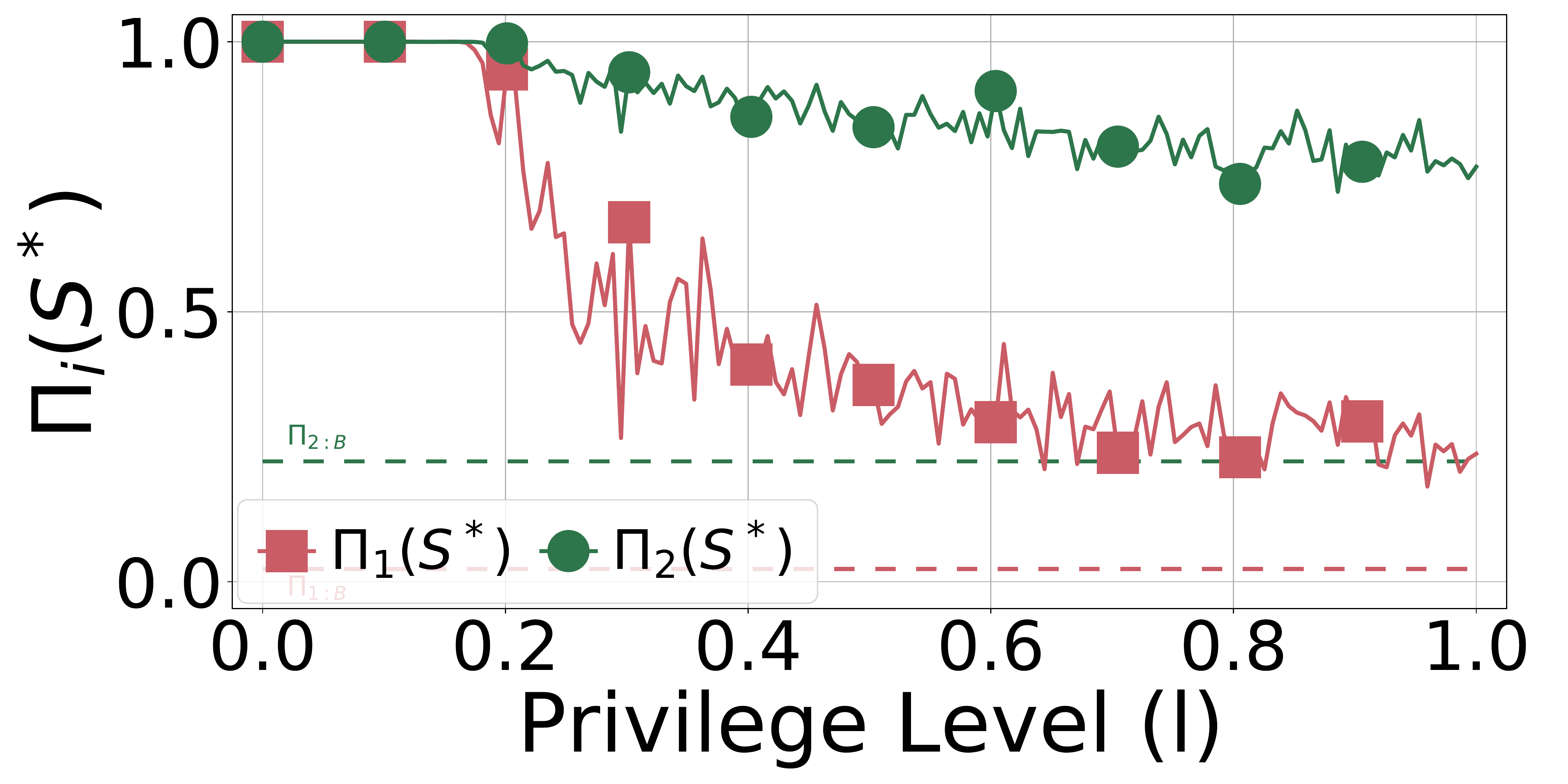}
        \caption{Bookshelf}
    \end{subfigure}%
    \par\centering
    \begin{subfigure}{0.495\columnwidth}
    \centering
        \includegraphics[width=0.95\linewidth]{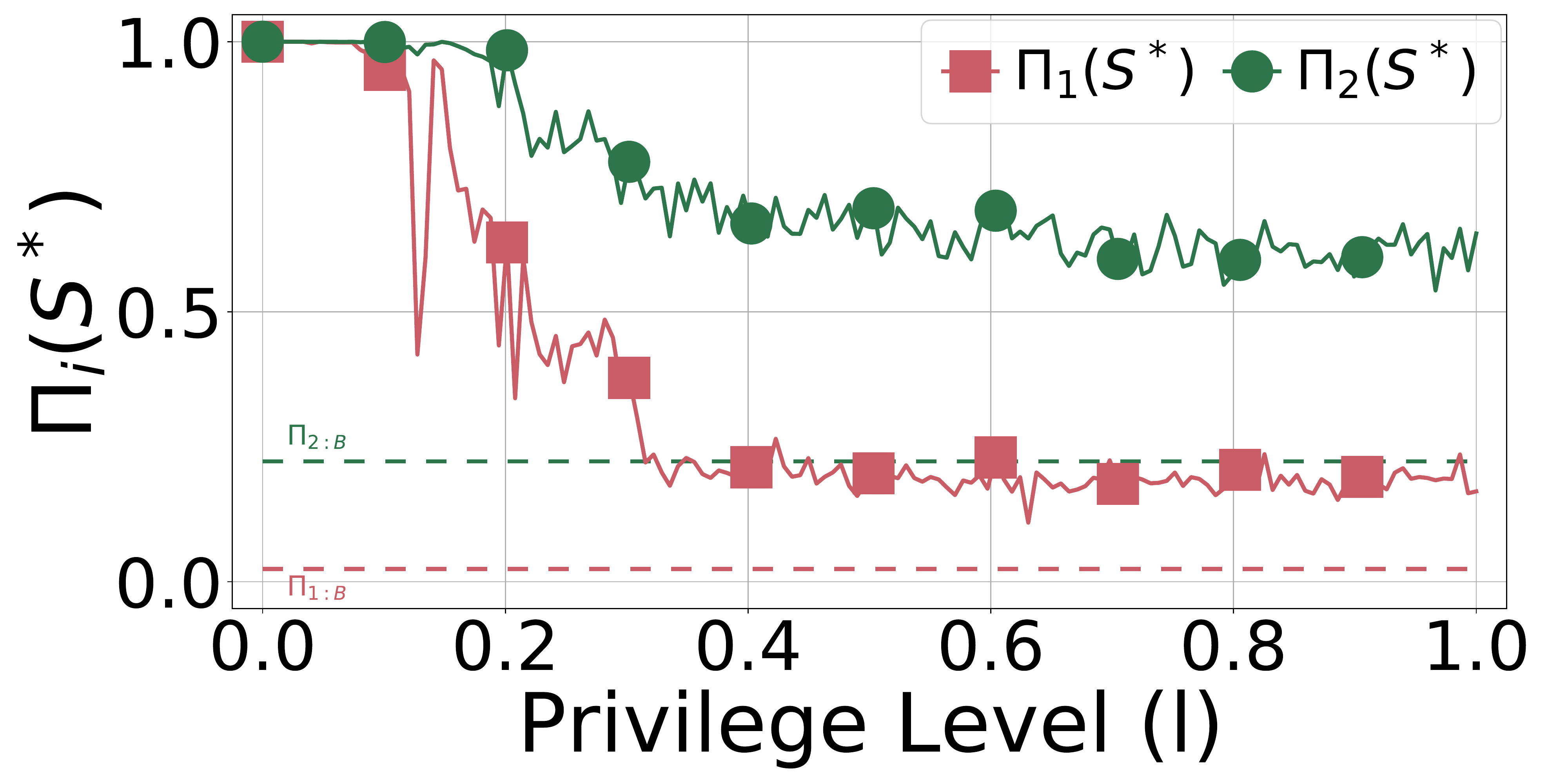}
        \caption{Cabinet}
    \end{subfigure}%
    \begin{subfigure}{0.495\columnwidth}
    \centering
        \includegraphics[width=0.95\linewidth]{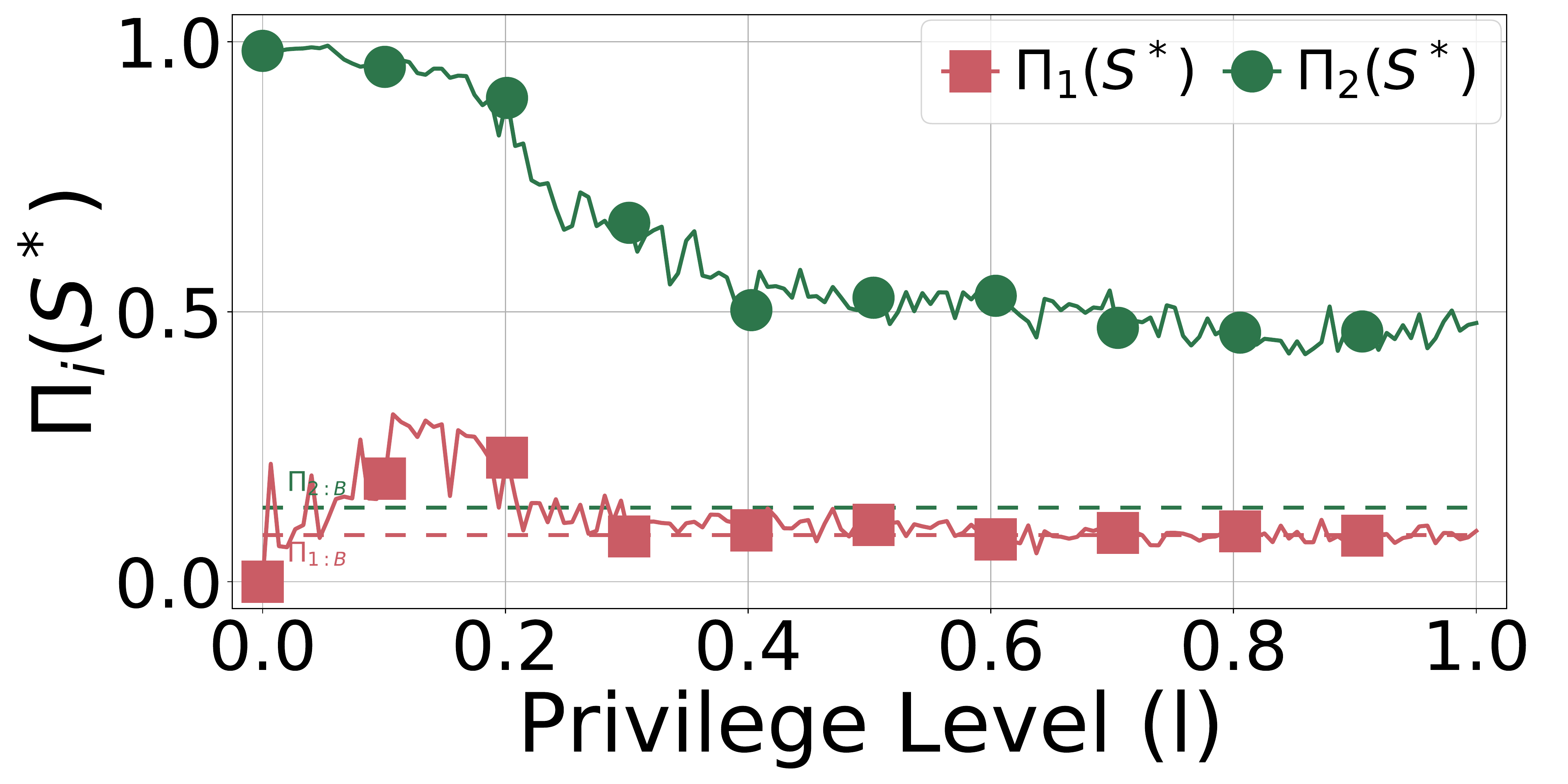}
        \caption{Lamp}
    \end{subfigure}%
    \par\centering
    \begin{subfigure}{0.495\columnwidth}
    \centering
        \includegraphics[width=0.95\linewidth]{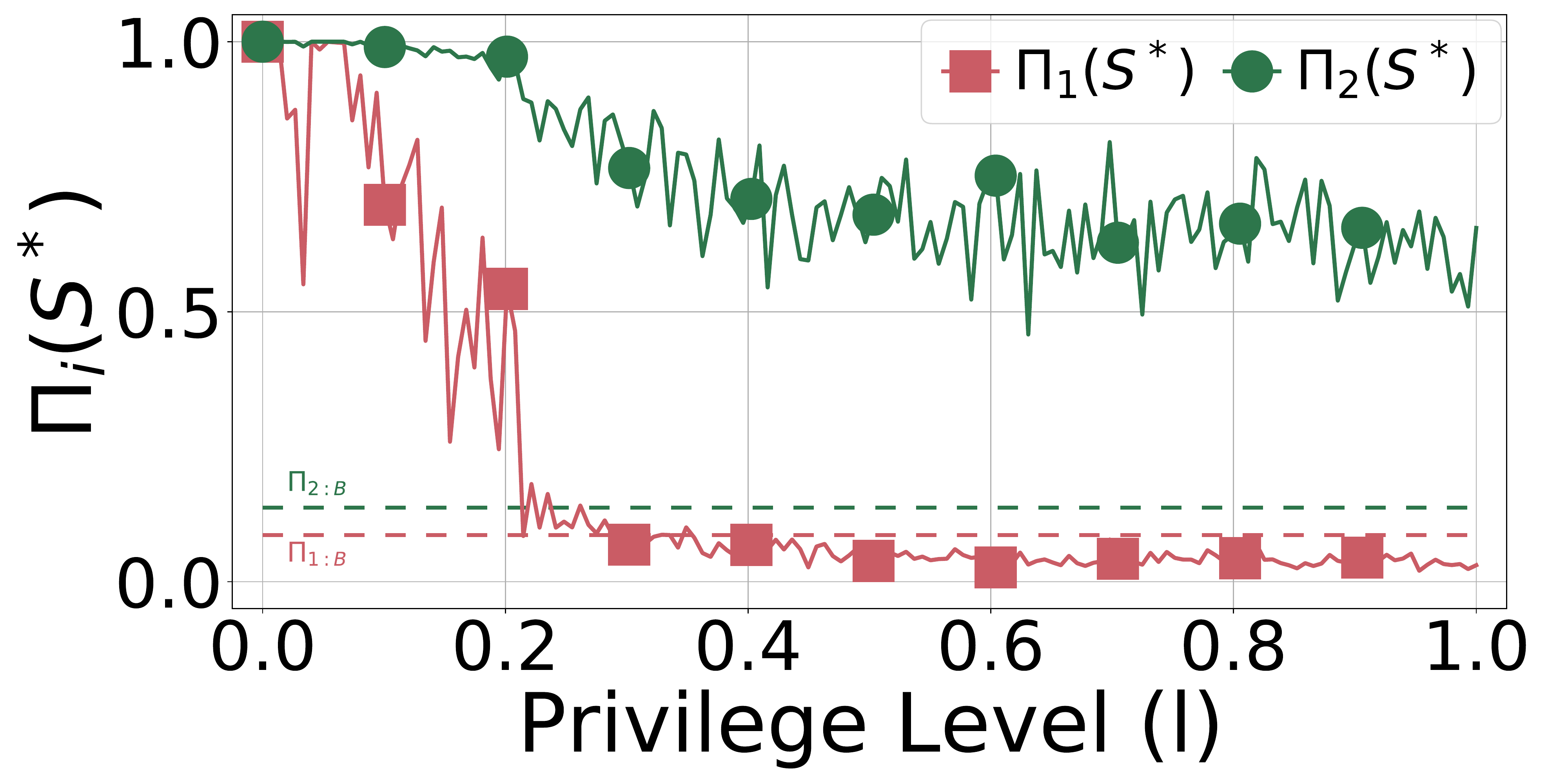}
        \caption{Monitor}
    \end{subfigure}%
    \begin{subfigure}{0.495\columnwidth}
    \centering
        \includegraphics[width=0.95\linewidth]{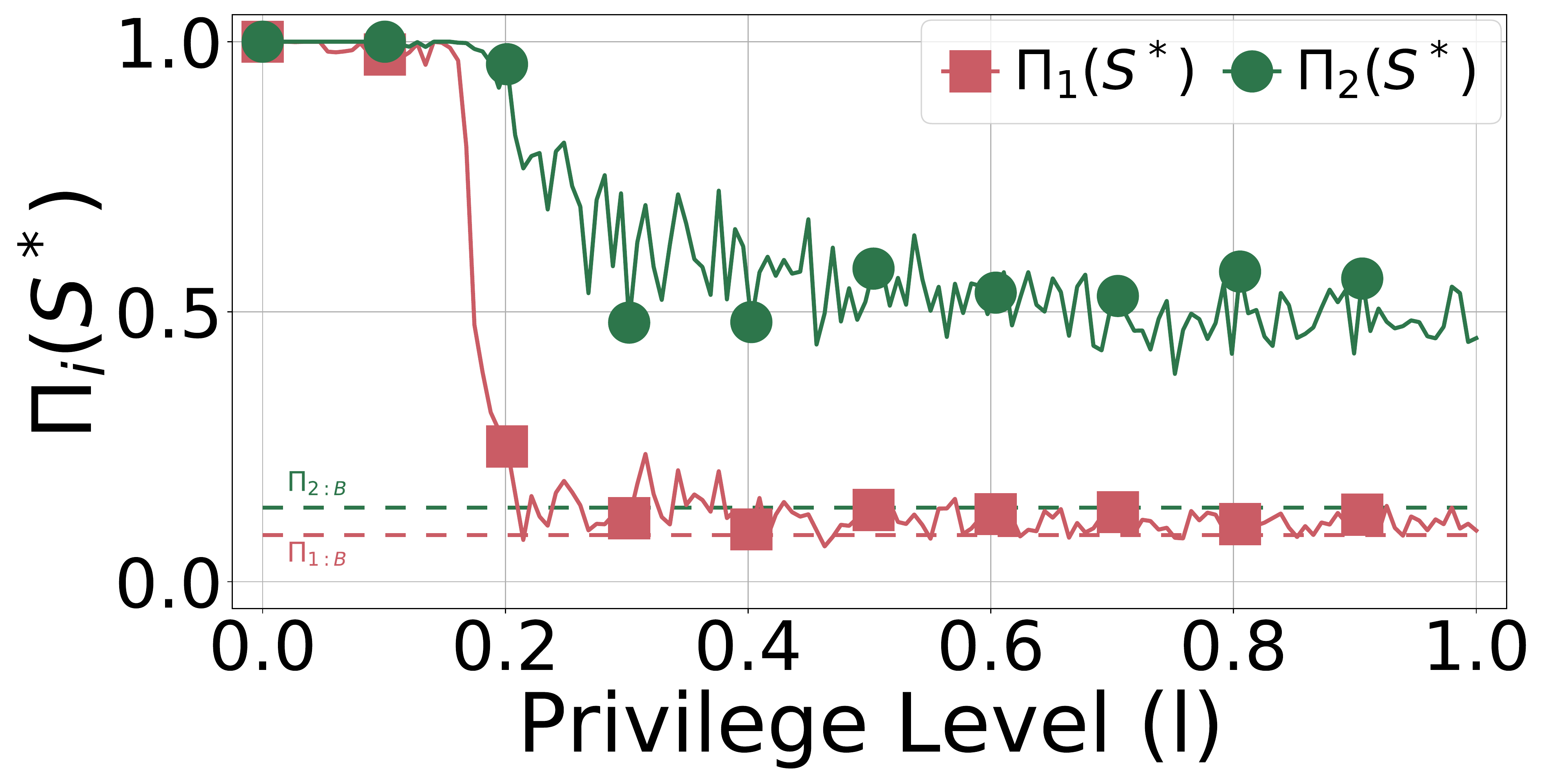}
        \caption{Sofa}
    \end{subfigure}%
    
    \caption{Super-class $\Pi_1$ and intra-class $\Pi_2$ privacy measures for the top-1 success of attacker 1 for some 3D objects: (a) bathtub, (b) bookshelf, (c) cabinet, (d) lamp, (e) monitor, and (f) sofa.}
    \label{fig:privacy-metric-n1-attacker1-appendix}
\end{figure}

For the completion of Fig. \ref{fig:privacy-metric-n1-attacker1}, here in Fig. \ref{fig:privacy-metric-n1-attacker1-appendix} we show the privacy metric results of Attacker $J_1$ top-1 case for the remaining 6 classes.

\subsection{Non-aggregated utility metrics for all super-classes}
\begin{figure}[H]
    \begin{subfigure}{0.5\columnwidth}
    \centering
        \includegraphics[width=0.95\linewidth]{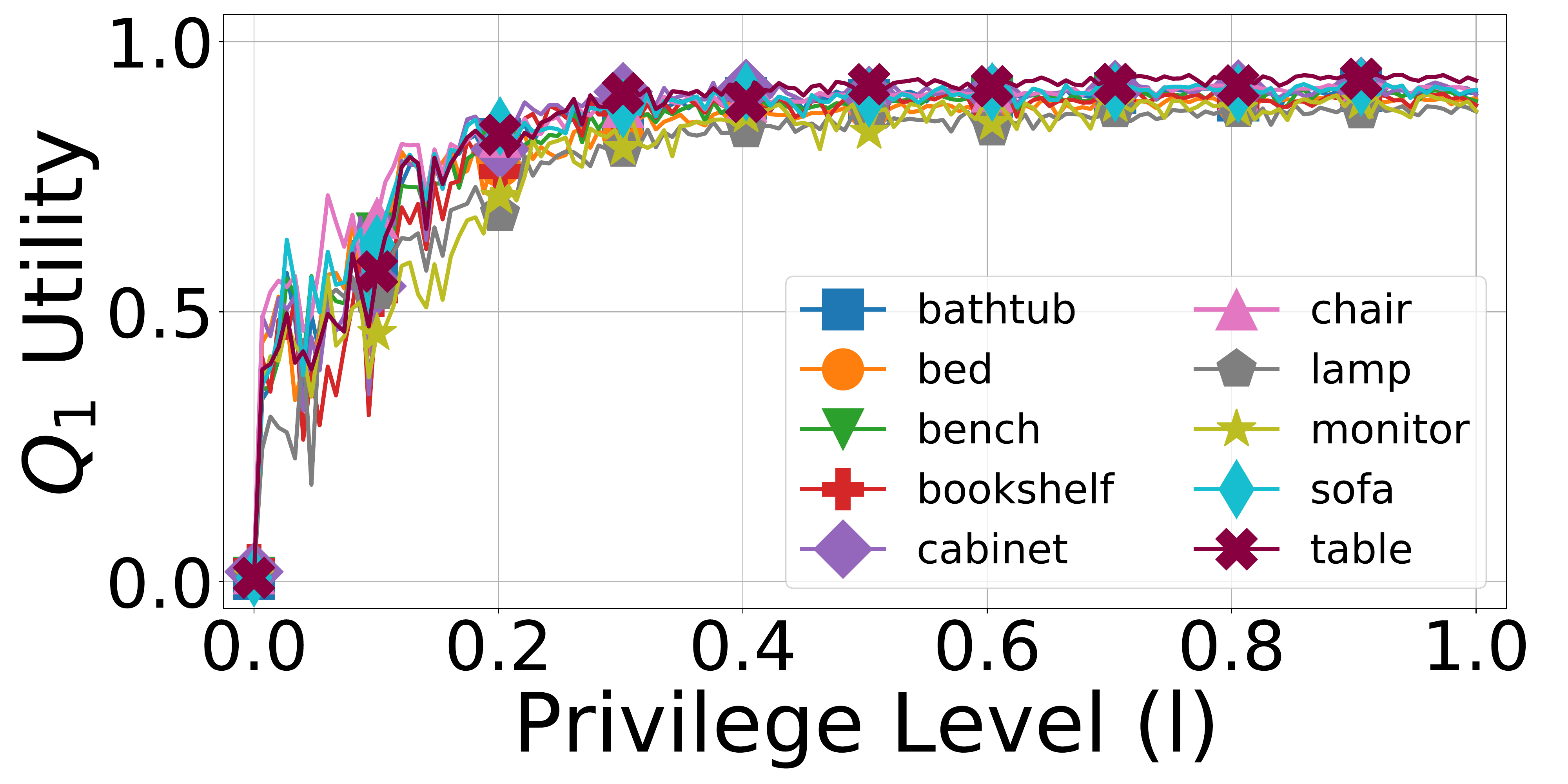}
        \caption{Bounding Box $\mathbf{Q_1}$}
    \end{subfigure}%
    \begin{subfigure}{0.5\columnwidth}
    \centering
        \includegraphics[width=0.95\linewidth]{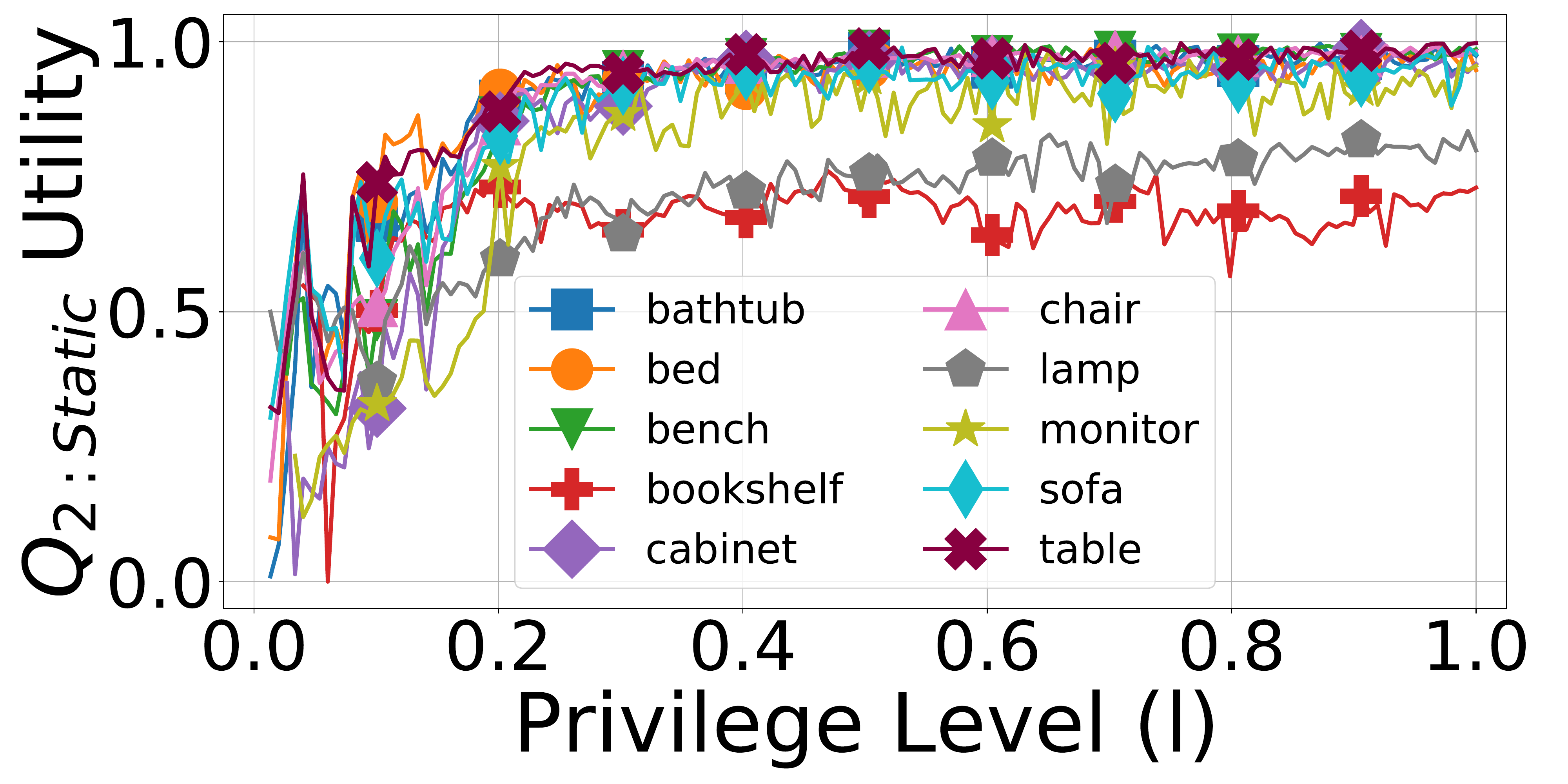}
        \caption{Static Anchoring $\mathbf{Q_2}$}
    \end{subfigure}%
    \par
    \begin{subfigure}{0.5\columnwidth}
    \centering
        \includegraphics[width=0.95\linewidth]{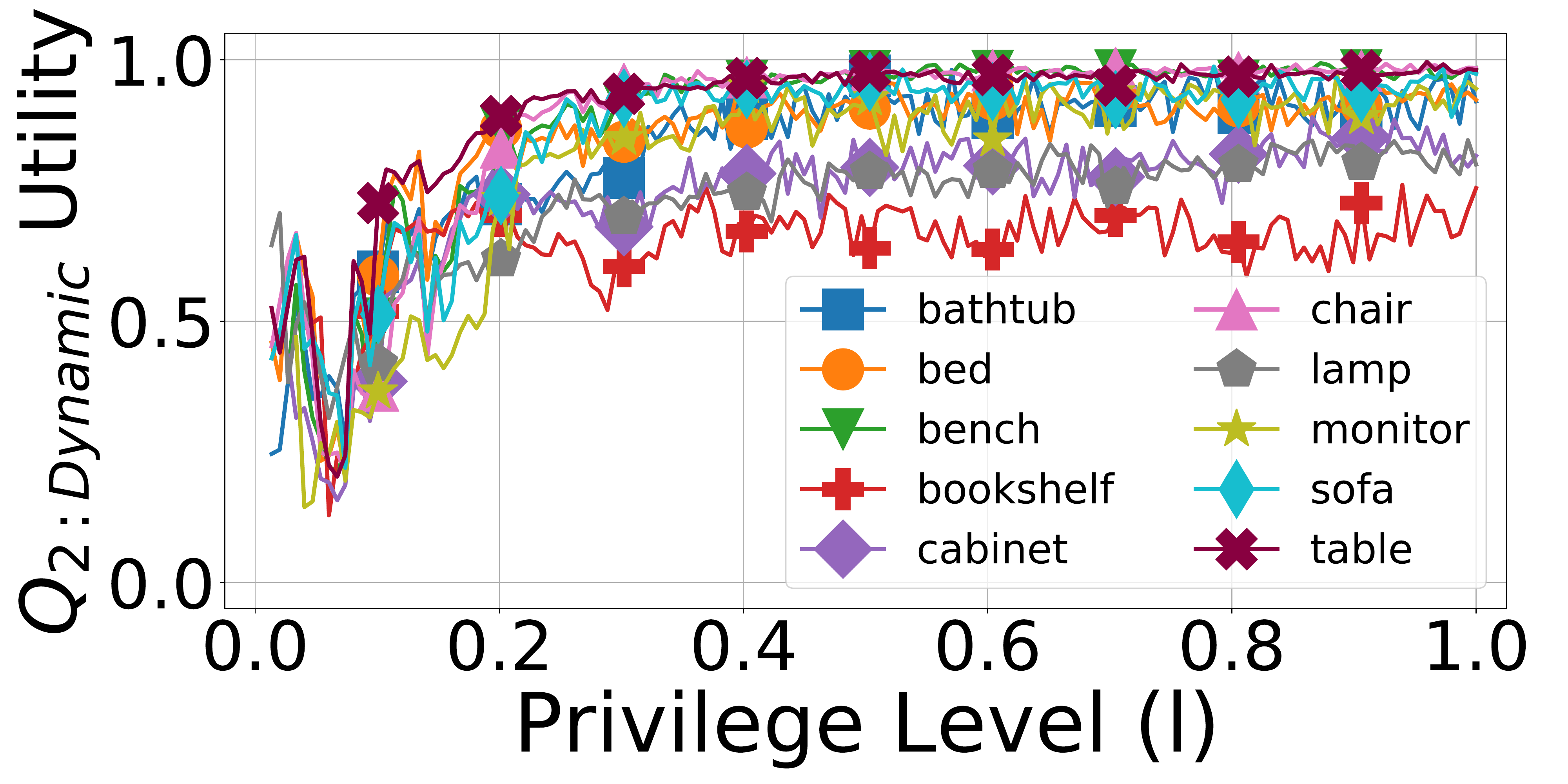}
        \caption{Dynamic Anchoring $\mathbf{Q_2}$}
    \end{subfigure}%
    \begin{subfigure}{0.5\columnwidth}
    \centering
        \includegraphics[width=0.95\linewidth]{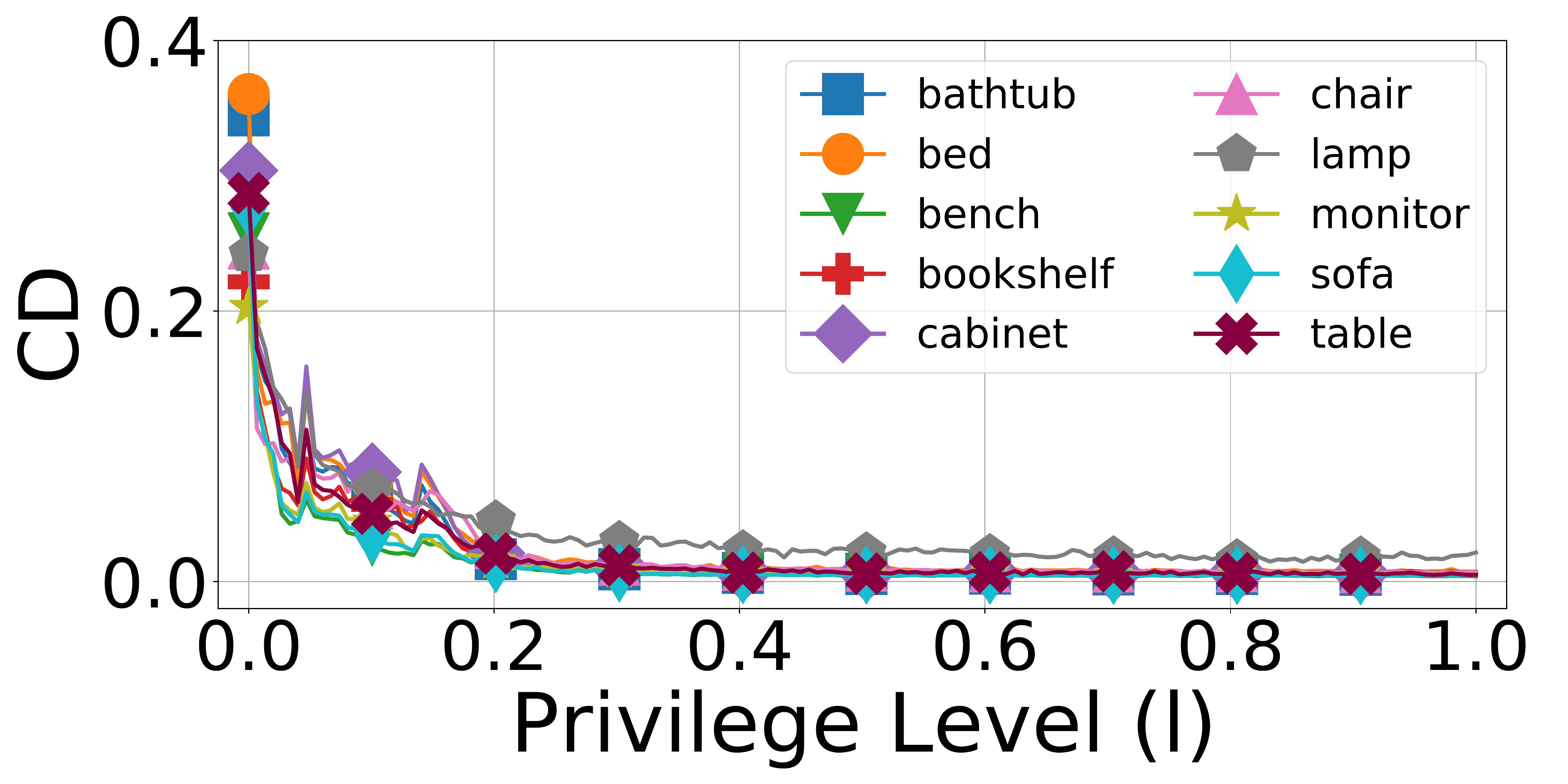}
        \caption{Chamfer Distance}
    \end{subfigure}%
    \caption{Utility Plots}
    \label{fig:util-non-agg}
\end{figure}

Fig.~\ref{fig:util-non-agg} shows all utility metrics including the Chamfer distance for every super-class.

\vspace{-4mm}
\begin{figure}[h]
    \begin{subfigure}{\columnwidth}
    \centering
        \includegraphics[width=0.7\columnwidth]{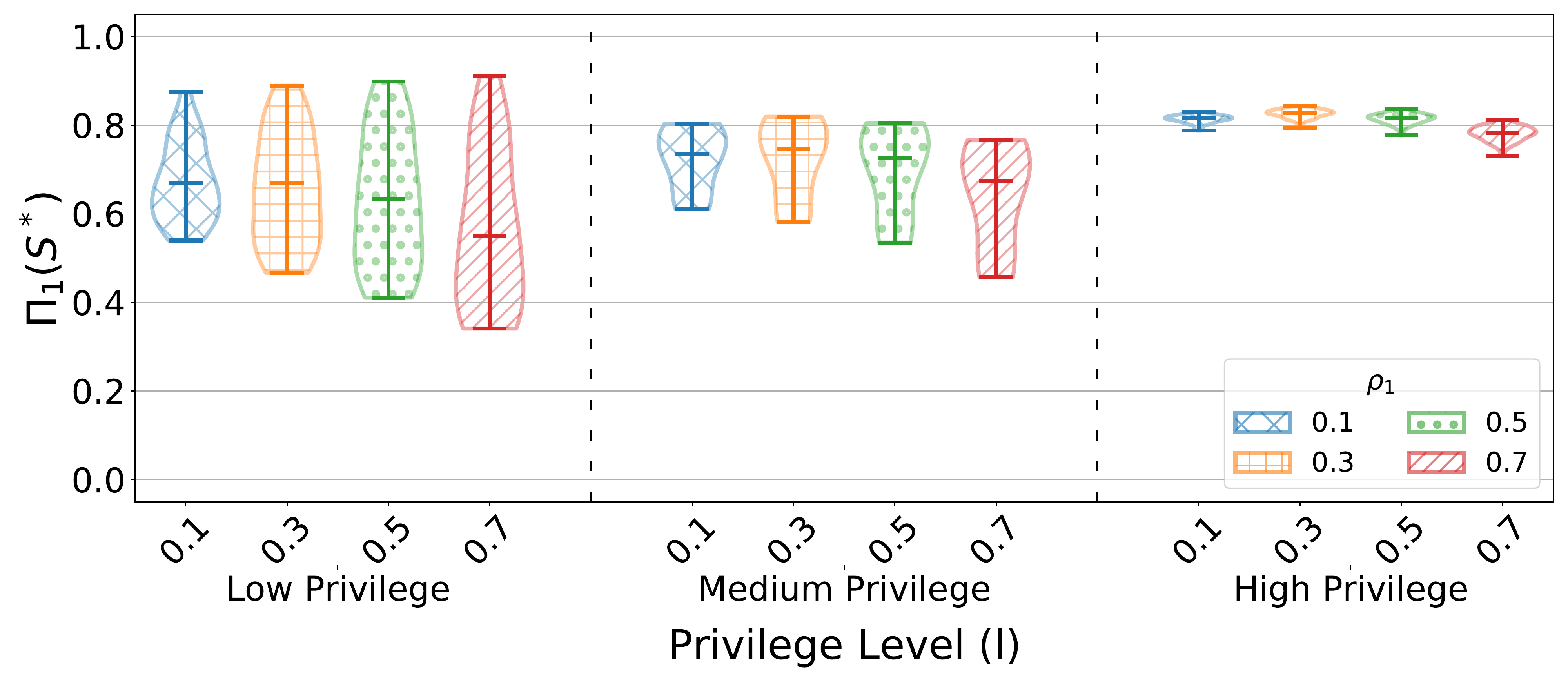}\vspace{-2mm}
        \caption{Super-class Privacy $\mathbf{\Pi_1(S^*)}$ with Attacker $J_2$}
    \end{subfigure}
    \begin{subfigure}{\columnwidth}
    \centering
        \includegraphics[width=0.7\columnwidth]{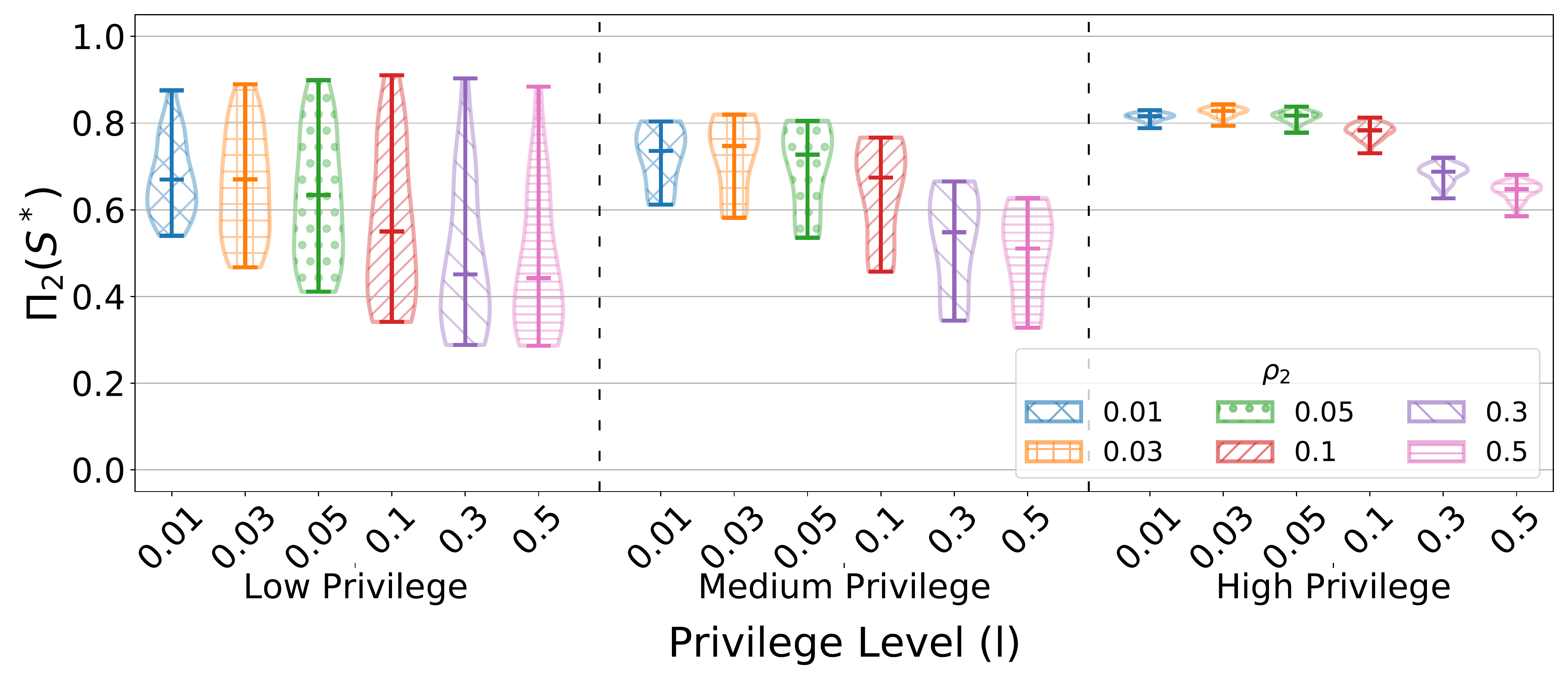}\vspace{-2mm}
        \caption{Intra-class Privacy $\mathbf{\Pi_2(S^*)}$ with Attacker $J_2$}
    \end{subfigure}
    

    \begin{subfigure}{\columnwidth}
    \centering
        \includegraphics[width=0.7\columnwidth]{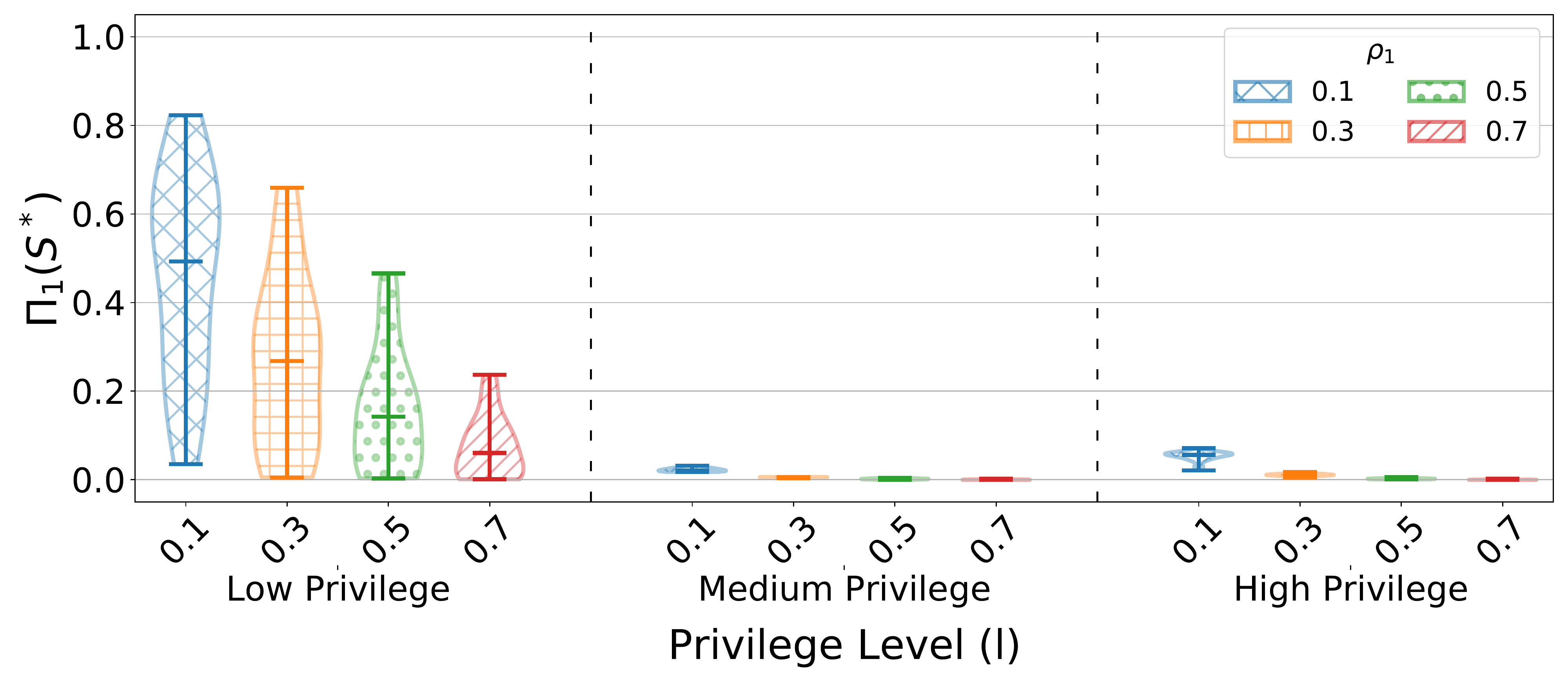}\vspace{-2mm}
        \caption{Super-class Privacy $\mathbf{\Pi_1(S^*)}$ with Attacker $J_3$}
    \end{subfigure}
    \begin{subfigure}{\columnwidth}
    \centering
        \includegraphics[width=0.7\columnwidth]{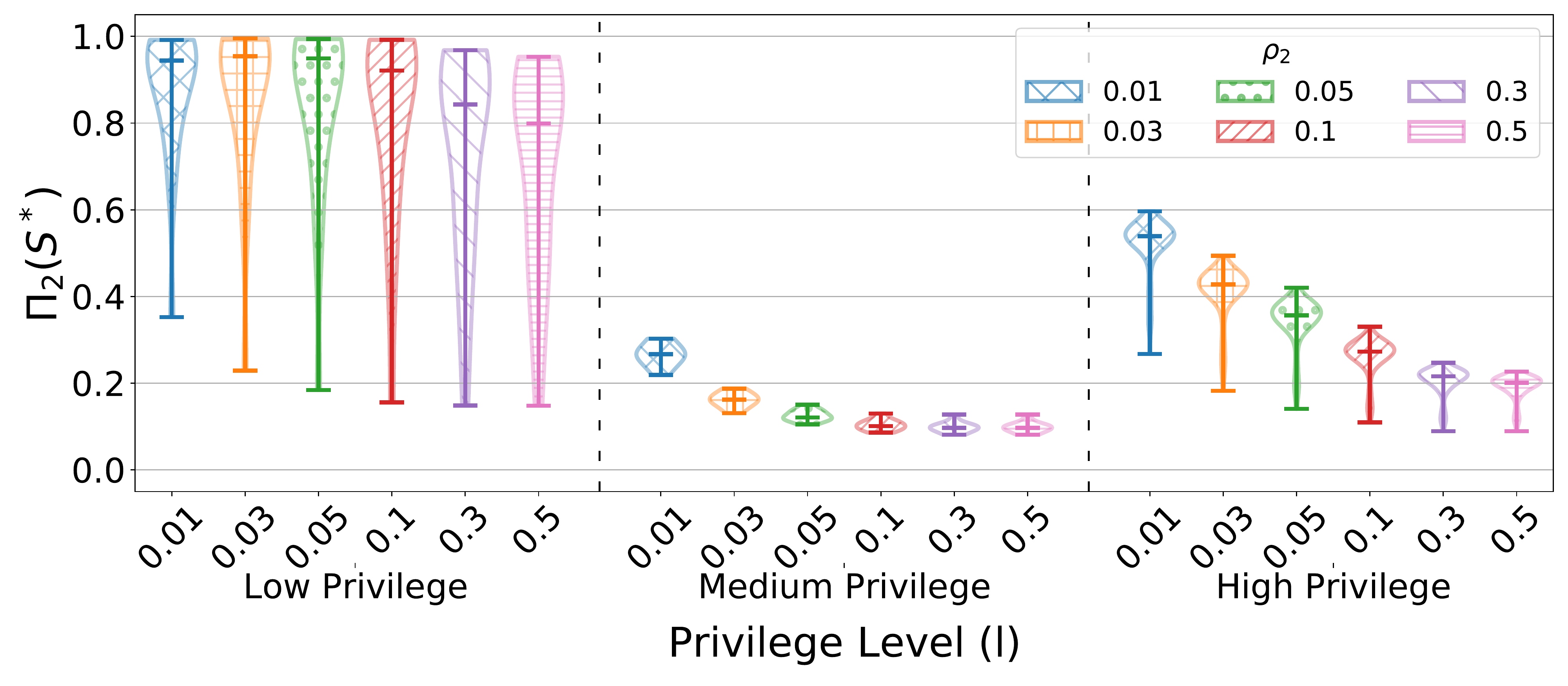}\vspace{-2mm}
        \caption{Intra-class Privacy $\mathbf{\Pi_2(S^*)}$ with Attacker $J_3$}
    \end{subfigure}

    \begin{subfigure}{\columnwidth}
    \centering
        \includegraphics[width=0.7\columnwidth]{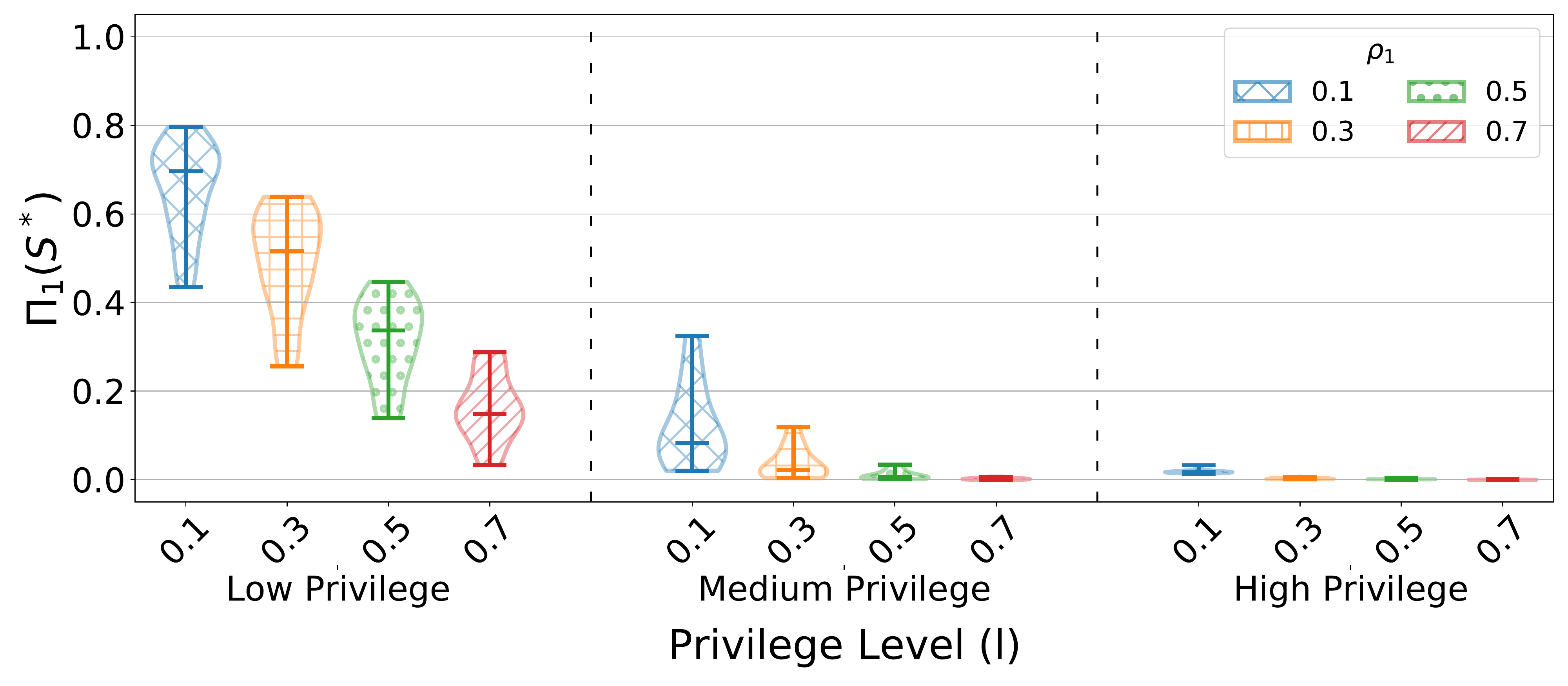}\vspace{-2mm}
        \caption{Super-class Privacy $\mathbf{\Pi_1(S^*)}$ with Attacker $J_4$}
    \end{subfigure}
    \begin{subfigure}{\columnwidth}
    \centering
        \includegraphics[width=0.7\columnwidth]{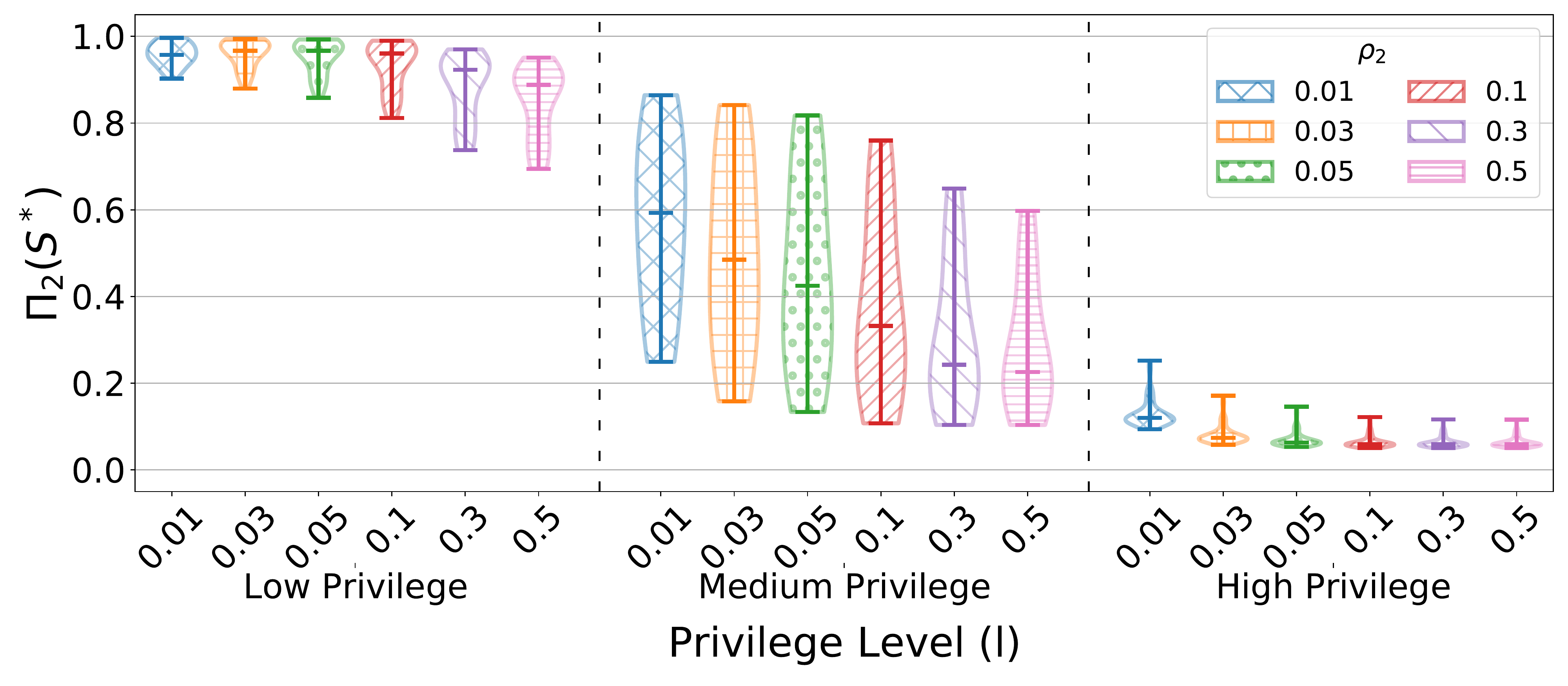}\vspace{-2mm}
        \caption{Intra-class Privacy $\mathbf{\Pi_2(S^*)}$ with Attacker $J_4$}
    \end{subfigure}\vspace{-3mm}
    \caption{$\Pi_1$ and $\Pi_2$ metrics aggregated at various hypothesis sizes for Attackers $J_2$, $J_3$ and $J_4$}\vspace{-3mm}
    \label{fig:priv-metric-vplot-attackers}
\end{figure}

\subsection{Privacy metric results for varying sizes of hypothesis sets for all the attackers}

In \S\ref{subsec:priv-vary-subset-size} (as shown in Fig.~\ref{fig:privacy-vs-subset-size}), we discussed the aggregated variations in privacy metric as we vary the size of hypothesis sets $\rho$ for Attacker $J_1$. In Fig. \ref{fig:priv-metric-vplot-attackers}, 
we display similarly aggregated $\Pi$ results for Attackers $J_2$, $J_3$ and $J_4$.

\begin{figure*}[t]
    \includegraphics[width=1\textwidth]{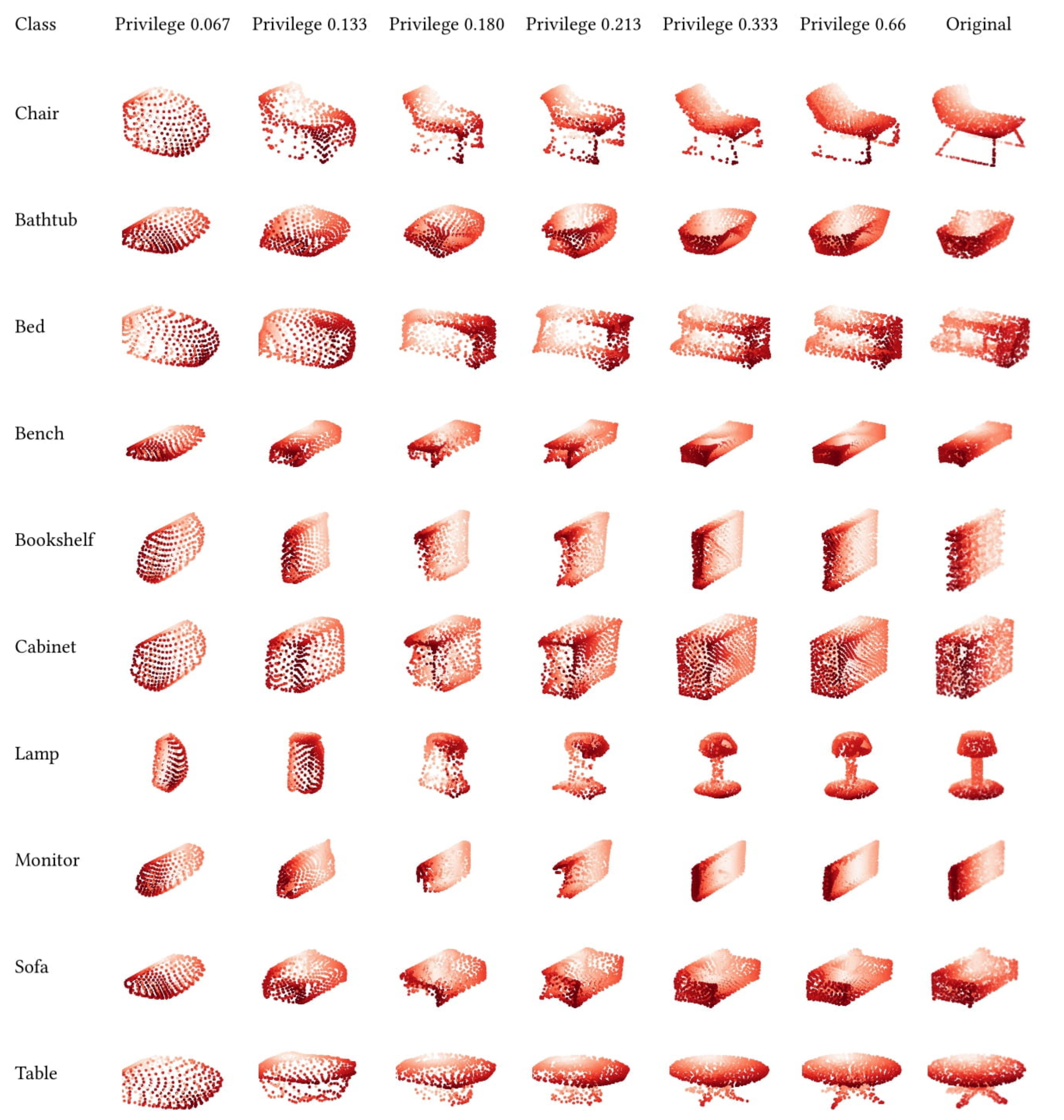}
    \caption{Sample 3D regenerations from each super-class at different privilege levels.}
    \label{fig:sample-regenerations}
\end{figure*}

\subsection{Example plots of 3D regenerations}
Fig.~\ref{fig:sample-regenerations} shows example 3D point clouds for all super-classes regenerated at different privilege levels.

\begin{figure*}[t]
    \includegraphics[width=0.9\textwidth]{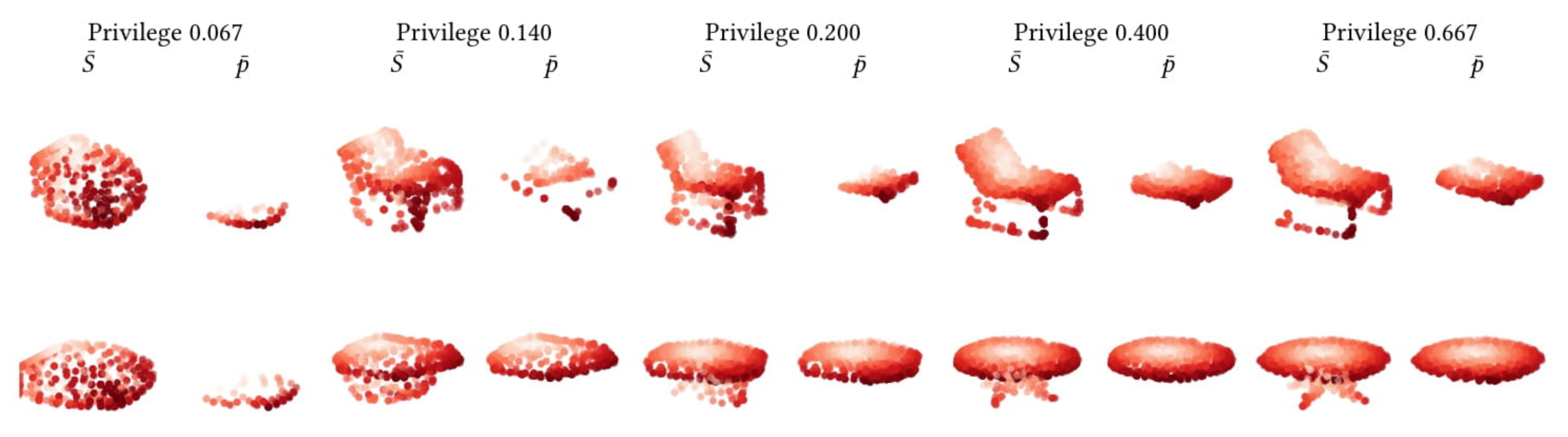}
    \caption{RANSAC sitting plane ($\bar{p}$) and regenerated point cloud ($\bar{S}$) for Chair and Table classes at different privilege levels.}
    \label{fig:ransac-examples}
\end{figure*}

\subsection{Example 3D plots for the RANSAC sitting plane for $Q_2$ utility}

Fig.~\ref{fig:ransac-examples} shows example 3D point clouds for the RANSAC-generated sitting plane at different privilege levels.



\end{document}